\documentclass[3p,twocolumn]{elsarticle}
\usepackage{xcolor}
\usepackage[utf8]{inputenc}
\usepackage{amsmath}
\usepackage{amsfonts}
\usepackage{array}
\usepackage{booktabs}
\usepackage{graphicx}
\usepackage{tabularx}
\usepackage{caption}
\usepackage{subcaption}
\usepackage{hyperref}

\usepackage{subfiles} 

    \newcommand{\paren}[1]{{\left( #1 \right)}}

\newcommand{\angbr}[1]{{\langle #1 \rangle}}

\newcommand{\partd}[2]{\frac{\partial{#1}}{\partial{#2}}}
\newcommand{\partdd}[3]{\frac{\partial^2{#1}}{\partial{#2}\partial{#3}}}

\newcommand{\partdds}[2]{\frac{\partial^2{#1}}{\partial{#2}^2}}

\newcommand{\const}[2]{{\paren{#1}_{#2}}}
\newcommand{\partdc}[3]{\const{ \partd{#1}{#2}}{#3} }
\newcommand{\partddc}[4]{\const{ \partdd{#1}{#2}{#3}}{#4}}
\newcommand{\partddsc}[3]{\const{ \partdds{#1}{#2}}{#3}}

\newcommand{\vfluct}[1]{{\angbr{#1^2}  - \angbr{#1}^2 }}
\newcommand{\xvfluct}[2]{{\angbr{#1#2}  - \angbr{#1}\angbr{#2} }}

\newcommand{\Nunit}{{N_{u}}}

\newcommand{\Lonsager}{{\pmb{L}}}
\newcommand{\Lonsagerij}{{L_{ij}}}
\newcommand{\Lonsagerji}{{L_{ji}}}
\newcommand{\Lonsagerijab}{{L_{ij,\alpha\beta}}}

\newcommand{\Traji}{{\Delta\vec{R}^{\zeta}_{i}}}

\newcommand{\Trajia}{{\Delta\vec{R}^{\zeta}_{i,\alpha}}}

\newcommand{\Trajjb}{{\Delta\vec{R}^{\zeta}_{j,\beta}}}

\begin{document}

\begin{frontmatter}

\title{CASM Monte Carlo: Calculations of the thermodynamic and kinetic properties of complex multicomponent crystals}

\author[add1]{Brian Puchala\corref{cor1}}
\address[add1]{Department of Materials Science and Engineering, University of Michigan Ann Arbor}
\ead{bpuchala@umich.edu}

\author[add2]{John C. Thomas\corref{cor1}}
\ead{johnct@umich.edu}

\author[add2]{Anton Van der Ven\corref{cor1}}
\address[add2]{Materials Department, University of California Santa Barbara}
\ead{avdv@ucsb.edu}

\cortext[cor1]{Corresponding author}

\date{September 20, 2023}

\begin{abstract}
Monte Carlo techniques play a central role in statistical mechanics approaches for connecting macroscopic thermodynamic and kinetic properties to the electronic structure of a material.
This paper describes the implementation of Monte Carlo techniques for the study multicomponent crystalline materials within the Clusters Approach to Statistical Mechanics (CASM) software suite, and demonstrates their use in model systems to calculate free energies and kinetic coefficients, study phase transitions, and construct first-principles based phase diagrams.
Many crystal structures are complex, with multiple sublattices occupied by differing sets of chemical species, along with the presence of vacancies or interstitial species.
This imposes constraints on concentration variables, the form of thermodynamic potentials, and the values of kinetic transport coefficients.
The framework used by CASM to formulate thermodynamic potentials and kinetic transport coefficients accounting for arbitrarily complex crystal structures is presented and demonstrated with examples applying it to crystal systems of increasing complexity. 
Additionally, a new software package is introduced, casm-flow, which helps automate the setup, submission, management, and analysis of Monte Carlo simulations performed using CASM.

\end{abstract}

\end{frontmatter}

\section{Introduction}
The thermodynamic and kinetic properties of a material determine how it behaves during heat treatments and when used as part of a device or a load-bearing structure. 
Thermodynamic and kinetic properties are also essential ingredients to meso-scale and continuum models that describe the temporal evolution of a material when taken out of thermal, chemical and/or mechanical equilibrium.\cite{cahn1961spinodal,hilliard1970spinodal,allen1979microscopic,voorhees1992ostwald,allen2005kinetics,chen2002phase,van2018first,teichert2019machine}
While intrinsic thermodynamic and kinetic properties are generally measured experimentally, they can also be calculated from first principles. 
A reliance on statistical mechanics is then essential due to the importance of entropy for most environmental conditions.\cite{de1994cluster,ceder1993derivation,van2002effect,van2018first,van2020rechargeable} 

Monte Carlo techniques \cite{landau2021guide,bishop2006pattern} play a central role in statistical mechanics schemes that seek to connect macroscopic thermodynamic and kinetic properties to the electronic structure of a material.\cite{wolverton1997ni,ozolicnvs1998cu,wolverton1998cation,kohan1998computation,wolverton2000short,wolverton2000first,barabash2008first,chan2008relative,hao2016quaternary,tepesch1996model,tepesch1995model,van1998first,han2005surface,zhou2006configurational,hinuma2007phase,das2017first,seko2006first,zhang2016cluster,barabash2008first,blum2004mixed,cao2018use,wen2021first,cordell2021probing,cordell2022simulation,burton2018first,zhang2019kinetically,kormann2017long,zarkevich2007first,van2005vacancies,van2008nondilute,puchala2013zro,Chen2015Calphad,thomas2013finite,belak2015effect,goiri2016phase,natarajan2016early,titus2016solute,gunda2018first,gunda2018resolving,kolli2018first,kolli2018controlling,natarajan2020crystallography,gunda2020understanding,bechtel2019finite,radin2020order,deng2020phase,van2008nondilute,bhattacharya2011first,goiri2019role,kolli2021elucidating,van2002self,van2002alloy,van2002first,adjaoud2009first,ravi2010cluster,ghosh2008first,gopal2012ab,nataraj2021systematic,zhu2023probing,kitchaev2018design,decolvenaere2019modeling,kitchaev2020mapping,zuo2021magnetoentropic,WOS:000610561300002,pilania2019distortion,aangqvist2017understanding,huang2017mechanism,aangqvist2019structurally,brorsson2021order,wang2020first,cheng2022atomic,wang2023generalization,cheng2023crystal}
They enable the numerical calculation of thermodynamic properties through the sampling of equilibrium atomic and/or electronic excitations.\cite{van2002self,zhou2006configurational,puchala2013zro,natarajan2017symmetry} 
Monte Carlo techniques are also invaluable in the calculation of kinetic properties. 
In crystalline materials where atomic hops are rare events, kinetic Monte Carlo simulations \cite{bortz1975new} can be used to calculate atomic transport coefficients in systems where the complexities due to correlated diffusion require simulated times that far exceed those typically possible using molecular dynamics.\cite{uebing1991monte,fichthorn1991theoretical,soisson1996monte,bulnes1998collective,belova2000collective,soisson2000monte,fichthorn2000island,van2001first,belova2001behaviour,tarasenko2001collective,van2005first,hartmann2005onsager,clouet2006kinetic,soisson2006kinetic,uberuaga2007defect,van2008nondilute,andersson2009role,van2010vacancy,soisson2010atomistic,bhattacharya2011first,van2013understanding,soisson2016radiation,goiri2019role,kolli2021elucidating,li2023predicting,abu2023barrier}

This paper describes implementations of Monte Carlo techniques within the Clusters Approach to Statistical Mechanics (CASM) software suite.\cite{puchala2023casm} 
The focus is on statistical mechanics schemes to calculate the thermodynamic and kinetic properties of multi-component crystals. 
The crystal structures of most materials are complex, often hosting multiple sublattices that each only accommodate a subset of chemical species.
These crystallographic complexities impose constraints on concentration variables and the form of thermodynamic potentials.\cite{larche1985overview} 
They also impose constraints on diffusion mechanisms and the mathematical form of diffusional flux expressions.\cite{cahn1983invariant,van2010vacancy} 
A generalized framework with which to formulate thermodynamic potentials and kinetic transport coefficients for arbitrarily complex crystal structures is developed. 
These quantities can be calculated from first principles using Monte Carlo sampling techniques. 
The approach relies on generalized cluster expansion Hamiltonians \cite{sanchez1984generalized,de1994cluster,drautz2004spin,van2018first,puchala2023casm} to interpolate expensive first-principles electronic structure calculations within Monte Carlo simulations of crystalline materials.

The paper is structured as follows. 
First, a general approach of tracking concentration within crystal structures that can host different sets of species on different sublattices is introduced. 
This is necessary to formulate semi-grand canonical free energies for arbitrarily complex crystal structures. 
The semi-grand canonical ensemble is especially convenient for Monte Carlo methods that calculate the thermodynamic properties of multi-component solids.\cite{van2002self,belak2015effect,goiri2019role} 
General statistical mechanics expressions are derived within the semi-grand canonical ensemble. 
The effect of crystallographic constraints on diffusional flux expressions within an arbitrarily complex crystal structure is derived next. 
This is followed by a brief overview of statistical mechanics principles that describe stochastic atomic hop events that are responsible for long-range diffusion in the crystalline state. 
A brief overview of cluster expansion techniques is provided to set the stage for illustrations of the types of results that can be calculated with Monte Carlo methods for a model system and a description of free energy integration techniques. 
The paper ends with a description of different Monte Carlo algorithms implemented within CASM and a summary of utilities to automate high throughput Monte Carlo simulations.

\section{Thermodynamics of alloyed crystals}
Thermodynamic descriptions of crystals require a careful consideration of the crystallographic constraints that limit the allowed variations in the concentration of the different chemical constituents of the crystal.\cite{larche1985overview,cahn1983invariant} 
This section introduces concentration variables that account for the crystallographic constraints that emerge when holding the number of crystal sites constant. 
Characteristic thermodynamic free energies are introduced next and a general definition of the semi-grand canonical free energy for an arbitrarily complex crystal is then formulated. 
Equations of state and response functions are derived from common characteristic potentials and the connection to the atomic and electronic scale is made using statistical mechanics.


\subsection{Concentration variables of crystals}
\label{sec:parametric_compositions}
\begin{figure}
    \centering
    \includegraphics[width=7cm]{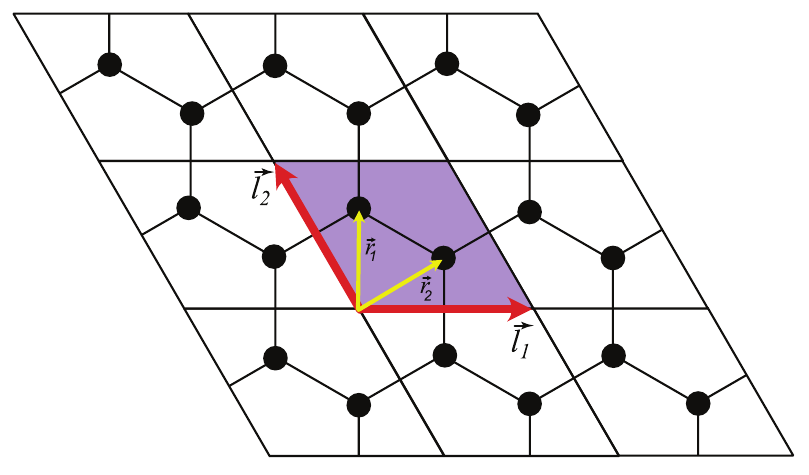}
    \caption{The unit cell of a 2-dimensional crystal, spanned by $\vec{l}_1$ and $\vec{l}_2$, and possessing basis atoms at positions $\vec{r}_1$ and $\vec{r}_2$.}
    \label{fig:crystal_basis}
\end{figure}

The structure of a crystal can be generated by periodically repeating a unit cell and a basis of atoms.
In three dimensions, the unit cell is defined by three vectors $\vec{l}_1$, $\vec{l}_2$ and $\vec{l}_3$. 
Integer linear combinations of the unit cell vectors generate the lattice of the crystal. 
The coordinates of the $n$ basis sites of the unit cell, denoted $\vec{r}_b$, $b=1,\dots,n$, can then be translated to each lattice site to form the full crystal. 
Each basis site within the unit cell defines a sublattice. 
Figure \ref{fig:crystal_basis} illustrates the unit cell of a 2-dimensional crystal, spanned by $\vec{l}_1$ and $\vec{l}_2$, and possessing basis atoms at positions $\vec{r}_1$ and $\vec{r}_2$. The resulting crystal is a two-dimensional honeycomb network. 

A crystal may host $s$ chemical species. 
The number of each chemical species, $i$, is tracked with a variable $N_i$. 
This quantity can be normalized by the number of unit cells in the crystal, $N_u$, yielding concentration variables $n_i=N_i/N_{u}$.
The concentrations of all chemical species in the crystal can be collected in the vector $\vec{n}^{\mathsf{T}}=[n_A,n_B,\dots]$.
In many crystals, different chemical species may segregate to only a subset of sublattices. 
For example, an oxynitride may consist of sublattices that host different transition metal cations and a separate set of sublattices that host oxygen and nitrogen anions. 
It is also common that a subset of sublattices host vacancies in appreciable numbers.

In many applications, it is convenient to define thermodynamic and kinetic quantities for a crystal with a fixed number of unit cells, $N_u$. 
This is a relevant thermodynamic boundary condition for many experimental situations where a solid maintains its crystal structure while undergoing an internal redistribution of chemical species through diffusional processes.\cite{van2010vacancy} 
Often thermodynamic quantities are normalized by the number of unit cells (or equivalently the number of sites) in a crystal.
A constant number of unit cells is also a common boundary condition in Monte Carlo simulations. 
The constraints that emerge when holding $N_u$ constant play an important role in determining the form of characteristic thermodynamic potentials and of diffusional flux expressions, as will be described in subsequent sections. 

\begin{figure}
    \centering
    \includegraphics[width=7cm]{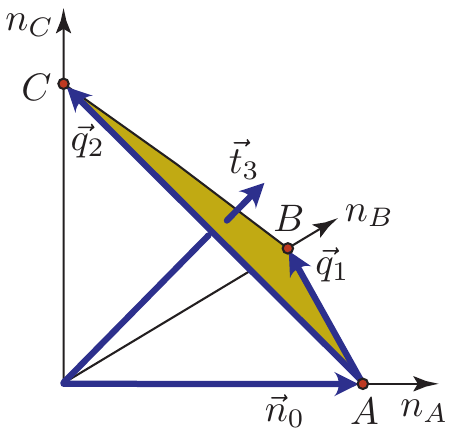}
    \caption{The space of allowed compositions in a ternary A-B-C alloy having a one-atom basis, spanned by vectors $\vec{q}_1$ and $\vec{q}_2$ that have their origin at $\vec{n}_0$. The vector $\vec{t}_3$ spans the null of the allowed composition space.}
    \label{fig:ternary_composition_axes}
\end{figure}

The concentration variables $n_{i}$, $i=1,\dots,s$ cannot be varied independently of each other in a crystal with a fixed number of unit cells, $N_u$.
This is illustrated for a ternary A-B-C alloy having a crystal with a one-atom basis (e.g., fcc or bcc). 
While the concentration variables $\vec{n}^{\mathsf{T}}=[n_{A},n_{B},n_{C}]$ reside in an $s=3$ dimensional space, their allowed variations in a crystal with a fixed number of unit cells are restricted to a two-dimensional space spanned by the vectors $\vec{q}_1$ and $\vec{q}_2$ that have their origin at $\vec{n}_0$.
This is illustrated in Figure \ref{fig:ternary_composition_axes} for a ternary alloy having a one atom unit cell. 
In Figure \ref{fig:ternary_composition_axes}, the chosen origin is the crystal in which all sites are occupied by A atoms. 
The spanning vectors $\vec{q}_1$ and $\vec{q}_2$ represent crystal preserving exchanges of A atoms with B atoms and A atoms with C atoms, respectively

In general, the subspace of allowed compositions for a crystal with a constant number of unit cells can be described mathematically as
\begin{equation}
    \vec{n}=\vec{n}_0+\sum_{i=1}^{k}x_i\vec{q}_i
    \label{eq:concentration_equation1}
\end{equation}
where $\vec{n}_0$ points to a chosen origin in the subspace. 
The $k$ variables $x_i$ are referred to as parametric concentrations and can be varied independently of each other.
By collecting the spanning vectors $\vec{q}_i$, $i=1,\dots,k$ as a $s\times k$ matrix $\pmb{Q}=[\vec{q}_1,\dots,\vec{q}_k]$, Eq. \ref{eq:concentration_equation1} can be expressed more compactly as
\begin{equation}
    \vec{n}=\vec{n}_0+\pmb{Q}\vec{x},
    \label{eq:concentration_equation2}
\end{equation}
where the parametric concentrations are collected as a $k$-dimensional vector $\vec{x}^{\mathsf{T}}=[x_1,\dots,x_{k}]$, which can be called the parametric composition.

In general, vectors that span the subspace of allowed compositions when holding $N_u$ constant do not form an orthonormal set. 
It will be useful to introduce the dual spanning basis, $\vec{r}_i$, which satisfy $\vec{r}^{\mathsf{T}}_i\vec{q}_{j}=\delta_{i,j}$, where $\delta_{i,j}$ is the Kronecker delta. By collecting these vectors in the $s\times k$ matrix $\pmb{R}=[\vec{r}_1,\dots,\vec{r}_k]$, the following identities hold $\pmb{R}^{\mathsf{T}}\pmb{Q}$=$\pmb{Q}^{\mathsf{T}}\pmb{R}$ = $\pmb{I}$, where $\pmb{I}$ is a $k\times k$ identity matrix. 
These vectors relate the parametric composition $\vec{x}$ to the concentration variables per unit cell according to
\begin{equation}
    \vec{x}=\pmb{R}^{\mathsf{T}}(\vec{n}-\vec{n}_0)
    \label{eq:unit_cell_to_parametric_composition}
\end{equation}

The composition vector $\vec{n}$ in a crystal with fixed $N_u$ is restricted to a $k$-dimensional subspace of the full $s$ dimensional space of concentration variables $[n_1,\dots,n_{s}]$.
The subspace that is orthogonal to the $k$ dimensional subspace of allowed compositions also plays an important role in the thermodynamic and kinetic formalism to be developed in the following sections. 
This subspace can be spanned with a set of $s-k$ orthonormal vectors $\vec{t}_{k+1},\dots,\vec{t}_{s}$. 
For the ternary A-B-C alloy of Figure \ref{fig:ternary_composition_axes}, this is a one dimensional space spanned by $\vec{t}_3^{\mathsf{T}}=(1/\sqrt{3})[1,1,1]$.
To ensure that $\vec{n}$ resides within the fixed crystal subspace, it is necessary for $\vec{t}^{\mathsf{T}}_j(\vec{n}-\vec{n}_0)=0$ for $j=k+1,\dots,s$. 
By collecting the $\vec{t}_{j}$ in the $s\times(s-k)$ dimensional matrix $\pmb{T}=[\vec{t}_{k+1},\dots,\vec{t}_{s}]$, these orthogonality criteria can be expressed compactly as
\begin{equation}
    \pmb{T}^{\mathsf{T}}(\vec{n}-\vec{n}_0)=\vec{0}
\end{equation}
where $\vec{0}$ is a $s-k$ vector of zeros. 
The appendix illustrates these concepts for a variety of multi-component crystals that are more complicated than the ternary alloy having a crystal with one basis site.


\subsection{Characteristic thermodynamic potentials}
\label{sec:characteristic_potentials}
The characteristic thermodynamic potential, often referred to as the free energy, embeds all the thermodynamic information about a system in the form of first and second derivatives. 
Furthermore, its minimum with respect to internal degrees of freedom determines the equilibrium state of the system, thereby providing information about the direction of spontaneous processes of unstable and metastable states. 

The specific form of the free energy is determined by the imposed thermodynamic boundary conditions.
Thermodynamic boundary conditions specify, for each pair of conjugate state variables, whether the extensive variable or the conjugate intensive variable is held constant. 
For example, if the volume of a system is held constant, then the system in equilibrium will adopt a particular pressure that can be measured.
The volume is then the control, or \emph{natural variable}. 
If however, the pressure is controlled and the volume is measured, then the pressure is the natural variable.

The characteristic free energy can be obtained by applying a Legendre transform to the internal energy, $U$, of the system for each intensive natural variable that is controlled as a thermodynamic boundary condition.\cite{Callen} 
When the only intensive natural variable is the temperature, $T$, and consequently all extensive variables except entropy, $S$, are also controlled, the characteristic potential is the Helmholtz free energy
\begin{equation}
F=U-TS.
\label{eq:helmholtz_free_energy}
\end{equation}
When the pressure, $P$, is a natural variable in addition to temperature, the characteristic potential is the Gibbs free energy
\begin{equation}
G=U-TS+PV.
\label{eq:gibbs_free_energy}
\end{equation}
Another common thermodynamic boundary condition occurs when the chemical potential of one of the $s$ chemical species is held constant, in addition to temperature and pressure. 
For example, the chemical potential of Li within an intercalation compound such as Li$_x$CoO$_2$ can be controlled in an electrochemical cell by fixing the open cell voltage.\cite{van2020rechargeable} 
The characteristic potential for this boundary condition is a grand canonical free energy
\begin{equation}
    \Lambda=U-TS+PV-\mu_{Li}N_{Li}
\end{equation}
The general rule to identify the characteristic free energy for a given set of boundary conditions is that a Legendre transform is applied to the internal energy for each intensive variable that is controlled. 

An important free energy for a crystal with a fixed number of unit cells is referred to as the semi-grand canonical potential \cite{van2002self,belak2015effect}. 
For arbitrarily complex, multi-component crystals, we generalize the definition of the semi-grand canonical potential as the characteristic potential whose intensive natural variables are conjugate to the parametric concentrations, $\vec{x}^{\mathsf{T}}=[x_1,\dots,x_{k}]$.
The parametric compositions are the independent composition variables that satisfy the constraints imposed upon fixing the number of unit cells of the crystal. 
The mathematical form of the semi-grand canonical potential can be identified by starting with Euler's theorem, which upon rearranging terms relates the Gibbs free energy, under conditions of constant temperature and pressure, to the sum of chemical potentials times the number of atoms of each species $i$ according to
\begin{equation}
    G=\sum^{s}_{i=1}\mu_i N_i = N_u\vec{\mu}^{T}\vec{n}
    \label{eq:Euler1}
\end{equation}
where $N_i=N_{u}n_i$ was used to obtain the second equality and where $\vec{\mu}^{\mathsf{T}}=[\mu_1,\dots,\mu_s]$. 
Inserting Eq. \ref{eq:concentration_equation2} into Eq. \ref{eq:Euler1}, yields
\begin{equation}
    G= N_u\vec{\mu}^{T}(\vec{n}_0+\pmb{Q}\vec{x})=N_u\left(\vec{\mu}^{T}\vec{n}_0+\vec{\tilde{\mu}}^{T}\vec{x}\right)
    \label{eq:Euler2}
\end{equation}
%
%
where 
\begin{equation}
    \vec{\tilde{\mu}}=(\pmb{Q}^{T}\vec{\mu})
    \label{eq:exchange_chemical_potentials}
\end{equation}
 are referred to as exchange chemical potentials. 
Equation \ref{eq:Euler2} shows that the intensive variables that are conjugate to $N_u\vec{x}$ are the elements of the exchange chemical potential vector $\vec{\tilde{\mu}}$.
The semi-grand canonical free energy is therefore defined as

\begin{equation}
    \Phi=G-N_u\vec{\tilde{\mu}}^{T}\vec{x} = N_u\vec{\mu}^{T}\vec{n}_0
    \label{eq:semi-grand_canonical_potential}
\end{equation}
where the second equality is derived using Eq. \ref{eq:Euler2}.
The semi-grand canonical potential can be normalized by the number of primitive unit cells in the crystal to yield
\begin{equation}
    \phi=g-\vec{\tilde{\mu}}^{T}\vec{x}=\vec{\mu}^{T}\vec{n}_0
    \label{eq:normalized_semi-grand_canonical_potential}
\end{equation}
where $\phi=\Phi/N_u$ and $g=G/N_u$.
The Appendix derives semi-grand canonical potentials for a variety of multi-component crystals having different crystal structures and sublattice constraints.

\subsection{Equations of state}

Equations of state relate thermodynamic variables to characteristic potentials in the form of derivatives of the characteristic potential with respect to its natural variables.\cite{Callen} 
The equations of state can be extracted from the differential of the characteristic potential. For example, the differential of the Gibbs free energy takes the form
\begin{equation}
    dG=-SdT+VdP+\sum_{i=1}^{s}\mu_{i}dN_{i}
\end{equation}
where the natural variables, $T$, $P$ and the $N_i$ ($i=1,\dots,s$), which are the thermodynamic boundary conditions, appear as differentials. 
The equations of state then become
\begin{align}
    S &= -\left(\frac{\partial G}{\partial T}\right)_{P,N_{i}} 
    \label{eq:eq_state_entropy_G} \\
    V &= \left(\frac{\partial G}{\partial P}\right)_{T,N_{i}}
    \label{eq:eq_state_volume_G} \\
    \mu_{i} &= \left(\frac{\partial G}{\partial N_{i}}\right)_{T,P,N_{j\ne i}} \label{eq:eq_state_Ni_G}
\end{align}
The equations of state relate each thermodynamic variable that is not controlled to the slope of the characteristic free energy with respect to its conjugate thermodynamic variable that is controlled as a boundary condition. 

When considering crystals, it is also useful to formulate equations of state in terms of the parametric composition $\vec{x}$ and the number of unit cells. 
By relating $N_i=N_un_i$ and using Eq. \ref{eq:concentration_equation2}, the differential of the Gibbs free energy can be expressed as
\begin{equation}
    dG=-SdT+VdP+N_u\vec{\tilde{\mu}}^{\mathsf{T}}d\vec{x}+\vec{\mu}^{\mathsf{T}}\vec{n}dN_u
\end{equation}
In this form, variations in the number of atoms can be performed either by holding the number of unit cells fixed while varying the relative amounts of the constituents, or by holding the composition fixed while varying the number of unit cells. 
The equations of state at constant $T$ and $P$ then become
\begin{align}
    \tilde{\mu}_i &= \left(\frac{\partial g}{\partial x_{i}}\right)_{T,P,x_{j\ne i},N_u} 
    \label{eq:eq_state_mu_G} \\
    g &= \left(\frac{\partial G}{\partial N_u}\right)_{T,P,x_{i}}
    \label{eq:eq_state_gibbs_G}
\end{align}
where $g=G/N_u=\vec{\mu}^{\mathsf{T}}\vec{n}$ is the Gibbs free energy per number of unit cells (Eq. \ref{eq:Euler1}). 

The differential form of the semi-grand canonical potential takes the form
\begin{equation}
    d\Phi=-SdT+VdP-N_u\vec{x}^{\mathsf{T}}d\vec{\tilde{\mu}}+\vec{\mu}^{\mathsf{T}}\vec{n}_0dN_u
    \label{eq:differential_semi-grand_canonical}
\end{equation}
where the only extensive natural variable is the number of unit cells $N_u$, which is conjugate to $\phi=\vec{\mu}^{\mathsf{T}}\vec{n}_0$, the semi-grand canonical potential normalized by the number of unit cells (Eq. \ref{eq:normalized_semi-grand_canonical_potential}).
As can be inferred from the differential form Eq. \ref{eq:differential_semi-grand_canonical}, the equations of state derived from the semi-grand canonical potential are
\begin{align}
    S &= -\left(\frac{\partial \Phi}{\partial T}\right)_{P,\tilde{\mu}_{i},\Nunit} 
    \label{eq:eq_state_entropy} \\
    V &= \left(\frac{\partial \Phi}{\partial P}\right)_{T,\tilde{\mu}_{i},\Nunit}
    \label{eq:eq_state_volume} \\
    x_{i} &= -\frac{1}{\Nunit}\left(\frac{\partial \Phi}{\partial \tilde{\mu}_{i}}\right)_{T,P,\tilde{\mu}_{j\ne i},\Nunit} \label{eq:eq_state_xi}\\
    \phi &= \left(\frac{\partial \Phi}{\partial \Nunit}\right)_{T,P,\tilde{\mu}_{ i}}.
    \label{eq:eq_state_xunit}
\end{align}
These equations of state are especially useful to obtain free energies from semi-grand canonical Monte Carlo simulations.

\subsection{Response functions}

The second derivatives of a characteristic potential with respect to its natural variables are related to response functions. 
A response function measures how a state variable that is not controlled as a boundary condition is affected by a variation of a natural variable. 
For $\Phi$, the following relations can be derived for the normalized heat capacity, compressibility, and chemical susceptibilities
\begin{align}
\begin{split}
    c_{P, \tilde{\mu}_{i}} &= -\frac{T}{N_u}\partddsc{\Phi}{T}{P, \tilde{\mu}_{i},N_u}
\end{split}
\end{align}
\begin{align}
\begin{split}
    \kappa_{T, \tilde{\mu}_{i}} &= -\frac{1}{V}\partddsc{\Phi}{P}{T, \tilde{\mu}_{i},N_u}
\end{split}
\end{align}
\begin{align}
\begin{split}
    \chi_{ij} &= -\frac{1}{N_u}\partddc{\Phi}{\tilde{\mu}_{i}}{\tilde{\mu}_{j}}{T, \tilde{\mu}_{{k \neq i,j}},N_u}
\end{split}
\end{align}
Similar to the heat capacity and the compressibility, the chemical susceptibilities, $\chi_{ij}$, are response functions in that they measure the change in the $i$-th parametric composition component upon an incremental change of the $j$-th exchange chemical potential component. 
The normalization by $N_u$ or $V$ is to ensure that the response functions are independent of the size of the crystal. 
The response functions listed above are a subset of the $m(m-1)/2$ possible response functions, one for each unique Hessian element of the characteristic potential with respect to its natural variables, with $m$ referring to the number of natural variables.

\subsection{Elements of Statistical Mechanics}

The atoms and electrons of a crystal posses a variety of degrees of freedom that can be excited at finite temperature.\cite{de1994cluster,ceder1993derivation,van2018first} 
Each collective excitation of the solid corresponds to a microstate. 
For example, a binary alloy consisting of A and B atoms can adopt one of many  possible arrangements over the sites of a parent crystal structure. 
The arrangement of A and B atoms varies spatially and evolves over time due to thermally activated atomic hops. 
Each arrangement of the A and B atoms is referred to as a configurational microstate. 
The atoms of a crystal also vibrate around their equilibrium sites and thereby produce vibrational microstates.\cite{van2002effect,fultz2010vibrational} 
Other common degrees of freedom are electronic in nature and include the order/disorder phenomena involving local magnetic moments of magnetic atoms or  different oxidation states of redox active atoms. 

The microstate of a solid can be tracked with a collection of variables that specify the state of each atom and localized electronic state within the crystal. 
The occupant at site $i$ of a binary A-B alloy, for example, can be specified by an occupation variable $\sigma_i$, which is +1 (-1) when occupied by B (A). 
The vibrational state of an atom at site $i$ can be tracked with a displacement vector $\vec{u}_i$, while the orientation of a magnetic moment at a site $i$ containing a magnetic atom is specified by a unit vector $\vec{m}_i$.
The microstate of a crystal containing N sites is then uniquely specified by the collection of all site variables $\mathbb{C}=(\sigma_1,...,\sigma_i,...,\sigma_N, \vec{u}_1,...\vec{u}_N,\vec{m}_1,...\vec{m}_N)$ when treating the degrees of freedom classically.

Statistical mechanics serves as a link between the electronic structure and the thermodynamic properties of a solid. 
The partition function plays a central role.\cite{hill2013statistical,jaynes1957information}
When treating discrete degrees of freedom (e.g. chemical configurational degrees of freedom) or when treating each microstate as a quantum mechanical eigenstate, the partition function can be expressed as ,\cite{hill2013statistical,jaynes1957information} 
\begin{align} \label{eq:partition_func}
    Z = \sum_{\mathbb{C}} e^{-\beta \Omega_{\mathbb{C}}},
\end{align}
with the sum in Eq. (\ref{eq:partition_func}) extending over all microstates, $\mathbb{C}$, consistent with the imposed boundary conditions ($\beta = 1 / k_{B} T$ with $k_{B}$ being Boltzmann's constant). 
The form of the generalized enthalpy $\Omega_\mathbb{C}$ depends on the externally imposed boundary conditions.\cite{hill2013statistical} 
For instance, when holding $T$, $V$, and the $x_i$ constant, $\Omega_{\mathbb{C}}$ is simply equal to the energy of the solid, $E_\mathbb{C}$. 
For constant $T$, $P$, and $x_i$, $\Omega_{\mathbb{C}} = E_{\mathbb{C}}+PV_{\mathbb{C}}$, where $V_{\mathbb{C}}$ is the volume of the solid in microstate $\mathbb{C}$.
When holding $T$, $P$ and the exchange chemical potentials, $\tilde{\mu}_{i}$, for $i=1,...,k$ constant, $\Omega_{\mathbb{C}}=E_{\mathbb{C}}+PV_{\mathbb{C}}-\Nunit\sum_{i}^{k}\tilde{\mu}_{i}x_{i,\mathbb{C}}$, where the $x_{i,\mathbb{C}}$ are the parametric concentrations in microstate $\mathbb{C}$.

It can be shown that the probability that a solid is in a microstate $\mathbb{C}$ at constant temperature takes the form \cite{hill2013statistical,jaynes1957information}
\begin{equation}
    P_\mathbb{C}=\frac{e^{-\beta \Omega_\mathbb{C}}}{Z}
\end{equation}
A key postulate of statistical mechanics states that measured thermodynamic properties are averages over their corresponding microscopic counterparts.\cite{hill2013statistical} 
The thermodynamic generalized enthalpy at constant temperature $T$, for example, is equal to
\begin{equation}
    \langle\Omega\rangle=\sum_{\mathbb{C}}P_\mathbb{C}\Omega_{\mathbb{C}}
\end{equation}
The postulate can be expressed more generally as
\begin{align}
    \angbr{X} = \frac{1}{Z} \sum_{\mathbb{C}} X_{\mathbb{C}} e^{-\beta\Omega_{\mathbb{C}}},
\end{align}
where $X_{\mathbb{C}}$ is some property of microstate $\mathbb{C}$. 
Ensemble averages can be calculated using Monte Carlo sampling techniques.

The characteristic thermodynamic potential, $\Phi$, is related to the partition function according to,
\begin{align} \label{eq:free_energy}
    \beta \Phi = -\mathrm{ln} Z.
\end{align}
Similar to the equations of state, Eq. (\ref{eq:eq_state_entropy}), (\ref{eq:eq_state_volume}) and (\ref{eq:eq_state_xi}), the following relations can be obtained for $\Phi$ by taking derivatives of Eq. (\ref{eq:free_energy}) and using (\ref{eq:partition_func})
\begin{align}
  \partdc{\beta \Phi}{\beta}{P,\tilde{\mu}_{i},\Nunit} 
  =& -\frac{1}{Z} \partd{Z}{\beta} 
  = \angbr{\Omega} \\
  \partdc{\beta \Phi}{P}{\beta, \tilde{\mu}_{i},\Nunit} 
  =& -\frac{1}{Z} \partd{Z}{P} 
  = \beta \angbr{V} \\
    \partdc{\beta \Phi}{\tilde{\mu}_i}{\beta, P, \tilde{\mu}_{j \neq i}, \Nunit}
  =& -\frac{1}{Z}\partd{Z}{\tilde{\mu}_i} 
  = -\beta \Nunit \angbr{\tilde{x}_i},
\end{align}
which relate partial derivatives of the characteristic potential with respect to its natural variables to ensemble averages of extensive properties. 

A similar approach can be followed to relate the thermodynamic response functions, which are determined by second derivatives of $\Phi$, to fluctuations in $\Omega_{\mathbb{C}}$, $V_{\mathbb{C}}$ and $x_{i,\mathbb{C}}$ according to
\begin{align}
\begin{split}
    c_{P, \tilde{\mu}_i,\Nunit}  &= \frac{\vfluct{\Omega}}{\Nunit kT^2} 
\end{split}
\label{eq:heatcap}
\end{align}
\begin{align} 
\begin{split}
    \chi_{ij} &= \frac{\xvfluct{x_{i}}{x_{j}}}{kT}\Nunit.
\end{split}
\label{eq:susc_x}
\end{align}
\begin{align}
\begin{split}
    \chi_{i,\Omega} &= \frac{\xvfluct{x_{i}}{\Omega}}{kT}.
\end{split}
\label{eq:susc_thermochem}
\end{align}



\section{Diffusion in multicomponent crystals}

The conservation of the number of crystal sites imposes constraints on diffusional fluxes and affects the thermodynamic driving forces for diffusion.\cite{cahn1983invariant,van2010vacancy} 
In this section, generalized diffusional flux expressions in crystals are derived that account for crystallographic constraints.
This is followed by a description of atomic hops that relies on transition state theory. 
A connection is then made between transport coefficients and fluctuations at the atomic scale using well-established Kubo-Green expressions.\cite{allnatt2003atomic} 

\subsection{Phenomenological description of diffusion}

Diffusional fluxes are driven by gradients in chemical potentials. 
For multicomponent, isotropic diffusion, the phenomenological equation that to first order relate the diffusional fluxes, $J_{i}$, to gradients in chemical potentials,  $\nabla \mu_i$, can be expressed in vector form as:
\begin{align}
\vec{J} = -\Lonsager \nabla \vec{\mu},
\label{eq:general_flux_expression}
\end{align}
where $\Lonsager$ is the matrix of Onsager kinetic coefficients and $\vec{J}^{\mathsf{T}}=[J_1,\dots,J_s]$. 
The element $\Lonsagerij$ gives the contribution of the gradient in chemical potential of species $j$, $\nabla \mu_j$, to the flux of species $i$, $J_i$. The elements of $\Lonsager$ must be symmetric ($\Lonsagerij=\Lonsagerji$).\cite{Allnatt1984}

Crystallographic constraints emerge when diffusion occurs within a single crystal away from extended defects such as dislocations, grain boundaries and interfaces.\cite{kehr1989mobility,van2010vacancy}
A redistribution of chemical species then occurs by preserving the underlying crystal structure. 
The conservation of crystal sites while atoms diffuse imposes constraints on the fluxes. 
For example, a conservation of crystal sites in an A-B alloy in which diffusion is mediated by a dilute concentration of vacancies requires that $J_A+J_B+J_{Va}=0$.\cite{kehr1989mobility,van2010vacancy}
More generally, as shown in \ref{sec:diffusion_appendix}, the constraints on the fluxes take the form
\begin{equation}
    \pmb{T}^{\mathsf{T}}\vec{J}=\vec{0}.
    \label{eq:orthogonal_constraints2_main}
\end{equation}
where $\pmb{T}$, the $s\times(s-k)$ dimensional matrix introduced in Section \ref{sec:parametric_compositions}, collects the vectors in composition space that is orthogonal to the composition space of a crystal with a fixed number of unit cells.  
The vector $\vec{0}$ is a $(s-k)$ dimensional vector of zeros. 
Substituting the flux expressions, Eq. \ref{eq:general_flux_expression}, into Eq. \ref{eq:orthogonal_constraints2_main} and using the fact that $\Lonsager$ is a symmetric matrix, leads to linear relationships between different Onsager transport coefficients that can be expressed compactly as
\begin{equation}
    \Lonsager\pmb{T}= \mathbf{0}   
    \label{eq:L_constraints1_main}
\end{equation} 
where $\mathbf{0}$ is a $s\times (s-k)$ matrix of zeros. 

Equations \ref{eq:orthogonal_constraints2_main} and \ref{eq:L_constraints1_main} enable a reduction in the number of independent flux expressions and independent thermodynamic driving forces for diffusion of the form 
\begin{equation}
    \vec{\tilde{J}}=-\tilde{\Lonsager}\nabla\vec{\tilde{\mu}}
    \label{eq:projected_flux_expressions_main}
\end{equation}
where the $k$ independent exchange fluxes are defined as
\begin{equation}
    \vec{\tilde{J}}={\pmb{R}}^{\mathsf{T}}\vec{J} 
    \label{eq:flux_projection_main}
\end{equation}
and where 
\begin{equation}
    \tilde{\Lonsager}={\pmb{R}}^{\mathsf{T}}\Lonsager\pmb{R}
    \label{eq:projected_L_matrix_main}
\end{equation}
is a $k\times k$ Onsager transport coefficient matrix for diffusion in a crystal with a constant number of unit cells. 
The $\vec{\tilde{\mu}}={\pmb{Q}}^{\mathsf{T}}\vec{\mu}$ appearing in Eq. \ref{eq:projected_flux_expressions_main} are the exchange chemical potentials that appear as natural variables of the semi-grand canonical potential.
The matrix $\pmb{R}^{\mathsf{T}}$, introduced in Section \ref{sec:parametric_compositions}, is the left pseudoinverse of $\pmb{Q}$. 
A full derivation of Equations \ref{eq:projected_flux_expressions_main}, \ref{eq:flux_projection_main} and \ref{eq:projected_L_matrix_main} is provided in \ref{sec:diffusion_appendix}. 

Chemical potentials are not as easily measured as composition gradients. 
It is therefore common to formulate the flux expressions in terms of gradients in concentration by applying the chain rule of differentiation to the gradients of chemical potentials in Eq. \ref{eq:projected_flux_expressions_main} to yield  
\begin{align}
\vec{\tilde{J}} = -\pmb{D} \nabla \vec{c},
\end{align}
where $\vec{c}$ is a $k$-dimensional vector of volumetric concentrations, $c_i = x_i/v_u$, with $v_u$ the volume per unit cell.
The chemical diffusion coefficient matrix, $\pmb{D}$, relates the $k$ exchange fluxes (Eq. \ref{eq:flux_projection_main}) to $k$ independent concentration variables and describes diffusion in a crystal in which the number of unit cells, $N_u$ is conserved. 

The matrix of chemical diffusion coefficients, $\pmb{D}$, can be factored into a product of a kinetic factor and a thermodynamic factor according to
\begin{equation}
    \pmb{D}=\tilde{\pmb{K}}\pmb{\Theta}
\end{equation}
where $\tilde{\pmb{K}}$ is a $k\times k$ matrix of kinetic coefficients with 
\begin{equation}
    \tilde{\pmb{K}}=v_uk_{B}T\pmb{\tilde{L}}
    \label{eq:kinetic_coefficients}
\end{equation} 
and where $\pmb{\Theta}$ is a $k\times k$ matrix of thermodynamic factors, with elements \cite{van2010vacancy}
\begin{align} \label{eq:thermofactor}
\Theta_{ij} = \frac{1}{k_{B}T}\partd{\tilde{\mu}_i}{x_j}=\frac{1}{k_{B}T}\frac{\partial^{2}g}{\partial x_j\partial x_i}.
\end{align}
While there is some arbitrariness as to how the matrix of diffusion coefficients can be factored, this particular factorization ensures that the elements of the kinetic factor matrix, $\tilde{\pmb{K}}$, have units of a diffusion coefficient (i.e. cm$^{2}$/s), while the elements of the thermodynamic factor matrix, $\pmb{\Theta}$, are unitless.

The extension to anisotropic diffusion is straightforward, and can be written  
\begin{align} \label{eq:L_aniso}
\tilde{J}_{i,\alpha} = -\sum_{j, \beta}D_{ij,\alpha\beta} \nabla_{\beta} c_j, \\
\tilde{J}_{i,\alpha} = -\sum_{j, \beta}\tilde{L}_{ij,\alpha,\beta} \nabla_{\beta} \tilde{\mu}_j,
\end{align}
where $D_{ij,\alpha\beta}$ and $\tilde{L}_{ij,\alpha\beta}$ are rank-four tensors, with $i$ and $j$ indicating species, $\alpha,\beta$ indicating spatial direction, and $\nabla_{\beta} c_j$ and $\nabla_{\beta} \tilde{\mu}_j$ are the gradients of $c_j$ and $\tilde{\mu}_j$, respectively, along spatial direction $\beta$.

\subsection{Atomistic description of diffusion}
\label{sec:atomistic_desc_diffusion}

Diffusion in crystals can often be modeled as an infrequent event system.  
The crystal evolves through relatively rare discrete events in which one or more chemical occupants change crystal sites, whether substitutional or interstitial, or reorient between discrete orientations. 
Between events, an infrequent event system is characterized by relative inactivity, in which the assignment of chemical occupants to crystal sites remains unchanged. 
The time evolution of the system can then be modeled using the master equation
\begin{align}
\partd{P_{\mathbb{C}'}(t)}{t} = \sum_{\mathbb{C} \neq \mathbb{C}'} \left(\Gamma_{\mathbb{C}\mathbb{C}'}P_\mathbb{C}(t) - \Gamma_{\mathbb{C}'\mathbb{C}}P_{\mathbb{C}'}(t) \right),
\end{align}
where $\Gamma_{\mathbb{C}\mathbb{C}'}$ is the rate at which the system transitions from state $\mathbb{C}$ to state $\mathbb{C}'$, and $P_{\mathbb{C}}(t)$ is the probability of being in state $\mathbb{C}$ at time $t$.

\begin{figure}
    \centering
    \includegraphics[width=7cm]{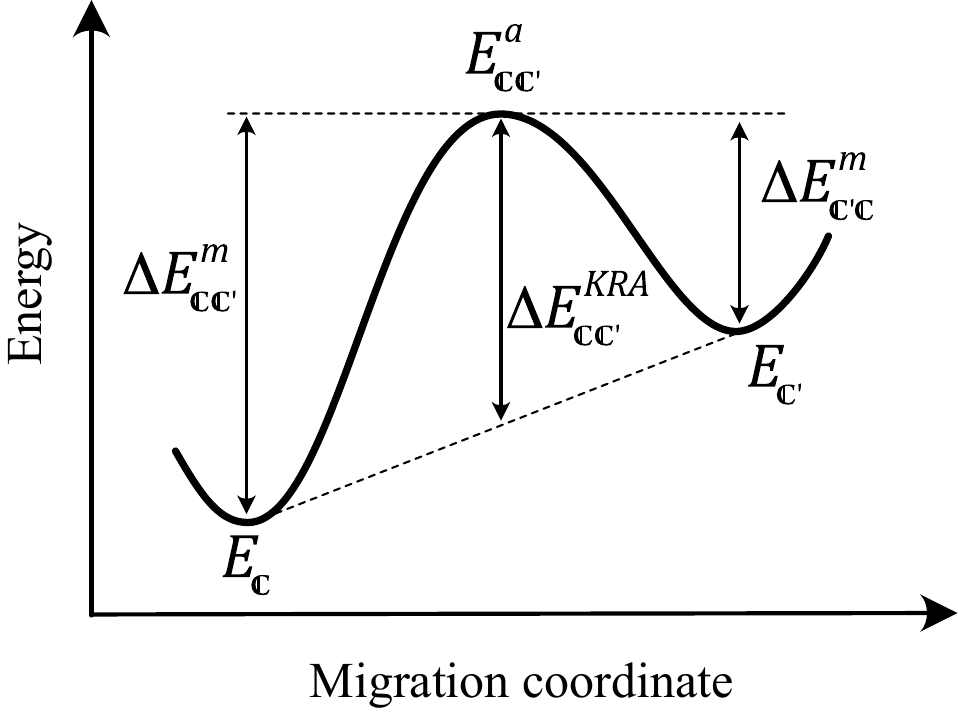}
    \caption{Schematic parameterization of the activation energy for diffusion $E_{\mathbb{C}\mathbb{C}'}^a$ in terms of the end state energies $E_{\mathbb{C}}$ and $E_{\mathbb{C}'}$ and a kinetically resolved activation energy $\Delta E_{\mathbb{C}\mathbb{C}'}^{KRA}$ for purposes of calculating the migration barrier $\Delta E_{\mathbb{C}\mathbb{C}'}^m$ and ensuring detailed balance is maintained in kinetic Monte Carlo simulations.}
    \label{fig:KRA}
\end{figure}

According to transition state theory, the rate at which a rare event transition from state $\mathbb{C}$ to state $\mathbb{C}'$ occurs is \cite{vineyard1957frequency}
\begin{align} \label{eq:kmc_rate}
\Gamma_{\mathbb{C}\mathbb{C}'} = \nu^{*}_{\mathbb{C}\mathbb{C}'} e^{ -\beta \Delta E_{\mathbb{C}\mathbb{C}'}^m },
\end{align}
where $\nu^{*}_{\mathbb{C}\mathbb{C}'}$ is the vibrational prefactor for the event. The vibrational prefactor typically has values on the order of $10^{12}-10^{13}$ s$^{-1}$ in solids.
The migration barrier $\Delta E_{\mathbb{C}\mathbb{C}'}$ is defined as the change in potential energy from the initial equilibrium state, $\mathbb{C}$, to the transition state between $\mathbb{C}$ and $\mathbb{C}'$
\begin{equation}
        \Delta E_{\mathbb{C}\mathbb{C}'}^m = E_{\mathbb{C}\mathbb{C}'}^a - E_{\mathbb{C}}
        \label{eq:migration_barrier}
\end{equation}
where $E_\mathbb{C}$ and $E_{\mathbb{C}\mathbb{C}'}^{a}$ refer to the energies in the initial state and the activated state of the hop, respectively (Figure \ref{fig:KRA}).

In a multi-component crystal, the initial and final microstates of atomic hop events will be a function of the local composition and the local degree of ordering.\cite{van2001first,van2005first,van2008nondilute,bhattacharya2011first,goiri2019role,van2018first,kolli2021elucidating} 
This often makes the values of $\nu^{*}_{\mathbb{C}\mathbb{C}'}$ and $\Delta E_{\mathbb{C}\mathbb{C}'}^m$ a function of the direction of the transition, such that $\nu^{*}_{\mathbb{C}\mathbb{C}'} \neq \nu^{*}_{\mathbb{C}'\mathbb{C}}$ and $\Delta E_{\mathbb{C}\mathbb{C}'}^m \neq \Delta E_{\mathbb{C}'\mathbb{C}}^m$. 
While the forward and reverse hops usually have a different migration barrier and prefactor, they nevertheless share the same transition state and therefore have the same energy in the activated state, $E_{\mathbb{C}\mathbb{C}'}^a = E_{\mathbb{C}'\mathbb{C}}^a$.
This fact can be used to calculate the migration barrier via the kinetically resolved activation (KRA) barrier, $\Delta E_{\mathbb{C}\mathbb{C}'}^{KRA}$, which is defined as the average barrier of the forward and reverse hops and is therefore direction independent.\cite{van2001first,van2018first} 
The energy of the activated state in terms of $\Delta E_{\mathbb{C}\mathbb{C}'}^{KRA}$ takes the form
\begin{align}
    E_{\mathbb{C}\mathbb{C}'}^a &= \Delta E_{\mathbb{C}\mathbb{C}'}^{KRA} + \frac{E_{\mathbb{C}'} + E_{\mathbb{C}}}{2}, 
    \label{eqn:activation_state_energy}
\end{align}
which can then be inserted into Eq. \ref{eq:migration_barrier} to calculate a migration barrier according to
\begin{equation}
    \Delta E_{\mathbb{C}\mathbb{C}'}^m=\Delta E_{\mathbb{C}\mathbb{C}'}^{KRA} + \frac{E_{\mathbb{C}}' - E_{\mathbb{C}}}{2}.
    \label{eqn:migration_barrier_in_terms_of_KRA}
\end{equation}
The relationships between the different quantities appearing in Equations \ref{eqn:activation_state_energy} and \ref{eqn:migration_barrier_in_terms_of_KRA} are shown in Figure \ref{fig:KRA}.

A benefit of working with a kinetically resolved activation barrier, $\Delta E_{\mathbb{C}\mathbb{C}'}^{KRA}$, is that its dependence on the local degree of ordering can be parameterized with a local cluster expansion \cite{van2001first,van2018first,goiri2019role}. 
The migration barrier for a hop in a disordered multi-component crystal can then be calculated by utilizing surrogate models such as cluster expansions that parameterize the configuration dependence of $\Delta E_{\mathbb{C}\mathbb{C}'}^{KRA}$, a local property, and the end state energies $E_\mathbb{C}$ and $E_{\mathbb{C}'}$ \cite{van2001first,van2018first,goiri2019role}.

The attempt frequency, which may also have a configuration dependence, is a function of the difference between the vibrational entropy in the activated state, $S^a_{\mathbb{C}\mathbb{C}'}$, and the vibrational entropy in the initial state, $S_{\mathbb{C}}$, according to \cite{Eyring1935, vineyard1957frequency}
\begin{equation}
    \nu^{*}_{\mathbb{C}\mathbb{C}'} = \frac{k_{B}T}{h}e^{\Delta S^m _{\mathbb{C}\mathbb{C}'} / k_{B}}
\end{equation}
where $\Delta S^m_{\mathbb{C}\mathbb{C}'} = S^a_{\mathbb{C}\mathbb{C}'} - S_{\mathbb{C}}$, and $h$ is Planck’s constant.
A similar approach as that used to parameterize energy barriers can also be used to parameterize $\Delta S_{\mathbb{C}\mathbb{C}'}$ in terms of the end state entropies $S_{\mathbb{C}}$ and $S_{\mathbb{C}'}$ and a kinetically resolved activation entropy, $\Delta S_{\mathbb{C}\mathbb{C}'}^{KRA}$
\begin{equation}
    \Delta S_{\mathbb{C}\mathbb{C}'}^m=\Delta S_{\mathbb{C}\mathbb{C}'}^{KRA} + \frac{S_{\mathbb{C}}' - S_{\mathbb{C}}}{2}.
    \label{eqn:migration_entropy_in_terms_of_KRA}
\end{equation}
With surrogate models for the end state entropies and kinetically resolved activation entropy, the dependence of $\nu^{*}_{\mathbb{C}\mathbb{C}'}$ on local configurational order can be included.
The use of kinetically resolved activation quantities as described above ensures consistency between purely thermodynamic properties and kinetic properties by preserving detailed balance.

\subsection{Kinetic transport coefficients}

The atoms of a solid constantly perform diffusive hops, even when the solid is in thermodynamic equilibrium.
Atomic hops can be sampled stochastically with kinetic Monte Carlo simulations given a catalogue of possible events and efficient methods for calculating $E_\mathbb{C}$, $\Delta E_{\mathbb{C}\mathbb{C}'}^{KRA}$, and $\nu^{*}_{\mathbb{C}\mathbb{C}'}$.
Each atom $\zeta$ of chemical type $i$ will wander through the crystal and end up at a position $\Delta\vec{R}^{\zeta}_{i}$ at time $t$ relative to its starting point at $t=0$.

An expression for $\Lonsager$ that connects to atomic hop events can be derived using Kubo-Green linear response methods \cite{allnatt2003atomic,Allnatt1984}, an approach that links fluctuations that occur at equilibrium to the macro-scale transport coefficients of linear kinetic rate equations. 
The calculated $\Lonsager$ are therefore suited to describe the evolution of a system that is out of equilibrium, but is nevertheless everywhere in local equilibrium.\cite{de2013non}
The Onsager transport coefficients of an isotropic solid, $\Lonsager$, can be calculated as an ensemble average over kinetic Monte Carlo trajectories according to
\begin{align}
\Lonsagerij = \frac{\bigg \langle \paren{\sum_{\zeta} \Delta\vec{R}^{\zeta}_{i}} \paren{\sum_{\zeta} \Delta\vec{R}^{\zeta}_{j}} \bigg \rangle }{2dtVk_{B}T}, 
\label{eq:Kubo_Green_Onsager1}
\end{align}
where $i$ and $j$ indicate species, $d$ is the dimension of space in which diffusion occurs, $t$ is the observation time, and $V$ is the volume of the Monte Carlo supercell. 
The kinetic coefficients defined according to Equation \ref{eq:kinetic_coefficients} then take the form
\begin{equation}
\tilde{\pmb{K}} = {\pmb{R}}^{\mathsf{T}}\pmb{K}\pmb{R},
\end{equation}
where the matrix $\pmb{K}$ has elements
\begin{equation}
K_{ij} = \frac{\bigg \langle \paren{\sum_{\zeta} \Delta\vec{R}^{\zeta}_{i}} \paren{\sum_{\zeta} \Delta\vec{R}^{\zeta}_{j}} \bigg \rangle }{2dtN_u}, 
\label{eq:Kubo_Green_Onsager2}
\end{equation}
and where as before, $N_u$ is equal to the number of unit cells in the crystal. 


Other quantities of interest that can be calculated using trajectories sampled with kinetic Monte Carlo simulations are collective and tracer diffusion coefficients. 
A {\it collective} diffusion coefficient, $D_i$, can be defined for each diffusing species $i$ according to
\begin{equation}
    D_{i}=\frac{K_{ii}}{n_i}=\frac{\bigg \langle \paren{\sum_{\zeta} \Delta\vec{R}^{\zeta}_{i}}^{2} \bigg \rangle }{2dtN_i}
    \label{eq:collective_diffusion}
\end{equation}
where $n_i$ is the number of atoms of type $i$ per unit cell and $N_i=n_iN_u$ is the number of atoms of type $i$ in the crystal in which the trajectories are sampled.
The collective diffusion coefficient averages the square of the displacement of the geometric center of mass of all the diffusing atoms of type $i$. 
It can be related to the tracer diffusion coefficient, $D^{*}_{i}$, upon expanding the square in Eq. \ref{eq:collective_diffusion} according to
\begin{equation}
     D_i=\frac{\bigg \langle \sum_{\zeta} \paren{\Delta\vec{R}^{\zeta}_{i}}^{2} \bigg \rangle}{2dt N_i}+\frac{\bigg \langle \sum_{\zeta}\sum_{\zeta'\ne\zeta} \Delta\vec{R}^{\zeta}_{i}\Delta\vec{R}^{\zeta'}_{i} \bigg \rangle}{2dt N_i}
     \label{eqn:collective_diffusion}
\end{equation}
where the first term corresponds to the standard definition of the tracer diffusion coefficient \cite{Murch2001}
\begin{align}
D^{*}_{i} = \frac{\bigg \langle \sum_{\zeta} \left(\Traji\right)^2 \bigg \rangle}{2dtN_i}.
\label{eqn:tracer_diffusion}
\end{align}
The tracer diffusion coefficient, $D^{*}_{i}$, is a measure of the mobility of individual atoms. 
The second term in Eq. \ref{eq:collective_diffusion} captures correlations between the trajectories of different atoms of the same chemical type $i$.
The Haven ratio \cite{Haven1954, Murch2001}, $(H_R)_i=D^{*}_i/D_i$, measures the degree with which trajectories of different atoms are correlated with each other, being equal to one when there are no correlations between diffusing atoms. 
The correlation factor, defined as 
\begin{equation}
    f_i=\frac{\bigg \langle  \left(\Delta\vec{R}^{\zeta}_{i} \right)^{2} \bigg \rangle}{h_i \left(\Delta \vec{r} \right)^2},
\end{equation}
in contrast, measures the degree with which successive hops of the same atom are correlated.\cite{Murch2001} 
$\Delta \vec{R}^{\zeta}_{i}$ represents a vector that connects the end points of a trajectory of species $i$, $h_i$ is the average number of hops by atoms of species $i$, and $(\Delta \vec{r})^2$ is the square of a hop vector.

While treated as scalars, the various atomic transport coefficients are second-rank tensors. 
In a crystal with cubic symmetry, the transport coefficient tensors only have diagonal non-zero elements, which by symmetry are all equal to each other. 
Anisotropic kinetic coefficients, $\Lonsagerijab$, can be calculated according to:
\begin{align}
\Lonsagerijab = \frac{\bigg \langle \paren{\sum_{\zeta} \Trajia} \paren{\sum_{\zeta} \Trajjb} \bigg \rangle }{2tVk_{B}T}, 
\end{align}
where $\Trajia$ is the $\alpha$-component of the vector connecting the beginning and ending points of the $\zeta$-th atom of species $i$.
Similar expressions hold for the other transport coefficients described above. 


\subsection{Calculating thermodynamic factors}

Similar to the Onsager transport coefficients, the thermodynamic factor matrix, with elements $\Theta_{i,j}$ defined according to Eq. \ref{eq:thermofactor}, is also related to fluctuations. 
For the thermodynamic factor, the fluctuations are of the parametric compositions, $\vec{x}$, within the semi-grand canonical ensemble where the exchange chemical potentials, $\vec{\tilde{\mu}}$, are held constant. 
This follows from the following relationship \cite{van2010vacancy}
\begin{equation}
    \sum_{k}\left(\frac{\partial \tilde{\mu}_i}{\partial x_{k}}\right)_{x_{l\neq k}}\left(\frac{\partial x_k}{\partial \tilde{\mu}_j}\right)_{\tilde{\mu}_{l\neq j}}=\delta_{i,j}
    \label{eq:identity_relation}
\end{equation}
Within the semi-grand canonical ensemble, the derivatives of the parametric concentrations $x_{i}$ with respect to the exchange chemical potentials $\tilde{\mu}_{j}$ is equal to fluctuations in the $x_{i}$ 
\begin{equation}
    \chi_{ij} = \left(\frac{\partial x_i}{\partial \tilde{\mu}_j}\right)_{\tilde{\mu}_{l\neq j}}
\end{equation}
which can be calculated using Eq.\ref{eq:susc_x}.
Because of Eq. \ref{eq:identity_relation}, the matrix of thermodynamic factor elements can be calculated from the inverse of the $k\times k$ matrix of $\chi_{ij}$ according to
\begin{equation}
    \Theta = \frac{1}{k_{B}T}\chi^{-1}.
    \label{eq:thermofactor_from_chi}
\end{equation}

\section{Surrogate models: cluster expansions}

An essential ingredient in the calculation of the thermodynamic properties of a solid with a statistical mechanics approach is the energy $E(\mathbb{C})$ of microstates $\mathbb{C}$. 
Formally these energies correspond to solutions to the Schr\"{o}dinger equation of the solid. 
There are now a multitude of state-of-the-art numerical approaches that can solve for the quantum mechanical energy spectrum of a solid by relying on approximations and extensions to first-principles density functions theory (DFT).\cite{lejaeghere2016reproducibility} 
The number of microstates that need to be sampled to perform statistical mechanical averages, however, are too large to be calculated directly from first principles. 
Instead, surrogate models are required to interpolate first-principles energies of a small number of microstates to predict the energies of microstates sampled in Monte Carlo simulations. 

The generalized cluster expansion approach provides guidance as to how to rigorously formulate a tunable expansion to represent properties of interacting atoms. 
First introduced by Sanchez et al for alloy degrees of freedom over the sites of a crystal\cite{sanchez1984generalized,de1994cluster,sanchez2010cluster}, it has been extended to describe the energy of crystals with non-collinear magnetic degrees of freedom\cite{drautz2004spin} and molecular orientational degrees of freedom \cite{thomas2018hamiltonians} and was recently generalized further to formulate rigorous descriptors of local atomic environments for machine-learned interatomic potentials \cite{drautz2019atomic}. 
The CASM Monte Carlo code base is designed to work with generalized cluster expansions for local degrees of freedom assigned to sites of a crystal. 

\subsection{Examples of cluster expansions}

As an illustration, consider a simple binary system of atoms and vacancies that share interstitial sites of a host material. 
Common examples include Li-vacancy disorder over the Li sites of intercalation compounds used as electrodes in Li-ion batteries, such as Li$_x$CoO$_2$, Li$_x$FePO$_4$ and Li$_x$Mn$_2$O$_4$,\cite{van2020rechargeable} and oxygen-vacancy disorder over the interstitial sites of refractory metals such as Ti, Zr and Nb, which are able to dissolve unusually high concentration of oxygen, nitrogen and carbon.\cite{puchala2013zro,gunda2018first,gunda2018resolving} 
Each interstitial site $i$ of the crystal can be assigned an occupation variable, $\sigma_i$, which is +1 if the site is occupied by an interstitial species and -1 if it is vacant. 
The configurational state of the crystal is then completely specified by the vector of all occupation variables $\vec{\sigma}=(\sigma_1,...,\sigma_i,...,\sigma_N)$.

The energy of the crystal will depend on how the guest atoms are arranged over the interstitial sites of the crystal and is therefore an explicit function of $\vec{\sigma}$.
The dependence of the energy of the crystal on $\vec{\sigma}$ can be parameterized with an alloy cluster expansion of the form\cite{sanchez1984generalized,de1994cluster}
\begin{equation}
    E(\vec{\sigma})=V_0+\sum_{i}V_{i}\sigma_i+\sum'_{i,j}V_{i,j}\sigma_{i}\sigma_{j}+\sum'_{i,j,k}V_{i,j,k}\sigma_{i}\sigma_{j}\sigma_{k}+...
\end{equation}
where the $V_0$, $V_i$, $V_{i,j}$, $V_{i,j,k}$, etc. are expansion coefficients that can be trained to a data set of first-principles energies. 
The sums extend over interstitial sites of the crystal, with the prime on the sums indicating that only distinct pairs, triplets etc. are summed over to avoid counting the same interactions multiple times. 
The cluster expansion can be expressed more compactly as\cite{sanchez1984generalized,de1994cluster}
\begin{equation}
    E(\vec{\sigma})=V_0+\sum_{\alpha}V_{\alpha}\Phi_{\alpha}(\vec{\sigma})
    \label{eq:binary_cluster_expansion}
\end{equation}
where 
\begin{equation}
    \Phi_{\alpha}=\prod_{i\in\alpha}\sigma_{i}
\end{equation}
are polynomial basis functions in terms of occupation variables of sites belonging to clusters of sites labeled with the index $\alpha$. 
The sum in Eq. \ref{eq:binary_cluster_expansion} extends over all distinct clusters of sites, including point clusters, pair clusters, triplet clusters etc. 
The $V_{\alpha}$ are adjustable expansion coefficients that are to be fit to a training set of first-principles data.\cite{van2002automating,hart2005evolutionary,mueller2009bayesian,nelson2013cluster,nelson2013compressive,goiri2018recursive} 
The symmetry of the underlying crystal structure imposes constraints on the expansion coefficients, dramatically reducing the number of independent expansion coefficients that need to be trained. 
The value of a cluster expansion is that, once parameterized, it can be evaluated rapidly within Monte Carlo simulations where microstates are sampled within a large supercell of the primitive cell of the crystal. 
A cluster expansion can also be used to describe local properties, such as the kinetically resolved activation barrier used to represent the dependence of migration barriers for diffusion on the local degree of order or disorder.\cite{van2001first,van2018first,goiri2019role} 

The cluster expansion formalism has been extended to represent the energy of a crystal as a function of continuous degrees of freedom. 
For example, the energy of a magnetic solid with non-collinear localized magnetic moments can be expressed according to \cite{drautz2004spin,kitchaev2020mapping}
\begin{equation}
    E(\vec{\bf{m}})=V_0+\sum_{\alpha,\vec{n}}V_{\alpha,\vec{n}}\Phi_{\alpha,\vec{n}}(\vec{\bf{m}})
\end{equation}
where $\Phi_{\alpha,\vec{n}}(\vec{\bf{m}})$ are composed of linear combinations of products of spherical harmonics, each a function of a local magnetic moment unit vector attached to a site belonging to cluster $\alpha$.
Cluster expansions can be formulated that couple multiple site degrees of freedom, as in a magnetic alloy, where magnetic degrees of freedom are coupled to chemical degrees of freedom.\cite{decolvenaere2019modeling} 
A similar cluster expansion has been formulated to describe the energy of a molecular crystal as a function of the relative orientations of the molecules of the crystal.\cite{thomas2018hamiltonians}
The CASM software package algorithmically constructs crystal-based cluster expansions for configurational, displacement and magnetic degrees of freedom and is able to formulate cluster expansions that couple multiple site degrees of freedom with homogeneous strain of the crystal.\cite{puchala2023casm,thomas2023}

\subsection{The CASM clexulator}

A large fraction of the computational expense of a Monte Carlo simulation is devoted to calculations of the energy using a cluster expansion. 
This requires the frequent evaluation of polynomial basis functions that are functions of site degrees of freedom that are then multiplied with expansion coefficients and summed.
The CASM clexulator is a unique feature of the CASM software pacakge designed to optimize the speed of Monte Carlo simulations.
CASM clexulators are C++ classes with member functions that have been algorithmically written by a CASM preprocessor to contain explicit expressions for evaluating the cluster expansion basis functions and changes in the basis functions due to changes in degrees of freedom.
The expressions involving site degrees of freedom are written on a per unit cell basis and CASM generates neighbor lists for each unit cell in a Monte Carlo simulation supercell to allow evaluation over an entire configuration. 
The functions are compiled with optimization and linked to the CASM Monte Carlo code at runtime.
This approach ensures that evaluations of energies and energy differences occur very rapidly.

\section{Monte Carlo simulations}

Monte Carlo methods for the calculation of the thermodynamic and kinetic properties of solids sample microstates explicitly. 
In a Monte Carlo simulation, the occupation variables of sites within a large supercell of the crystal are held in computer memory. 
Periodic boundary conditions are usually imposed. 
When modeling configurational degrees of freedom for a binary alloy, for example, each site $i$ has an occupation variable $\sigma_i$ that tracks the occupant of that site in the current microstate.
A chain of microstates are successively sampled by applying small perturbations to the $n^{th}$ microstate to generate the $(n+1)^{th}$ microstate. 
The sampled microstates are used to collect quantities whose statistical mechanical averages yield macroscopic thermodynamic quantities. 

In this section, we illustrate how Monte Carlo methods can be used to calculate thermodynamic properties and transport coefficients of multicomponent crystals. 
A distinction is made between the thermodynamic state variables that are imposed as boundary conditions and the conjugate thermodynamic state variables that are calculated with a Monte Carlo method. 
The most common boundary conditions for multicomponent crystals are those of the semi-grand canonical ensemble, where the number of unit cells $N_u$, the temperature, $T$, and the exchange chemical potentials, $\tilde{\mu}_i$ (conjugate to the parametric concentration $\tilde{x}_i$) are held constant. 
Calculated properties are then the semi-grand canonical generalized enthalpy $\langle \Omega \rangle$, the  parametric concentrations of the different species in the solid $\langle x_i \rangle$ along with response functions $C_{\mu}$, $\chi_{ij}$ and $\chi_{i\Omega}$. 
Semi-grand canonical Monte Carlo simulations are generally run over a dense grid of temperatures and chemical potentials to determine the functional relationship between  $\langle\Omega\rangle$, the various $\langle x_i \rangle$ and the response functions on the natural variables of the semi-grand canonical ensemble $T,\tilde{\mu}_1,...\tilde{\mu}_{k}$.
In some contexts, it may be more convenient to impose the boundary conditions of the canonical ensemble, which fixes the temperature and the composition of the crystal, $\vec{x}$. 

\begin{figure}
\centering
\includegraphics[width=3in]{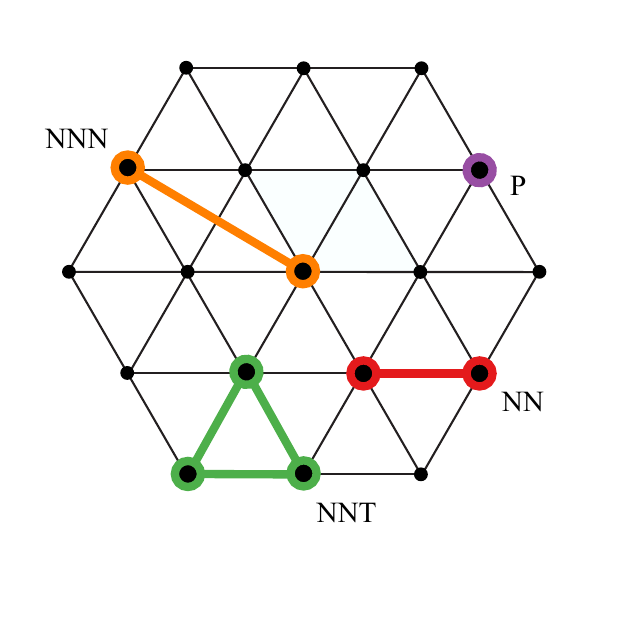}
\vspace{-0.5cm}
\caption{\label{fig:clusters} The point cluster (P), nearest neighbor pair cluster (NN), second nearest neighbor pair cluster (NNN), and nearest neighbor triplet cluster (NNT) on the two-dimensional triangular lattice with unit cell shown in light blue.}
\end{figure}

\begin{figure*}
\centering
\begin{subfigure}{6in}
    \centering
    \includegraphics[height=2in]{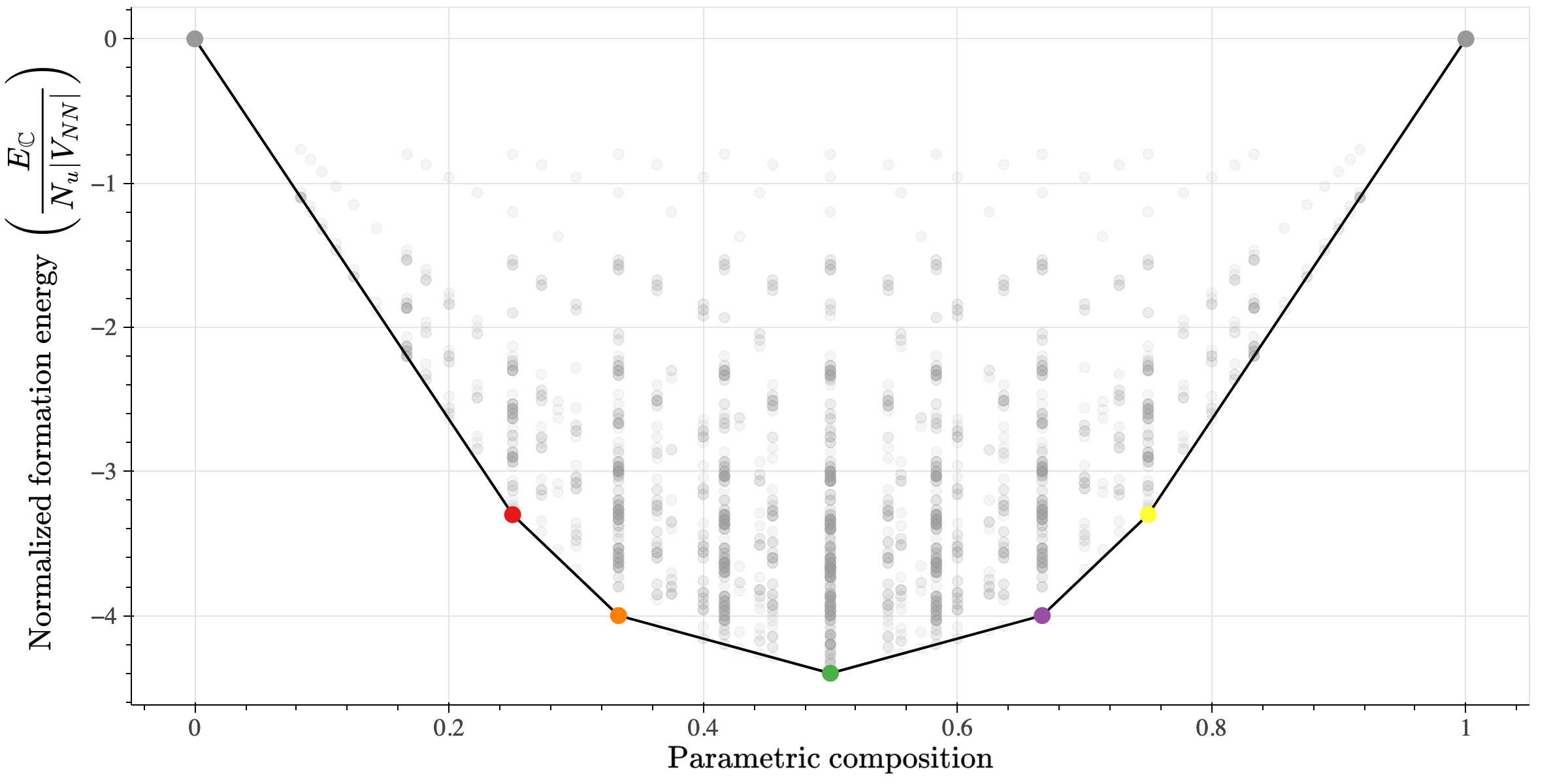}
    \label{fig:hull_coeff2}
    \caption{}
\end{subfigure}
\begin{subfigure}{6in}
    \centering
    \includegraphics[width=4in]{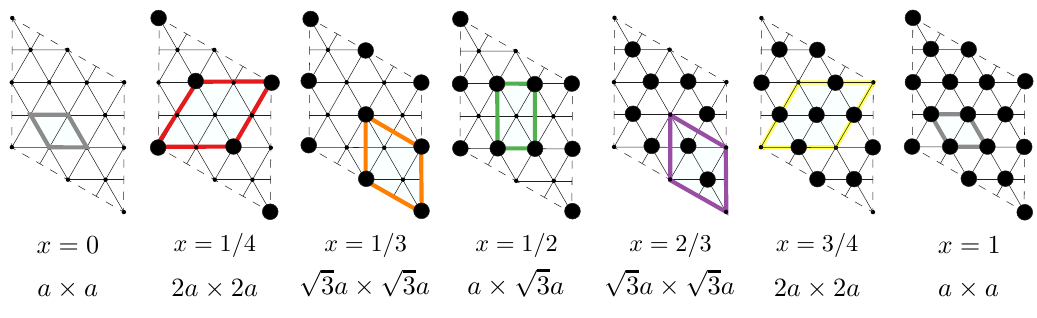}
    \label{fig:gs_configs_coeff2}
    \caption{}
\end{subfigure}
\begin{subfigure}{6in}
    \centering
    \includegraphics[height=2in]{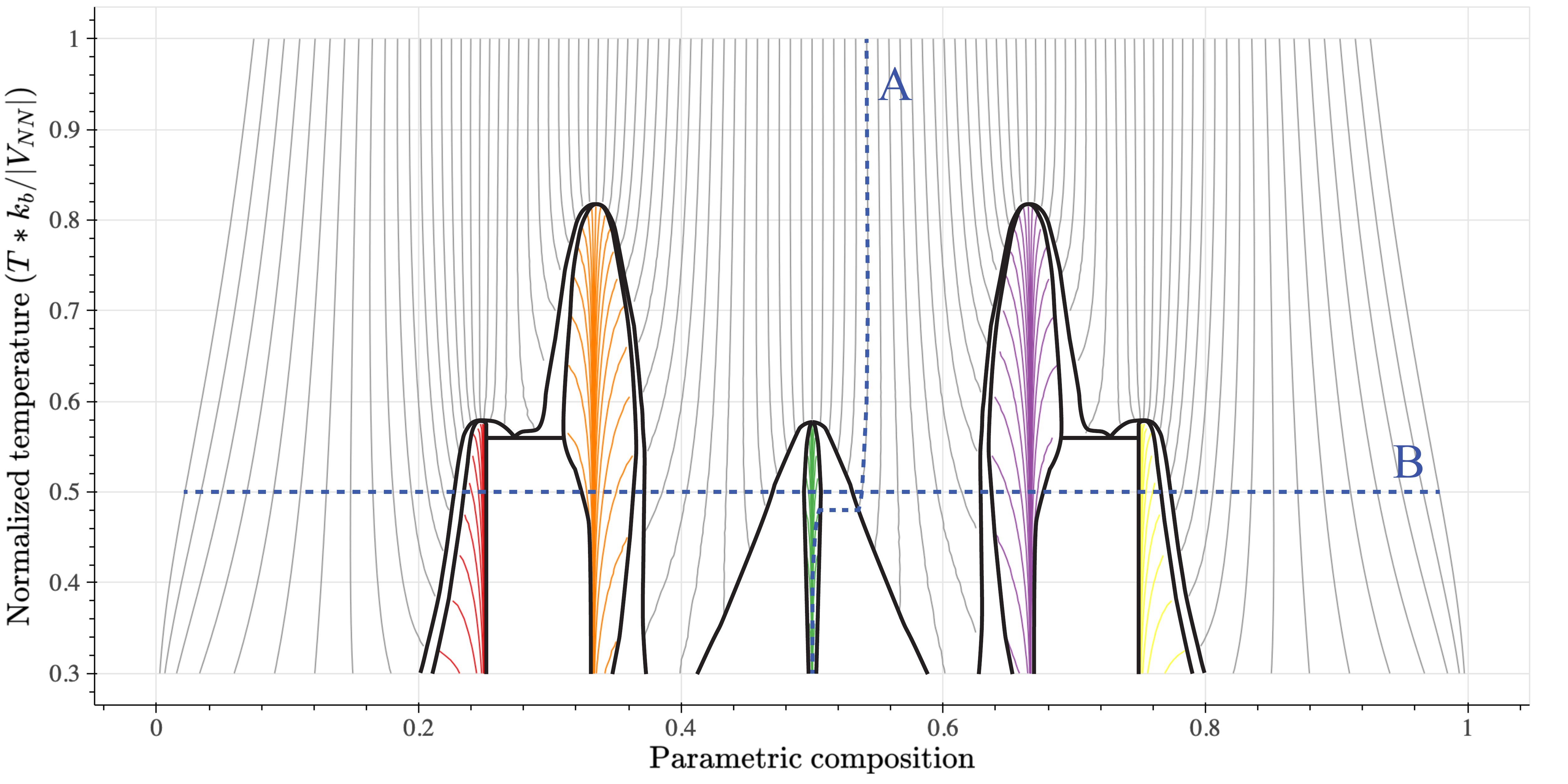}
    \label{fig:pd_comp}
    \caption{}
\end{subfigure}
\begin{subfigure}{6in}
    \centering
    \includegraphics[height=2in]{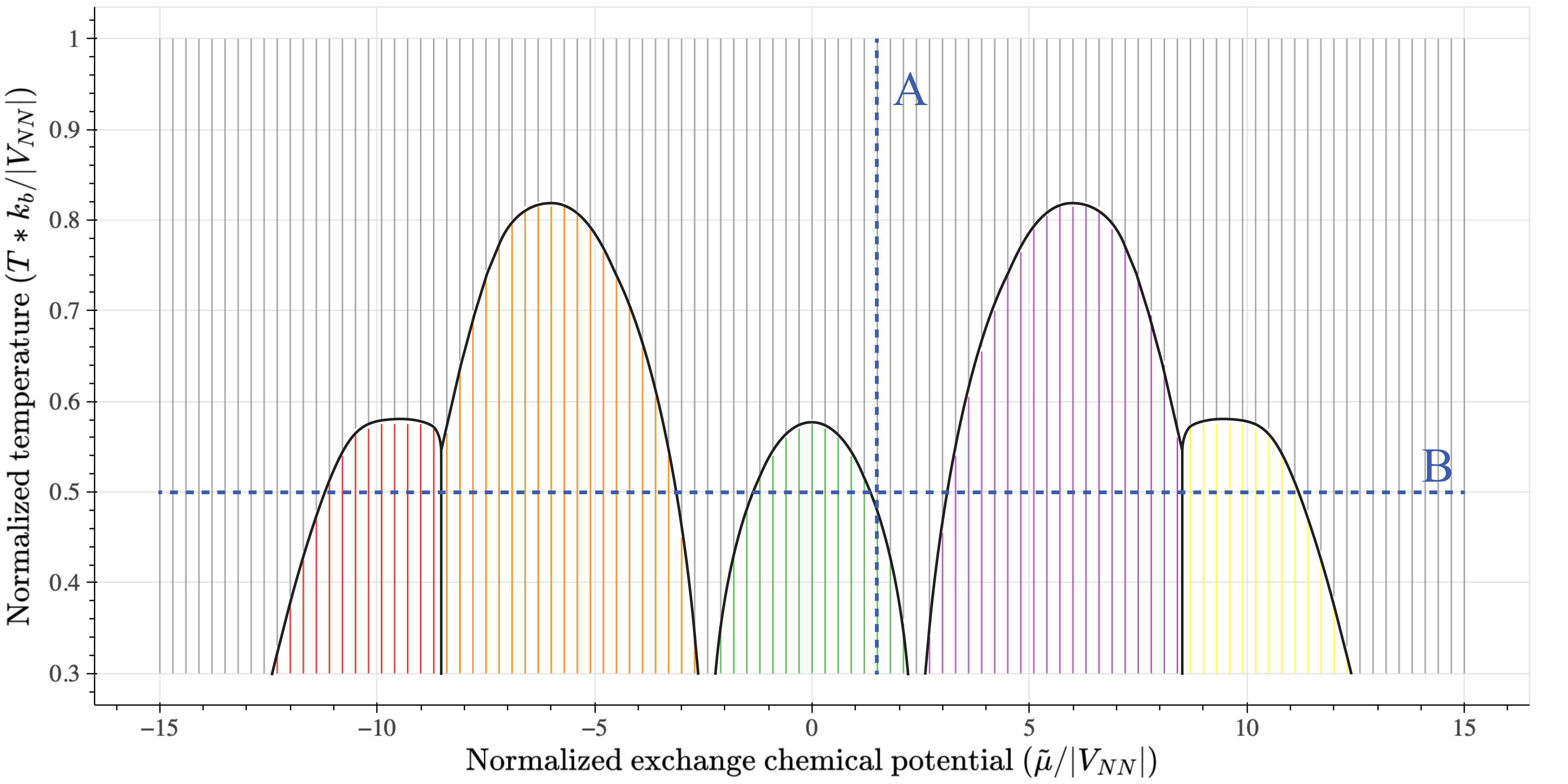}
    \label{fig:pd_mu}
    \caption{}
\end{subfigure}
\vspace{0.2cm}
\caption{\label{fig:pd_hex} Thermodynamic properties of a two-dimensional triangular lattice model cluster expansion with $V_{NNN}/V_{NN} = 1/10$ and $V_{NNT} = 0$. (a) The formation energies and convex hull as predicted with the model cluster expansion and (b) ground state orderings having formation energies on the convex hull. Phase diagrams as determined by minimizing free energies obtained with Monte Carlo simulations plotted versus (c) the parametric composition and (d) the exchange chemical potential. The vertical lines correspond to paths of constant chemical potential.} 
\end{figure*}

\subsection{Example 1: First-order phase transitions}
\label{sec:triangular_lattice_first_order_transition}





As a first model system, we consider an atom-vacancy mixture on a two-dimensional triangular lattice. For the energy of the system, a cluster expansion of the form
\begin{align}
    E(\vec{\sigma})={}&V_{0} + \sum_{(i,j)= NN}V_{NN}\sigma_i\sigma_j \nonumber\\
    +{}&\sum_{(i,j)= NNN}V_{NNN}\sigma_i\sigma_j
\end{align}
is used.
The occupation variables, $\sigma_i$, assigned to each lattice site $i$ are equal to 1 when the site is occupied and -1 when the site is vacant.
Examples of clusters on the two-dimensional triangular lattice are illustrated in Figure \ref{fig:clusters}. 
The stable ordered phases at low temperature and the elevated temperature thermodynamic properties are sensitive to the values of the interactions coefficients. 
Since the model is a simple atom vacancy system, the composition is fully specified with only one parametric concentration, $x$, and one independent exchange chemical potential, $\tilde{\mu}$.
We choose interaction coefficients $V_{NNN}/V_{NN} = 1/10$, and the constant interaction coefficient, $V_0 = -3(V_{NN} + V_{NNN})$, is chosen to ensure that the $x=0$ (all vacancies) and the $x=1$ (all sites of the triangular lattice occupied) states have a zero energy.

The energies (normalized by $|V_{NN}|$) of a large number of different atom-vacancy configurations as calculated with the model cluster expansion are shown as a function of composition in Figure~\ref{fig:pd_hex}(a). 
The orderings with formation energies on the convex hull are the stable ground state phases at zero Kelvin. 
Figure~\ref{fig:pd_hex}(a) shows that there are five ground state orderings with compositions $x=$ 1/4, 1/3, 1/2, 2/3 and 3/4 for this model cluster expansion.  
The orderings and their superlattices are illustrated in Figure~\ref{fig:pd_hex}(b) along with the fully vacant ($x=0$) and fully occupied ($x=1$) configurations.


Figure~\ref{fig:pd_hex}(c) shows the calculated temperature versus composition phase diagram for the model cluster expansion. 
The ground state ordered phases are stable at low temperatures and within narrow composition ranges. 
Each ordered phase under goes an order-disorder phase transitions at elevated temperatures. 
In this model system, each ordered phase disorders by means of a first-order phase transition. 
Superposed on the temperature versus composition phase diagram are lines of constant chemical potential. 
These lines exhibit discontinuities upon passing through two-phase regions since the chemical potentials of coexisting phases at a constant temperature are equal to each other. 
At low temperatures, all the constant chemical potential lines converge to one of the stoichiometric ground state compositions. 

Phase stability can also be displayed in a temperature versus chemical potential phase diagram as illustrated in Figure~\ref{fig:pd_hex}(d), where the horizontal axis corresponds to the parametric chemical potential $\tilde{\mu}$. 
Two-phase regions in the temperature versus composition phase diagram become lines in a temperature versus chemical potential phase diagram. 
Each ground state ordering at low temperature is stable in a wide chemical potential window. 

\begin{figure}
    \centering
    \vspace{0.2cm}
    \includegraphics[width=3in]{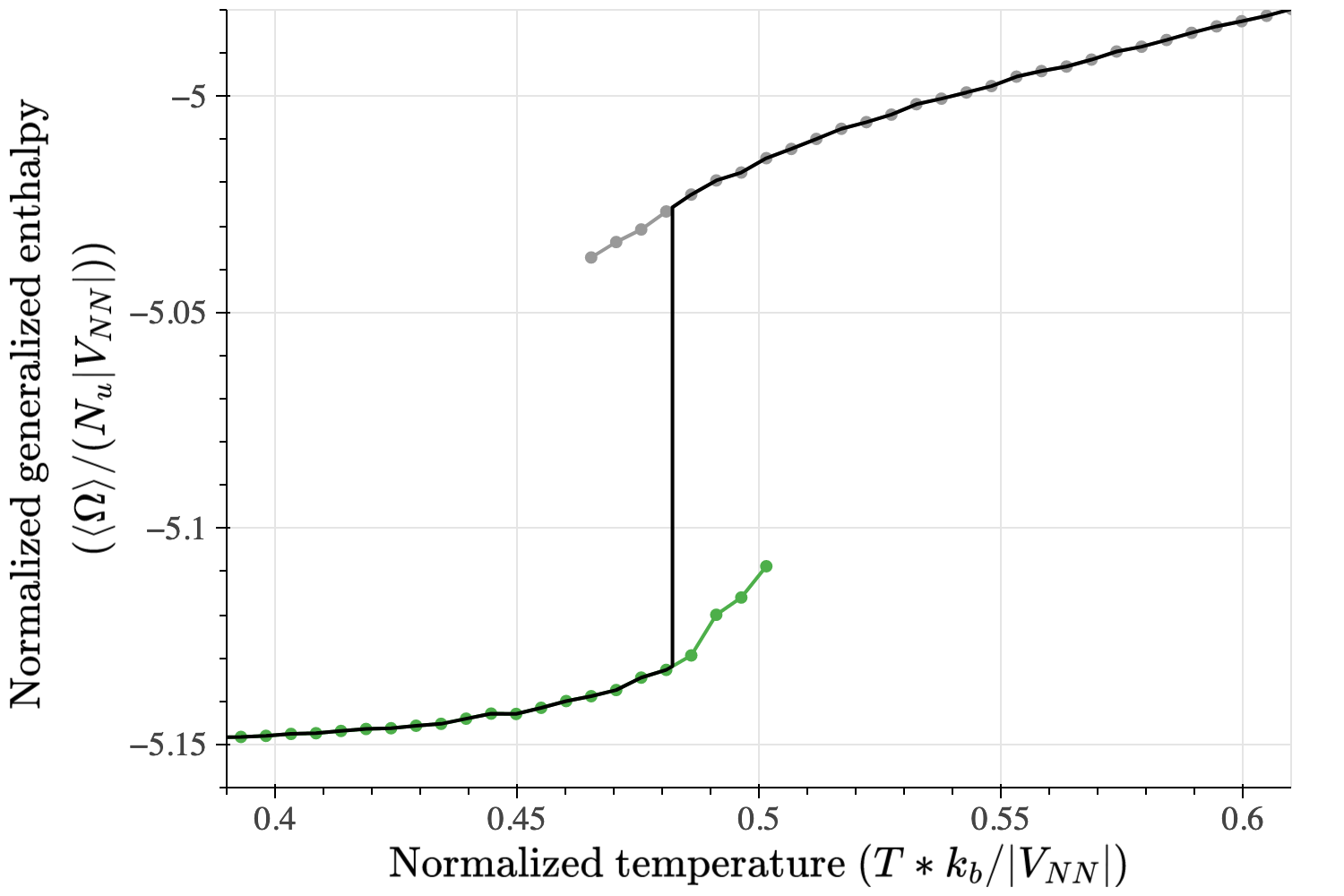}
    \label{fig:energy_T_A_mu0p06}
\caption{\label{fig:generalized_enthalpy_vs_T} A first order phase transition exhibits hysteresis in $\langle \Omega \rangle$ between heating (green) and cooling (gray) runs.
The value of $\langle \Omega \rangle$ in the phase with minimum semi-grand canonical potential $\phi$ is shown with black line. Results correspond to line A in Figure~\ref{fig:pd_hex}.}
\end{figure}

It is instructive to inspect other thermodynamic quantities at constant chemical potential as a function of temperature. 
Figure \ref{fig:generalized_enthalpy_vs_T} shows the average semi-grand canonical generalized enthalpy $\langle\Omega\rangle$ as a function of temperature along the constant chemical potential line labeled A in Figures \ref{fig:pd_hex}(c) and (d). 
The semi-grand canonical generalized enthalpy $\langle\Omega\rangle$ exhibits a discontinuous jump upon crossing the two-phase region. 
The discontinuity is the latent heat associated with the first-order phase transition. 
First-order phase transitions exhibit hysteresis. 
The ordered phase can remain metastable above the transition temperature upon passing through a first-order phase transition from low temperatures, while the disordered phase can be super cooled below the transition temperature upon crossing the first-order transition from high temperatures. 
This form of hysteresis can be replicated within Monte Carlo simulations when the last microstate sampled at $T$ and $\tilde{\mu}$ is used as the first microstate at the next temperature $T$+$\Delta T$ and $\tilde{\mu}+\Delta\mu$.
Figure \ref{fig:generalized_enthalpy_vs_T} shows the hysteresis in the semi-grand canonical generalized enthalpy $\langle\Omega\rangle$ from a heating run and a cooling run.

\begin{figure}
\centering
\begin{subfigure}{3in}
    \centering
    \vspace{0.2cm}
    \includegraphics[width=3in]{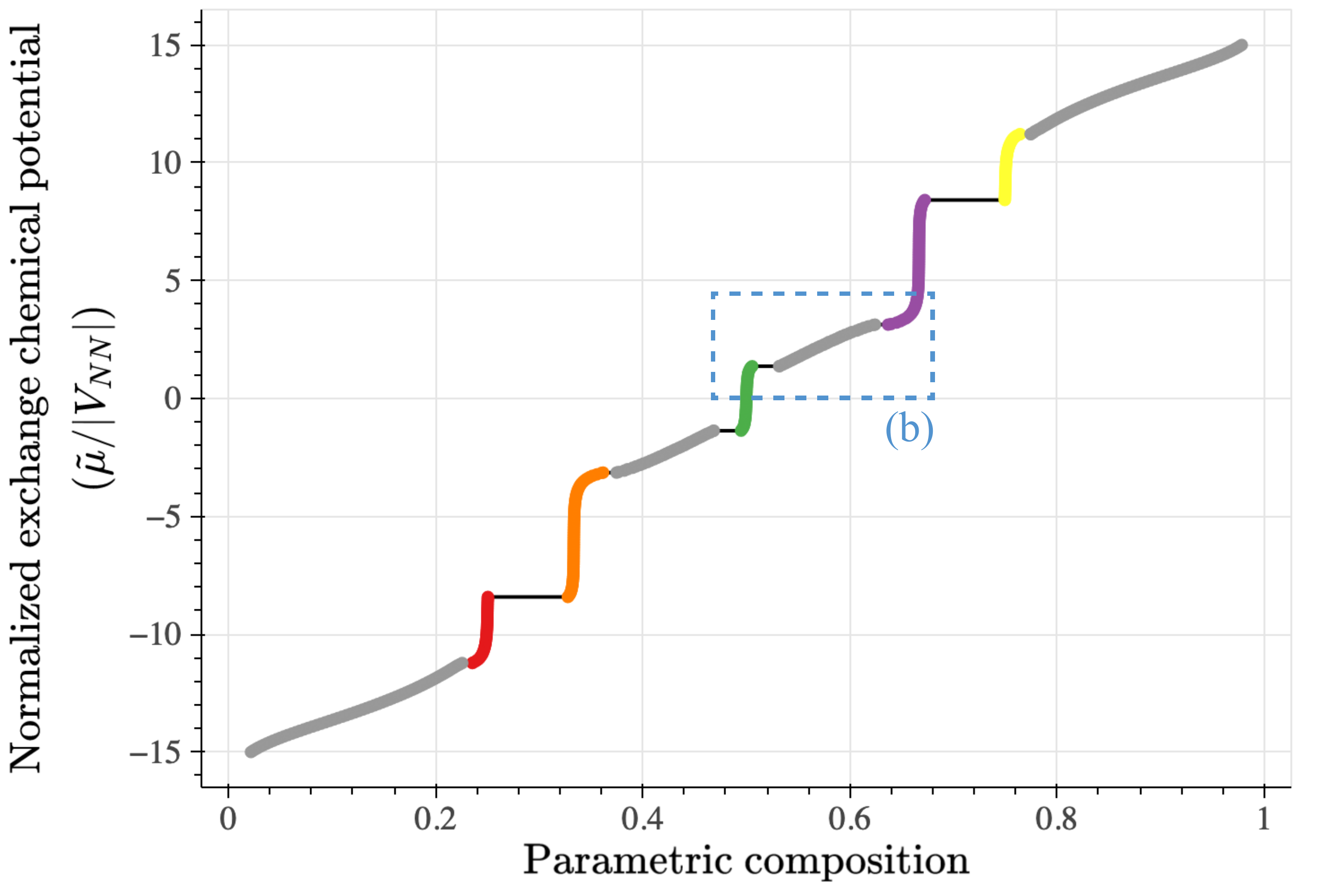}
    \label{fig:comp_vs_mu_A_175T_gs_only}
    \vspace{-0.5cm}
    \caption{}
\end{subfigure}
\begin{subfigure}{3in}
    \centering
    \vspace{0.2cm}
    \includegraphics[width=3in]{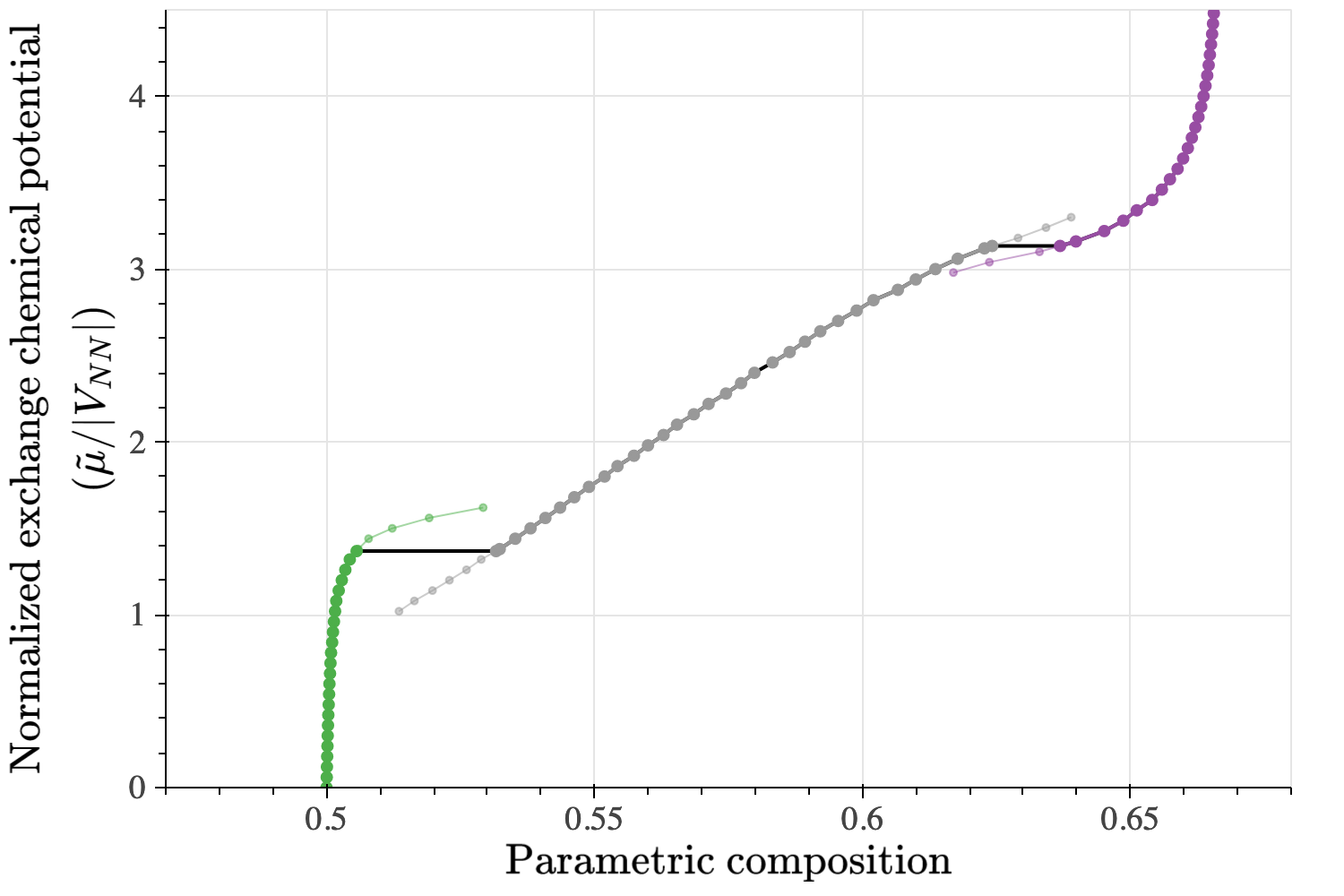}
    \label{fig:comp_mu_A_175T_metastable}
    \vspace{-0.5cm}
    \caption{}
\end{subfigure}
\caption{\label{fig:comp_line} 
The parametric composition $x$ for the phase with the minimum semi-grand canonical potential $\phi$ at each $\tilde{\mu}$ value is shown in (a) across the entire  $\tilde{\mu}$ range.
Hysteresis in the calculated parameteric composition between increasing / decreasing $\tilde{\mu}$ runs is shown in (b) for the limited $\tilde{\mu}$ range indicated by the box in (a).
Results correspond to line B in Figure~\ref{fig:pd_hex}.
}
\end{figure}

It is also often useful to inspect the variation of $\tilde{\mu}$ as a function of $x$ at constant temperature.
Figure \ref{fig:comp_line}(a) shows the variation of the chemical potential as a function of the composition $x$ along the constant temperature line labeled B in Figures \ref{fig:pd_hex}(c) and (d). 
The chemical potential varies continuously in single phase regions, sloping gently in disordered solid solutions, but varying strongly over a small composition interval in ordered phases. 
The plateaus in the chemical potential versus composition plots correspond to two-phase regions. 
This is because the coexisting phases within the two-phase regions have the same chemical potential. 
Each plateau in a constant temperature plot of $\tilde{\mu}$ versus $x$ corresponds to a first-order phase transition due to a discontinuous variation of the composition at the chemical potential of the plateau.
These first-order phase transitions are also accompanied by hysteresis in Monte Carlo simulations. 
Figure \ref{fig:comp_line}(b) shows hysteresis as it emerges in Monte Carlo simulations at constant temperature. 
The curve connected by the solid black line in Figure \ref{fig:comp_line}(b) corresponds to the true equilibrium chemical potential versus composition curve. 
However, Monte Carlo simulations can exhibit a path dependence when passing through a thermodynamic first-order phase transition at constant temperature, as is clear by the metastable extensions of the chemical potential versus composition curves in Figure \ref{fig:comp_line}(b).
The true equilibrium curve can be determined by minimizing the free energy as described in Section \ref{sec:free_energy_integration}.

The Monte Carlo simulations used to generate Figures \ref{fig:pd_hex}, \ref{fig:generalized_enthalpy_vs_T} and \ref{fig:comp_line} were performed at varying temperatures, $T$, and exchange chemical potential values, $\tilde{\mu}$, along paths where one or the other boundary condition was fixed.
For the calculated phase diagrams in Figure~\ref{fig:pd_hex}(c) and (d), Monte Carlo simulation cells were chosen so that they are commensurate with the ground state ordering at the value of $\tilde{\mu}$ where the path began.
For illustrating hysteresis in Figures~\ref{fig:generalized_enthalpy_vs_T} and~\ref{fig:comp_line}, a supercell was chosen to be fully commensurate with all symmetrically equivalent variants of the ground state orderings shown in Figure~\ref{fig:pd_hex}(b).
All supercells contain at least $N=10^{5}$ sites.
At least $10^{3}N$ Monte Carlo steps were performed, and a cutoff was enforced to stop the calculations if a maximum of $10^{7}N$ Monte Carlo steps was reached.
Calculations were run with a requested precision of $\pm10^{-3}|V_{NN}|$ for the average semi-grand canonical generalized enthalpy $\langle\Omega\rangle$ and $\pm10^{-3}$ for the average composition $\langle x\rangle$, using the method described in Section~\ref{sec:avg_converge_output} to estimate when a run was equilibrated and calculate the precision in the sample means.

\begin{figure}
\centering
\begin{subfigure}{3in}
    \centering
    \vspace{0.2cm}
    \includegraphics[width=3in]{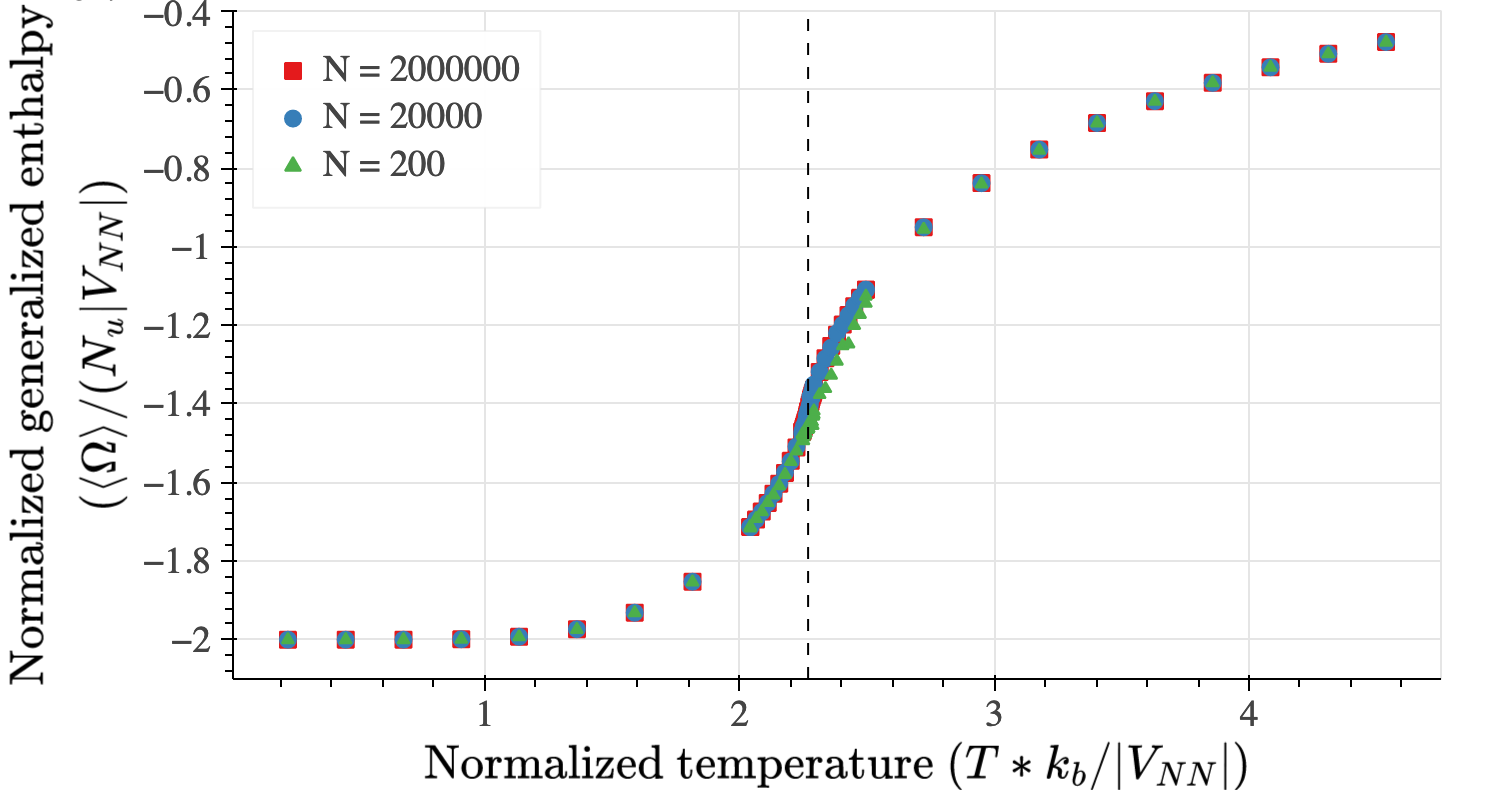}
    \label{fig:ising_generalized_enthalpy}
    \vspace{-0.5cm}
    \caption{}
\end{subfigure}
\begin{subfigure}{3in}
    \centering
    \vspace{0.2cm}
    \includegraphics[width=3in]{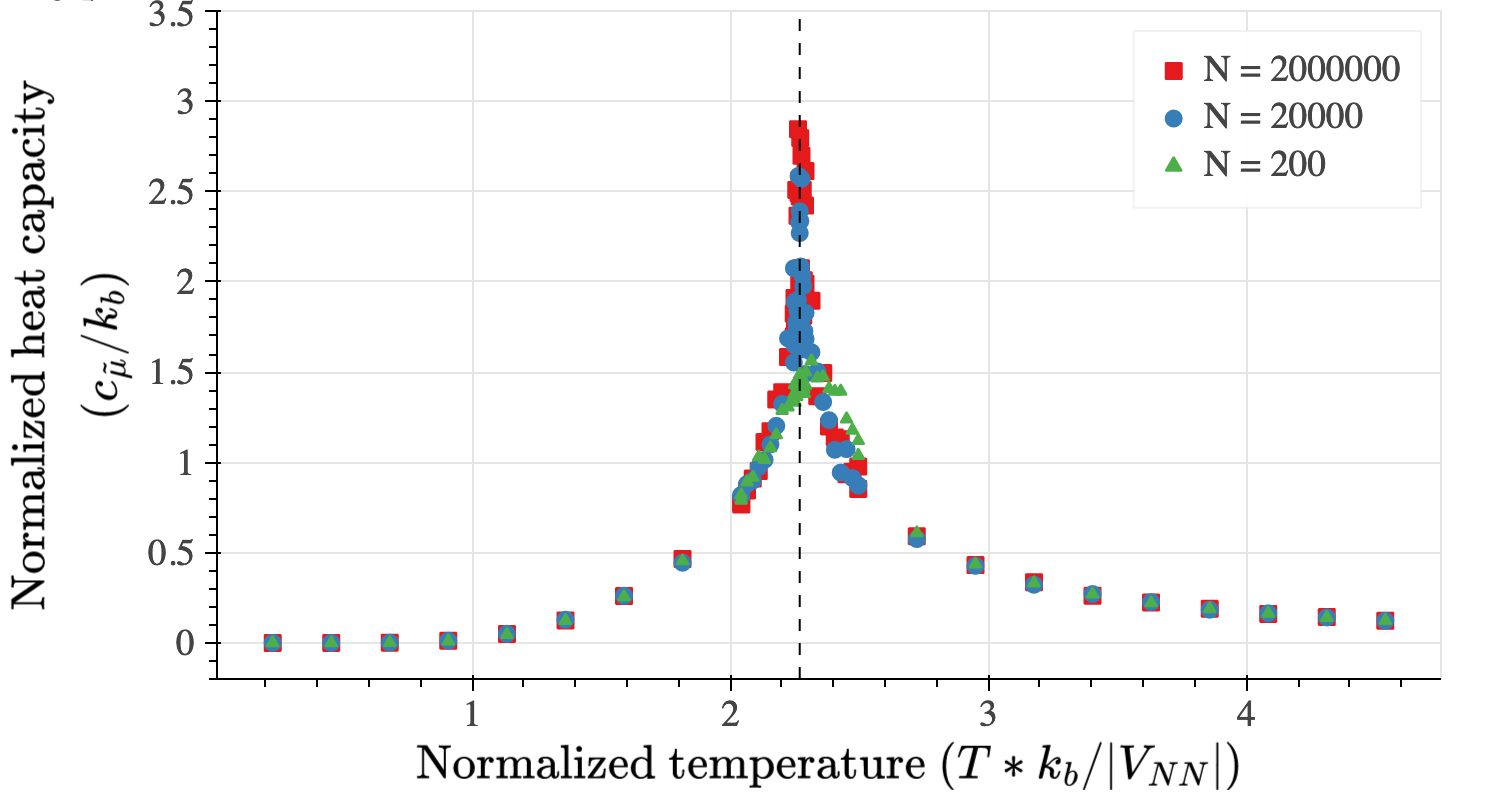}
    \label{fig:ising_heat_capacity}
    \vspace{-0.5cm}
    \caption{}
\end{subfigure}
\begin{subfigure}{3in}
    \centering
    \vspace{0.2cm}
    \includegraphics[width=3in]{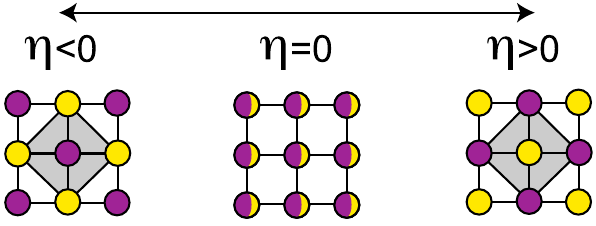}
    \label{fig:checker_board_order_parameters}
    \vspace{-0.5cm}
    \caption{}
\end{subfigure}
\begin{subfigure}{3in}
    \centering
    \vspace{0.2cm}
    \includegraphics[width=3in]{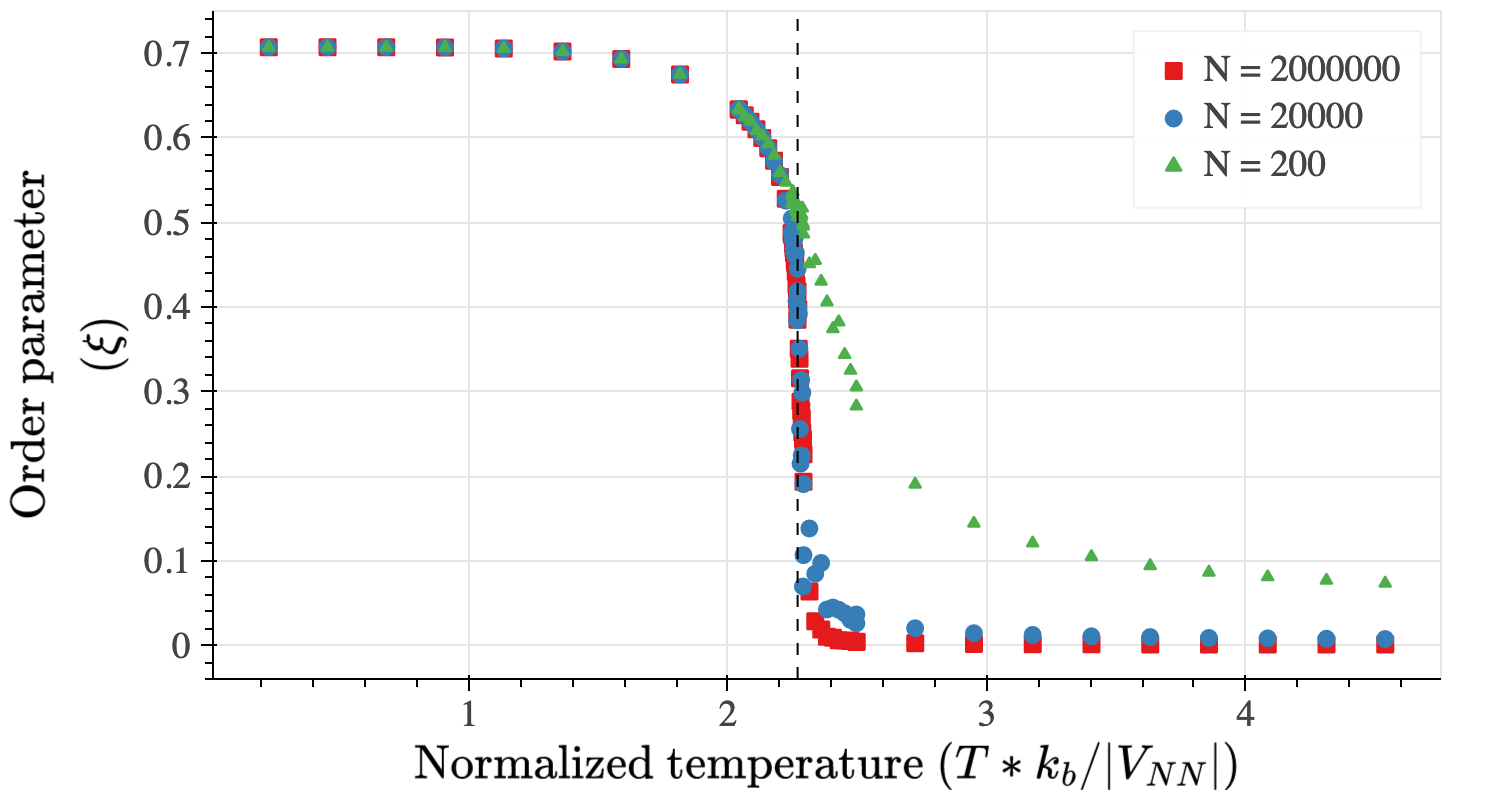}
    \label{fig:ising_order_parameter}
    \vspace{-0.5cm}
    \caption{}
\end{subfigure}
\caption{{\label{fig:ising}} Monte Carlo simulations of the second-order phase transition in the Ising model with $V_{NN}>0$ on a 2-dimensional square lattice with varying number of sites, $N$, shows (a) continuous variation in $\langle \Omega \rangle$ with temperature, (b) a strong peak in $c_{\tilde{\mu}}$ with increasing system size, and (c) an order parameter, $\xi$ as defined in Eq.~\ref{eq:checkerboard_order_parameter_xi}, which smoothly and continuously decreases to zero. The exact value of the critical temperature in the infinite system, $T_c$, is shown with a dashed line.}
\end{figure}

\newpage

\begin{figure*}
\centering
\begin{subfigure}{2in}
    \vspace{0.2cm}
    \includegraphics[width=2in]{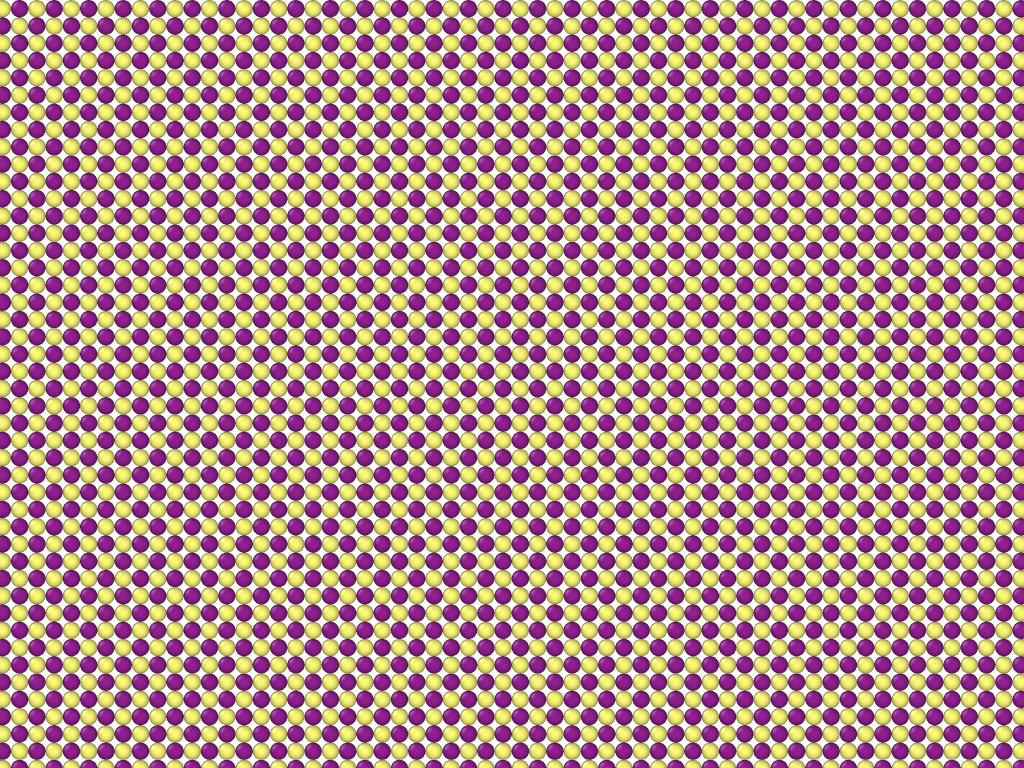}
    \label{fig:snapshot_1}
    \vspace{-0.5cm}
    \caption{$T= 0.1 * T_c$, $\xi = 1.000 * \xi_0$}
\end{subfigure}
\begin{subfigure}{2in}
    \vspace{0.2cm}
    \includegraphics[width=2in]{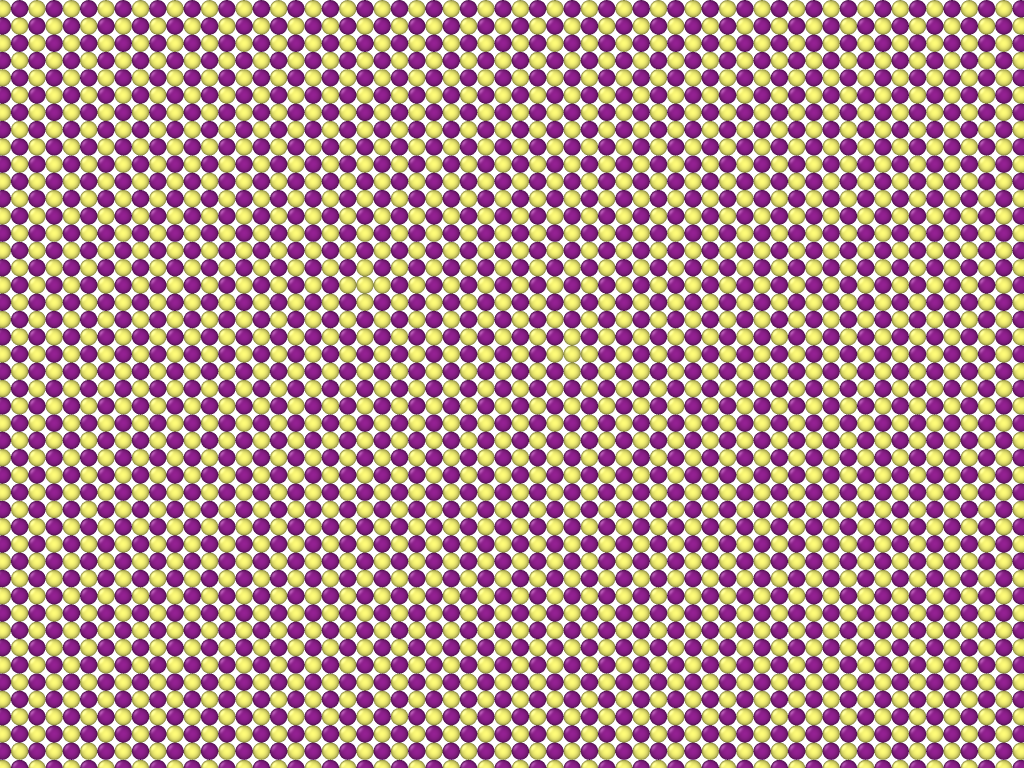}
    \label{fig:snapshot_5}
    \vspace{-0.5cm}
    \caption{$T= 0.5 * T_c$, $\xi = 0.998 * \xi_0$}
\end{subfigure}
\begin{subfigure}{2in}
    \vspace{0.2cm}
    \includegraphics[width=2in]{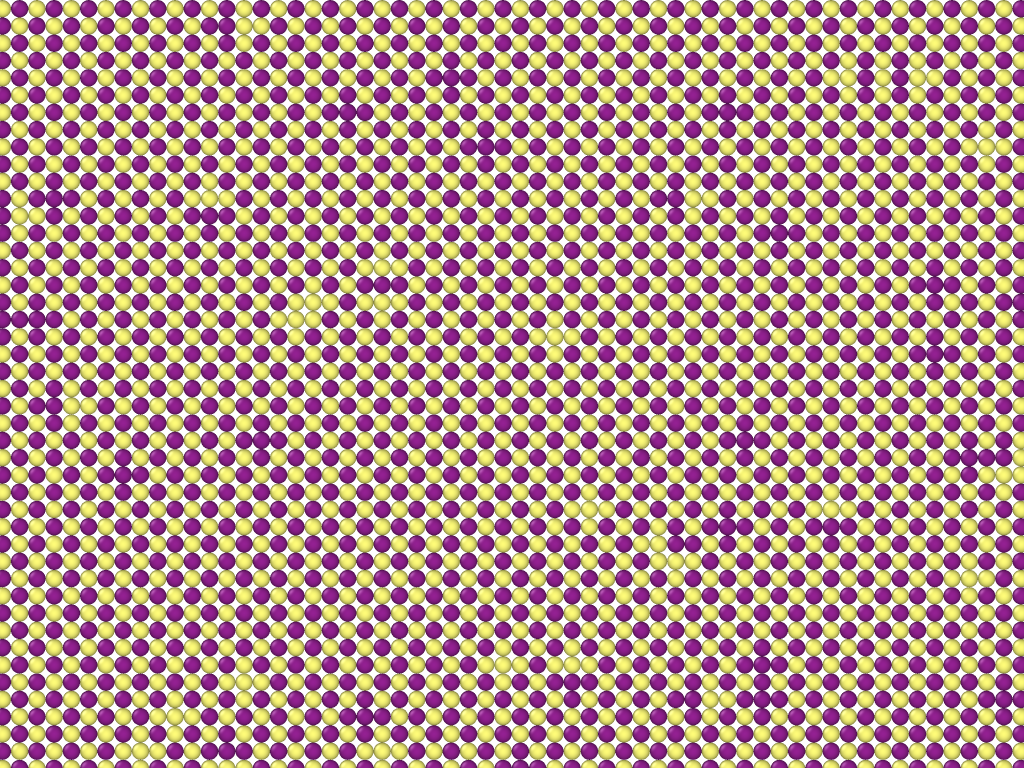}
    \label{fig:snapshot_7}
    \vspace{-0.5cm}
    \caption{$T = 0.8 * T_c$, $\xi = 0.954 * \xi_0$}
\end{subfigure}

\begin{subfigure}{2in}
    \vspace{0.2cm}
    \includegraphics[width=2in]{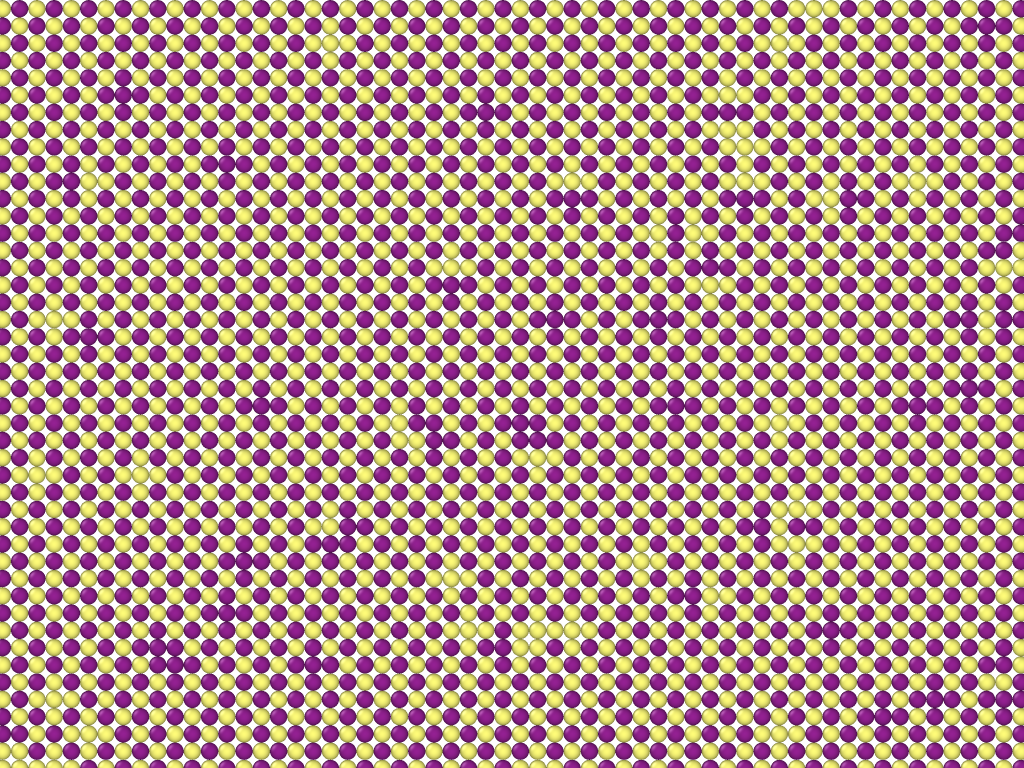}
    \label{fig:snapshot_9}
    \vspace{-0.5cm}
    \caption{$T = 0.9 * T_c$, $\xi = 0.896 * \xi_0$}
\end{subfigure}
\begin{subfigure}{2in}
    \centering
    \vspace{0.2cm}
    \includegraphics[width=2in]{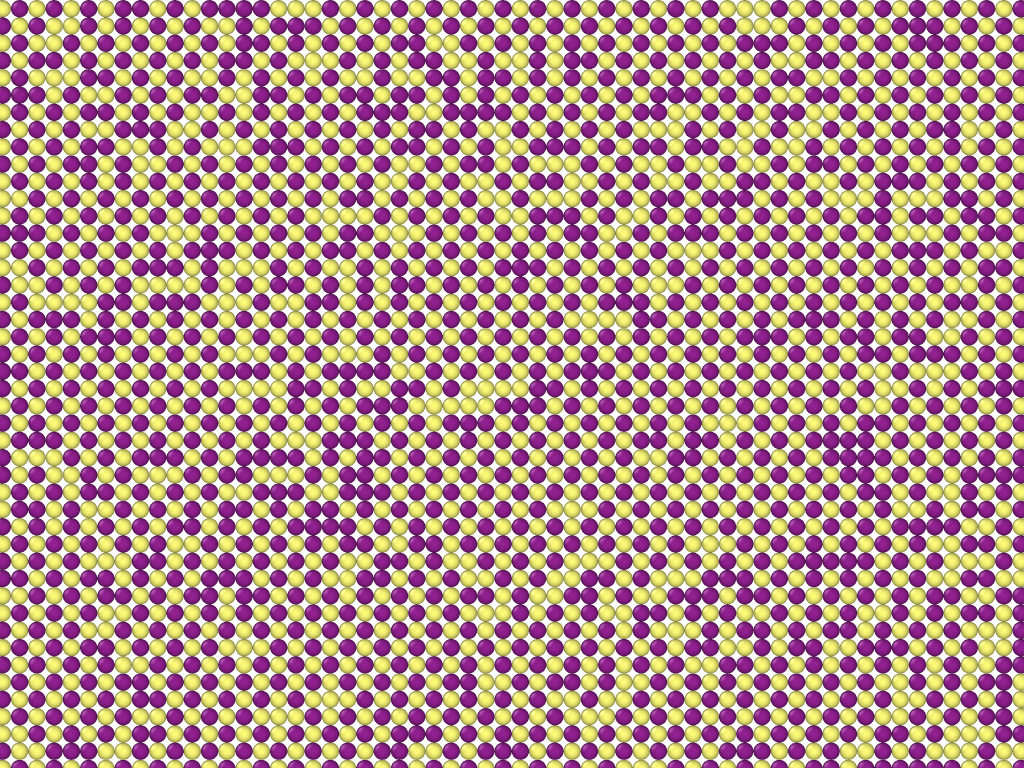}
    \label{fig:snapshot_10}
    \vspace{-0.5cm}
    \caption{$T= 1.0 * T_c$, $\xi = 0.592 * \xi_0$}
\end{subfigure}
\begin{subfigure}{2in}
    \centering
    \vspace{0.2cm}
    \includegraphics[width=2in]{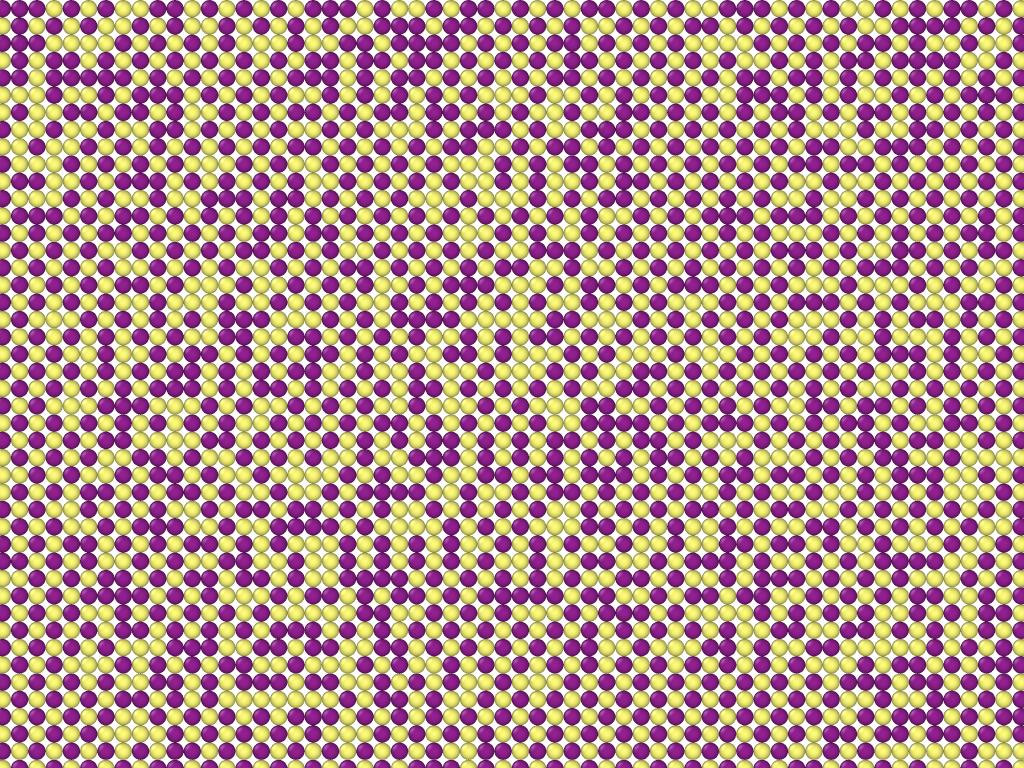}
    \label{fig:snapshot_11}
    \vspace{-0.5cm}
    \caption{$T = 1.1 * T_c$, $\xi = 0.005 * \xi_0$}
\end{subfigure}

\begin{subfigure}{2in}
    \centering
    \vspace{0.2cm}
    \includegraphics[width=2in]{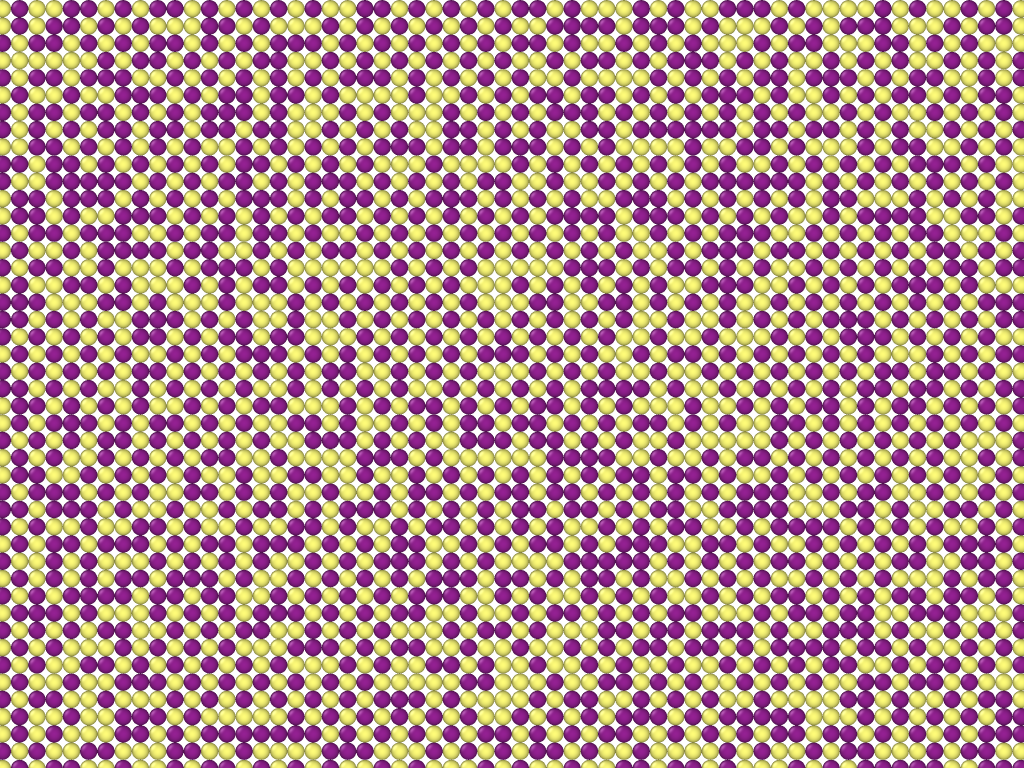}
    \label{fig:snapshot_12}
    \vspace{-0.5cm}
    \caption{$T= 1.2 * T_c$, $\xi = 0.003 * \xi_0$}
\end{subfigure}
\begin{subfigure}{2in}
    \centering
    \vspace{0.2cm}
    \includegraphics[width=2in]{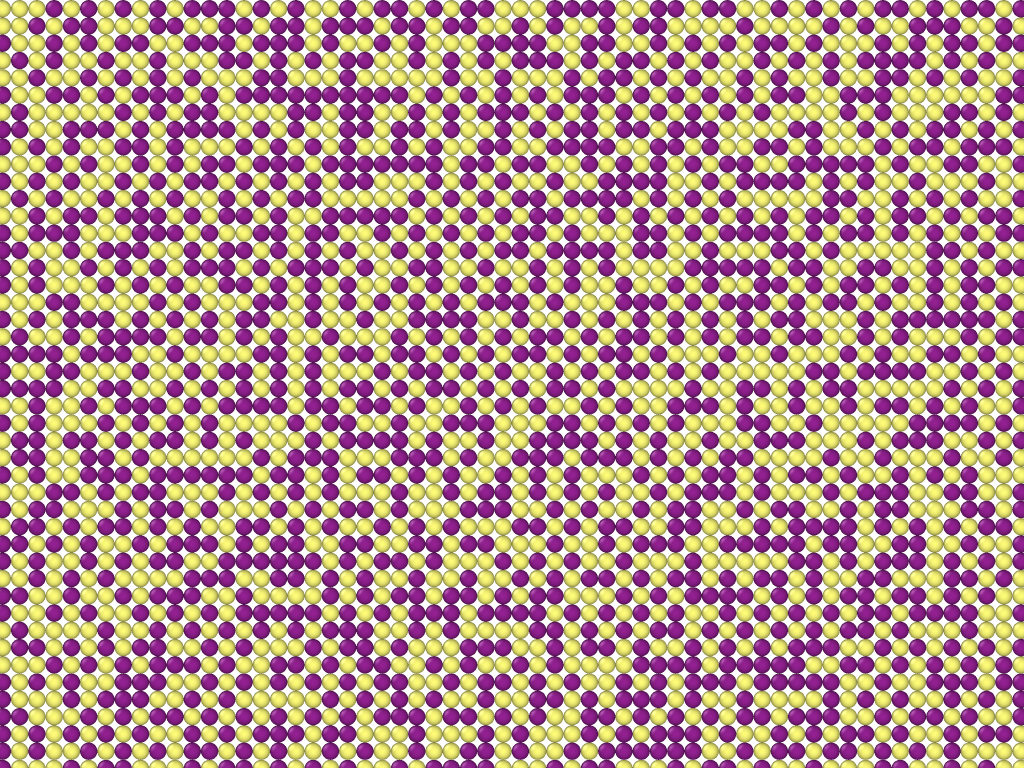}
    \label{fig:snapshot_15}
    \vspace{-0.5cm}
    \caption{$T = 1.5 * T_c$, $\xi = 0.001 * \xi_0$}
\end{subfigure}
\begin{subfigure}{2in}
    \centering
    \vspace{0.2cm}
    \includegraphics[width=2in]{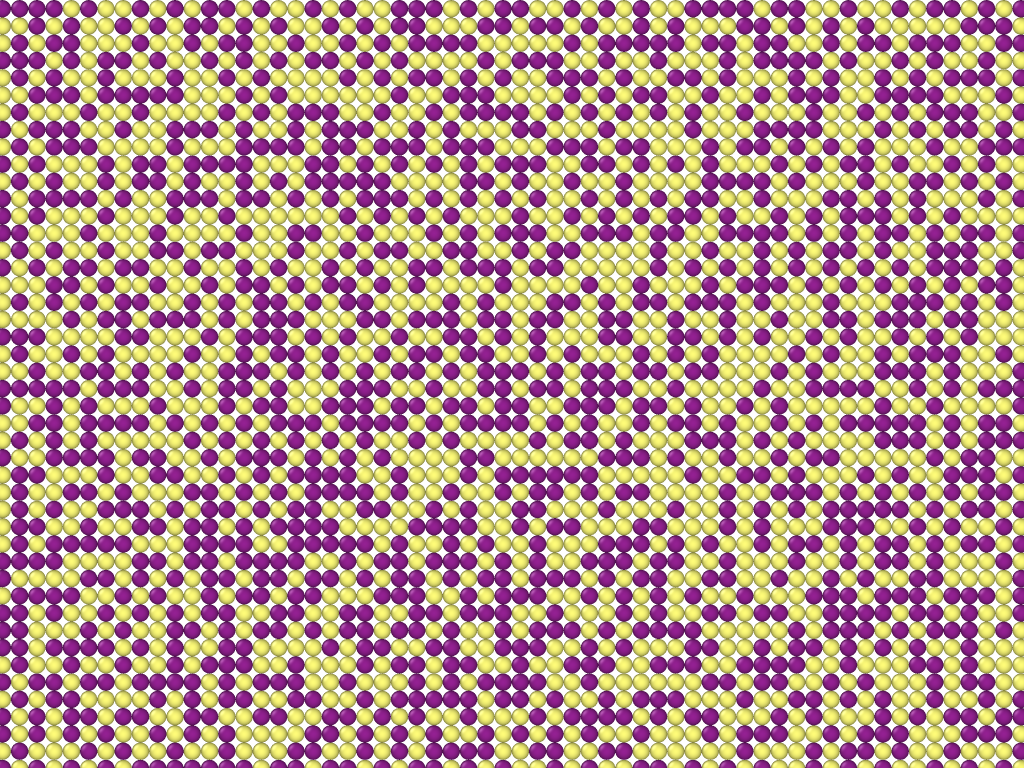}
    \label{fig:snapshot_20}
    \vspace{-0.5cm}
    \caption{$T = 2.0 * T_c$, $\xi = 0.001 * \xi_0$}
\end{subfigure}
\caption{\label{fig:snapshots_top}Sections of the final state of the $N=2*10^6$ Monte Carlo supercells (red squares in Fig.~\ref{fig:ising}) and calculated order parameter, $\xi$, as a fraction of the order parameter for the perfect checkerboard ordering, $\xi_0 = 1/\sqrt{2}$, at varying temperatures relative to the exact critical temperature, $T_c$, for the second order transition in the infinitely large system. Images were generated using OVITO \cite{ovito}.}
\end{figure*}

\subsection{Example 2: Second order-phase transitions}

A first-order phase transition is characterized by discontinuities in extensive thermodynamic variables. 
As illustrated in the previous section, a first-order phase transition within the semi-grand canonical ensemble exhibits a discontinuity in the generalized enthalpy, $\langle\Omega\rangle$, and in the composition $\langle x \rangle$. 
These thermodynamic quantities are related to first derivatives of the characteristic free energy of the system and change discontinuously when the system transitions between phases residing on separate free energy surfaces. 
Phase transitions can also be of second order. 
In contrast to a first-order phase transition, the extensive thermodynamic variables related to first derivatives of the characteristic free energy vary continuously upon passing through a second order transition.
While this can make it difficult to detect a second order transition, a stronger signature is manifested by the response functions, some of which exhibit a divergence at a second order transition. 
Response functions, such as the heat capacity and the generalized susceptibilities, are related to second derivatives of the characteristic free energy. 

To illustrate the characteristics of a second order phase transition, we consider a simple nearest-neighbor pair cluster expansion model for a binary system of A and B atoms on a two-dimensional square lattice
\begin{align}
    E(\vec{\sigma})=\sum_{(i,j)= NN}V_{NN}\sigma_i\sigma_j.
\end{align}
The occupation variables, $\sigma_i$, assigned to each lattice site $i$ are equal to 1 or -1 depending whether the occupying species is A or B, respectively.
This model, with a positive nearest-neighbor interaction coefficient $V_{NN}$, favors a checkerboard ordering pattern at low temperature and undergoes a disordering reaction at elevated temperature that is of second order.

Monte Carlo simulations were performed in supercells commensurate with the checkerboard ordering with varying number of sites, $N$, at varying temperatures, and at a fixed exchange chemical potential value, $\tilde{\mu}=0.0$.
At least $10^{4}N$ Monte Carlo steps were performed, and a cutoff was enforced to stop the calculations if a maximum of $10^{5}N$ Monte Carlo steps was reached.
Calculations were run with a requested precision of $\pm10^{-3}|V_{NN}|$ for the average semi-grand canonical generalized enthalpy $\langle\Omega\rangle$ and $\pm10^{-4}$ for the parametric composition $\langle x\rangle$, using the method described in Section~\ref{sec:avg_converge_output} to estimate when a run was equilibrated and calculate the precision in the sample means.

Figure \ref{fig:ising}(a) shows the variation of the average semi-grand canonical generalized enthalpy $\langle\Omega\rangle$ as a function of temperature. 
In contrast to a first order phase transition, it does not exhibit a discontinuity at the transition temperature, but rather continuously passes through an inflection point. 
Similarly, other thermodynamic variables that are related to extensive quantities, such as the composition, vary continuously through a second order phase transition. 

The occurrence of a second order transition is more evident upon inspection of the heat capacity.
Figure \ref{fig:ising}(b) shows the calculated heat capacity as a function of temperature. 
There is a clear tendency of the heat capacity to diverge at the second order transition temperature. 
The heat capacity is proportional to the variance in the fluctuating semi-grand canonical generalized enthalpy $\Omega$ according to Eq. \ref{eq:heatcap}.
Since the correlation length of spatial fluctuations in energy and composition diverges at a second order phase transition, the full spectrum of fluctuations that contribute to the divergence of a response function cannot be captured in a finite sized Monte Carlo cell. 
Both the peak of the divergence and the temperature at which the divergence occurs will therefore vary with the size of the Monte Carlo cell. 
This is evident in Figure \ref{fig:ising}(b), which shows the heat capacity as calculated using different sizes of the Monte Carlo simulation cell. 
Finite size scaling approaches have been developed to estimate the true second order transition temperatures.\cite{binder2002monte} 




Order parameters are another set of variables that aid the identification of a second-order transition. 
A useful order parameter to detect the checkerboard ordering on a square lattice is defined as a difference of the two sublattice concentrations, $n^1_B$ and $n^2_B$, according to 
\begin{equation}
    \eta=\frac{1}{\sqrt{2}}(n^1_B-n^2_B)
    \label{eq:checkerboard_order_parameter}
\end{equation}
where $n^1_B$ and $n^2_B$ track the concentration of species B on the sublattices of the checkerboard supercell as illustrated in Figure \ref{fig:ising}(c). 
This order parameter is zero in the fully disordered state when $n^1_B=n^2_B$ and has a non-zero value $\pm 1/\sqrt{2}$ for the perfect checkerboard ordering.  
It is also able to distinguish the two translational variants of the checkerboard ordering, with positive values of $\eta$ signifying one translational variant and negative values the other translational variant. 
In the thermodynamic limit, order parameters such as $\eta$ decrease continuously to zero upon approaching a second order transition from low temperatures when maintaining one particular variant of the ordered phase. 
In finite sized Monte Carlo cells, however, the cell may fluctuate between different translational and/or orientational variants of the same ordered phase, resulting in a mean value of $0$ for $\langle \eta \rangle$ when averaging over long computational times, even below the transition temperature.
To measure the degree of ordering irrespective of the translational variant, the norm 
\begin{equation}
    \xi = \sqrt{\eta^2}
    \label{eq:checkerboard_order_parameter_xi}
\end{equation}
can be used.
Figure \ref{fig:ising}(d) shows the calculated order parameter $\xi$ as a function of temperature. 
While not going to zero at the true transition temperature, the average value of $\xi$ approaches values that are close to zero as long as a large enough supercell size is used.

Portions of the final state of Monte Carlo simulations in supercells with $N=2 \times 10^6$ sites are shown in Figure~\ref{fig:snapshots_top} at varying temperatures. 
For this model, the exact critical temperature, $T_c$, of the second order transition in the infinite system can be calculated exactly as $T_c k_{B}/|V_{NN}|=2/\ln(1+\sqrt{2})$ \cite{Onsager1944}. 
As the temperature is increased to $0.9 * T_c$, increasing numbers of anti-site defects in the checkerboard ordering can be observed without the formation of distinct domains and $\xi$ remains close to the value for perfect ordering.
At $1.0 * T_c$, the two translational variants of the checkerboard ordering can be observed as distinct domains and the order parameter $\xi$ is sharply reduced.
As high as $1.5 * T_c$, local regions with the checkerboard ordering are still clearly noticeable, but since both variants exist in larger supercells the order parameter $\xi$ rapidly drops to zero.
As the temperature increases further, short range order diminishes and the site occupation approaches that of a random alloy.

\begin{figure}
\centering
\includegraphics[width=1.5in]{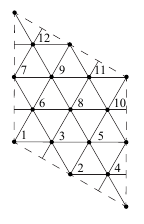}
\vspace{-0.5cm}
\caption{Sublattices for order parameter basis definition in the $2\sqrt{3}a \times 2\sqrt{3}a$ supercell.}
\label{fig:orderparameter_sublattices}
\end{figure}

\subsection{Example 3: Tracking long-range order}

As was illustrated for the checkerboard ordering in the previous section, order parameters are invaluable in tracking the degree of long-range order within Monte Carlo simulations. 
The order parameter, $\eta$, Eq. \ref{eq:checkerboard_order_parameter}, used to track the presence of the checkerboard ordering is especially simple. 
It is equal to zero in the absence of long-range order and has either a positive or negative value when the atoms adopt one of the two translational variants of the checkerboard pattern. 
Long-range order parameters can be identified for any ordered phase such that they distinguish all the symmetrically equivalent variants of the ordering and become equal to zero in the disordered state. 
Algorithmic approaches to formulate order parameters for any ordered phase are described in Natarajan et al \cite{natarajan2017symmetry} and Walsh et al \cite{walsh2022order} and have been implemented in CASM.
In general, order parameters that track an ordered phase on a particular parent crystal structure are not as simple as the checkerboard order parameter, Eq. \ref{eq:checkerboard_order_parameter}. 
To enable a distinction between the disordered phase and all the symmetrically equivalent variants of a particular ordering, long-range order parameters are defined as symmetry-adapted linear combinations of sublattice concentrations within the supercell that defines the periodicity of the ordered phase.\cite{natarajan2017symmetry} 
The ability to distinguish multiple symmetrically equivalent variants of a particular ordered phase often requires two or more order parameter variables. 
Furthermore, a parent crystal may host multiple distinct ordered phases, which each require their own set of order parameters. 

To illustrate how long-range ordering can be tracked in a system that hosts several ordered phases we revisit the triangular lattice cluster expansion of Section \ref{sec:triangular_lattice_first_order_transition}. 
This model system favors five ordered phases at low temperatures (Figure \ref{fig:pd_hex}(b)) that exhibit three distinct superlattice periodicities. 
The orderings at $x=1/4$ and $x=3/4$ form in a $2a\times 2a$ supercell of the underlying parent triangular lattice unit cell and each have four symmetrically equivalent translational variants. 
The orderings at $x=1/3$ and $x=2/3$ have a $\sqrt{3}a\times\sqrt{3}a$ supercell and have three translational variants each. 
The ordering at $x=1/2$, consisting of rows of atoms alternated by rows of vacancies, has a $a\times\sqrt{3}a$ supercell, and due to its lower symmetry has both symmetrically equivalent translation variants as well as orientational variants.  
The row ordering can adopt three orientations on the parent triangular lattice and each orientation has two symmetrically equivalent translational variants. 

\begin{figure}
\centering
\includegraphics[width=3in]{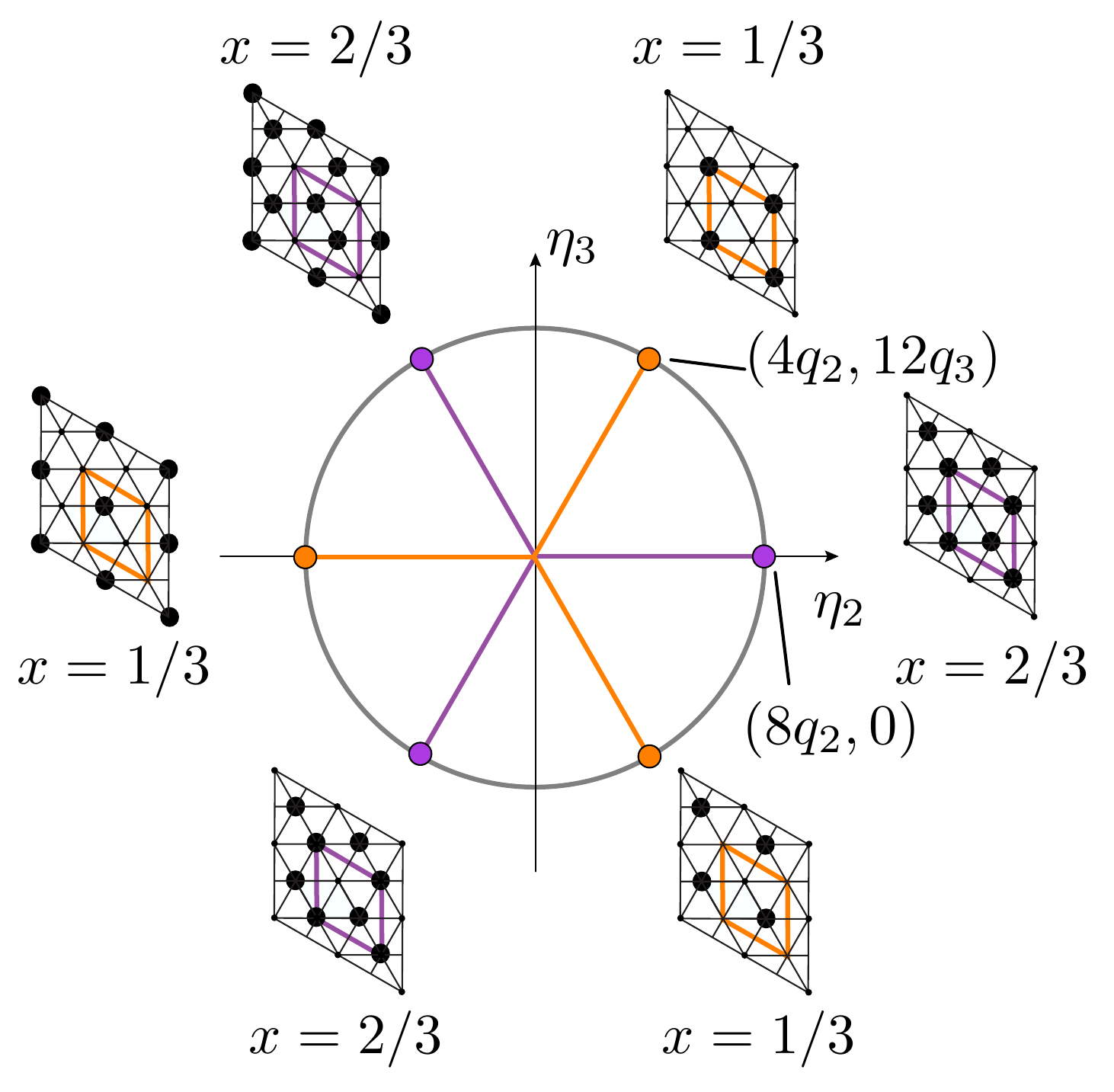}
\caption{\label{fig:orderparameter1} The coordinates of the translational variants of the $\sqrt{3}a\times\sqrt{3}a$ orderings at $x=1/3$ and $x=2/3$ in the order parameter space spanned by $\eta_2$ and $\eta_3$.}
\end{figure}

A suitable set of order parameters should be able to distinguish the disordered state, the different orderings and the different translational and orientational variants of each particular ordered phase. 
The approach to identifying order parameters starts with a determination of the smallest supercell on the parent crystal that is commensurate with all orderings and all their translational and orientational variants.\cite{natarajan2017symmetry} 
For the orderings of Figure \ref{fig:pd_hex}(b), this is the 12 sublattice $2\sqrt{3}a\times2\sqrt{3}a$ supercell shown in Figure \ref{fig:orderparameter_sublattices}. 
Similar to the formulation of the checkerboard order parameter, sublattice concentrations are introduced for each of the 12 sublattices of the mutually commensurate supercell of Figure \ref{fig:orderparameter_sublattices}.
Group theoretical techniques as described by Natarajan et al \cite{natarajan2017symmetry} and Thomas et al \cite{thomas2017exploration} then enable the identification of symmetry adapted linear combinations of the sublattice concentrations. 
For these particular orderings, 6 useful order parameters emerge that can be divided into three subspaces. 
The first, defined as the average of all sublattice concentrations of species B, $n^i_B$, $i$ being the sublattice index, within the $2\sqrt{3}a\times2\sqrt{3}a$ supercell of Figure \ref{fig:orderparameter_sublattices} according to
\begin{equation}
    \eta_0=q_1\sum_{i=1}^{12}n^i_B,
\end{equation}
with $q_1=\frac{1}{\sqrt{12}}$, is invariant to the symmetry of the parent lattice and simply tracks the overall concentration of the solid. 
The next two order parameters together are able to distinguish the orderings at $x=1/3$ and $x=2/3$ and are defined in terms of the sublattice concentrations according to
\begin{align}
\begin{bmatrix} \eta_2 & \eta_3 \end{bmatrix} = \vec{n}_B^\mathsf{T}\begin{bmatrix} 
2q_2 & 0 \\
2q_2 & 0 \\
-q_2 & q_3 \\
-q_2 & q_3 \\
-q_2 & -q_3 \\
-q_2 & -q_3 \\
2q_2 & 0 \\
2q_2 & 0 \\
-q_2 & q_3 \\
-q_2 & q_3 \\
-q_2 & -q_3 \\
-q_2 & -q_3 
\end{bmatrix},
\end{align}
where $q_2=1/(2\sqrt{6})$, $q_3=1/(2\sqrt{2})$, and $\vec{n}_B^\mathsf{T} = \left[n^1_B, n^2_B, \dots, n^{12}_B \right]$.
Figure \ref{fig:orderparameter1} shows the coordinates of the translational variants of the $x=1/3$ and $x=2/3$ orderings within the $\sqrt{3}a\times\sqrt{3}a$ supercell within the two-dimensional $\eta_2$ and $\eta_3$ order parameter subspace.  
The three translational variants of a particular $\sqrt{3}a\times\sqrt{3}a$ ordering reside on a circle at $120^{o}$ intervals. 
The fully disordered state resides at the origin in the supspace spanned by $\eta_2$ and $\eta_3$.

\begin{figure}
\centering
\includegraphics[width=3in]{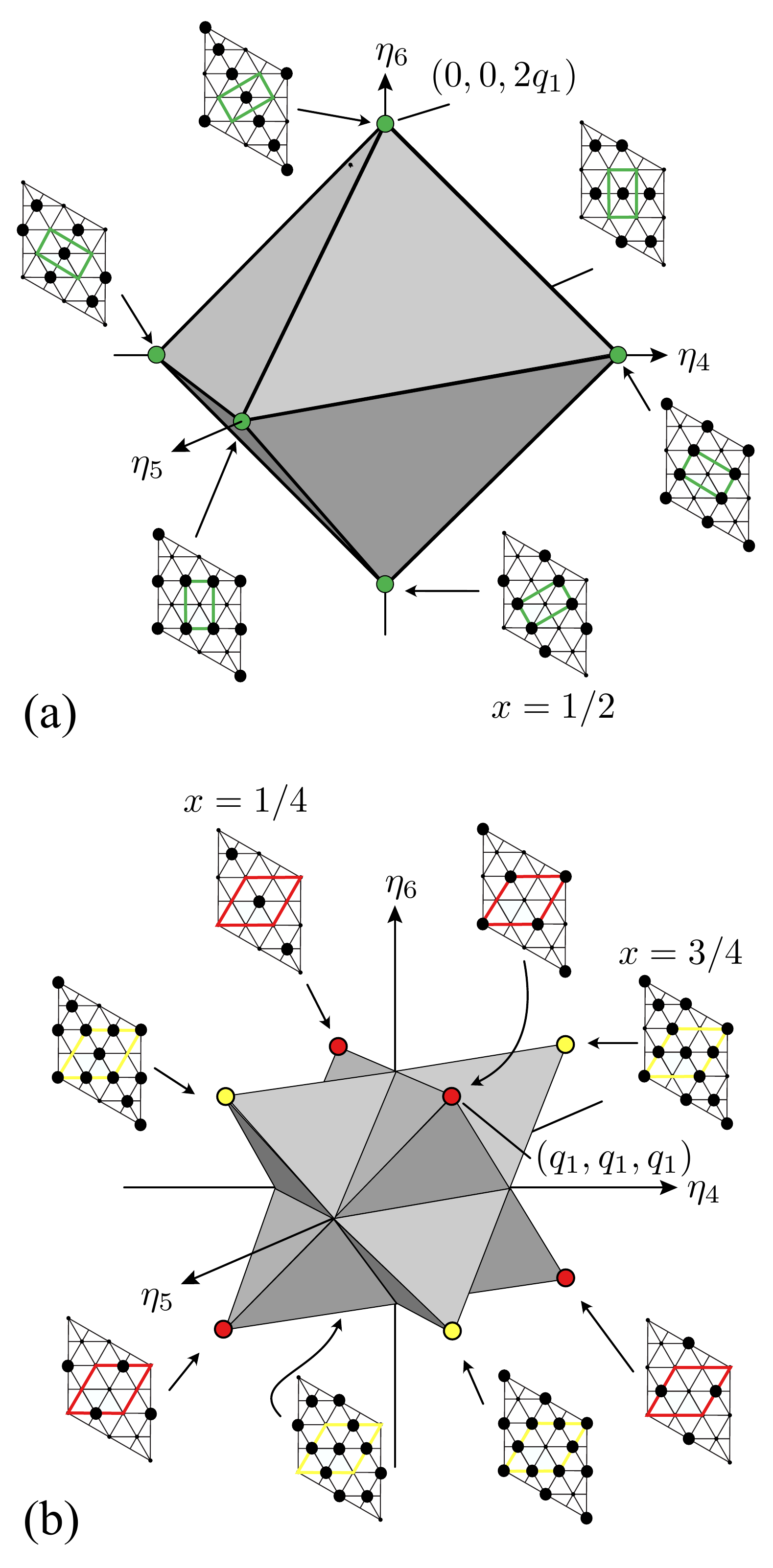}
\caption{\label{fig:orderparameter2} The coordinates of the translational and orientational variants of the $x=1/4$, $x=1/2$ and $x=3/4$ in the order parameter space spanned by $\eta_4$, $\eta_5$ and $\eta_5$.}
\end{figure}

The third category of order parameters span a three dimensional subspace and track the $x=1/4$, $x=3/4$ and $x=1/2$ ordered phases. 
These are defined in terms of the sublattice concentrations of the $2\sqrt{3}\times2\sqrt{3}a$ supercell of Figure \ref{fig:orderparameter_sublattices} according to
\begin{align}
\begin{bmatrix} \eta_4 & \eta_5 & \eta_6 \end{bmatrix} = \vec{n}_B^\mathsf{T}\begin{bmatrix}
-q_1 & -q_1 & -q_1 \\
-q_1 & q_1 & q_1 \\
q_1 & -q_1 & q_1 \\
q_1 & q_1 & -q_1 \\
-q_1 & -q_1 & -q_1 \\
-q_1 & q_1 & q_1 \\
q_1 & -q_1 & q_1 \\
q_1 & q_1 & -q_1 \\
-q_1 & -q_1 & -q_1 \\
-q_1 & q_1 & q_1 \\
q_1 & -q_1 & q_1 \\
q_1 & q_1 & -q_1 \end{bmatrix}.
\end{align}
Figure~\ref{fig:orderparameter2} illustrates the coordinates of the symmetrically equivalent variants of the orderings at $x=1/4$, $x=1/2$ and $x=3/4$.
The four translational variants of the $x=1/4$ and $x=3/4$ orderings form tetrahedra in the order parameter space spanned by $\eta_4$, $\eta_5$ and $\eta_6$. 
The row ordering at $x=1/2$, which has both orientational and translational variants reside at the corners of an octahedron in the same three dimensional subspace.\cite{natarajan2017symmetry} 

\begin{figure}
\centering
\begin{subfigure}{3in}
    \centering
    \vspace{0.2cm}
    \includegraphics[width=3in]{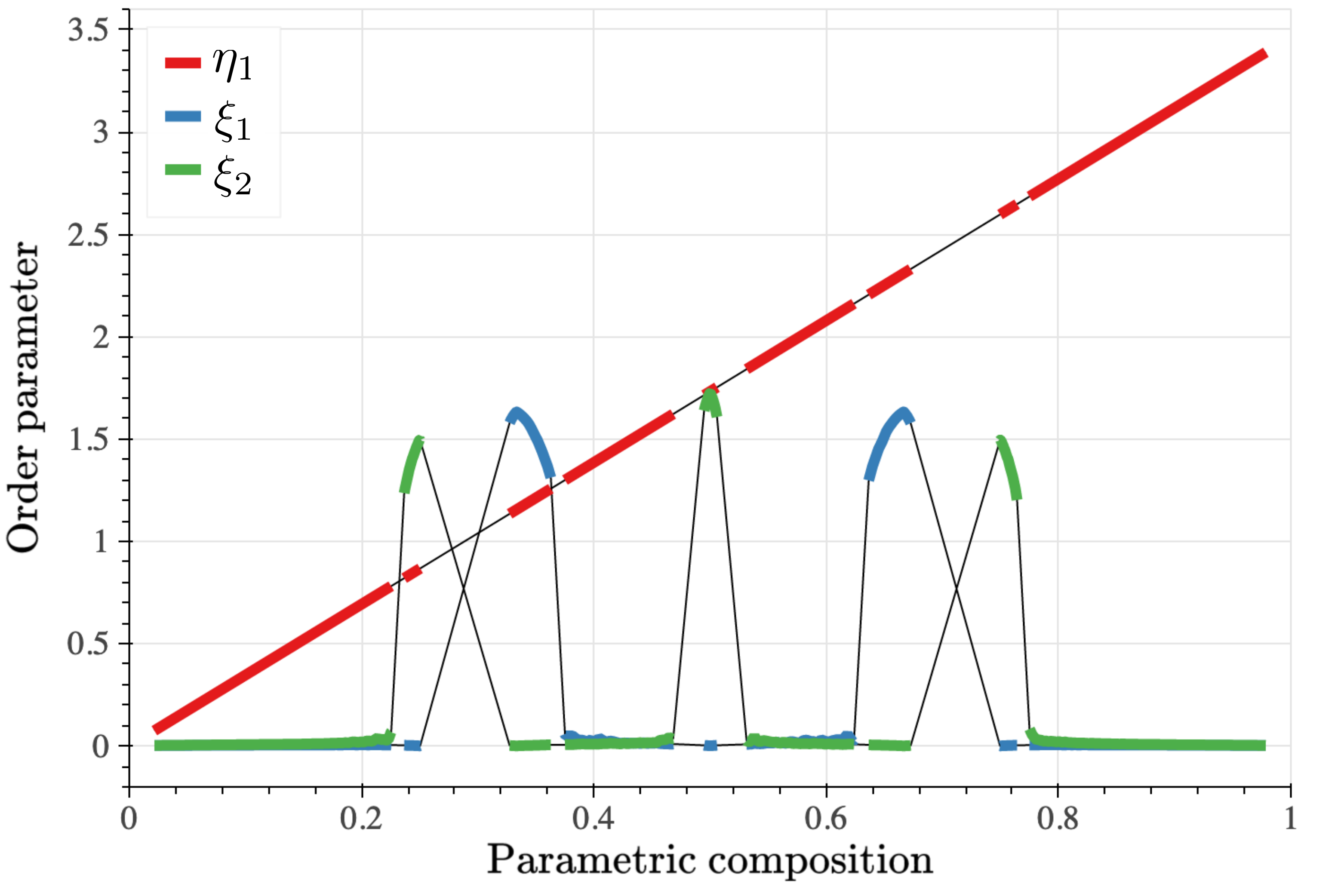}
    \label{fig:op_comp_by_irrep}
    \vspace{-0.5cm}
    \caption{}
\end{subfigure}
\begin{subfigure}{3in}
    \centering
    \vspace{0.2cm}
    \includegraphics[width=3in]{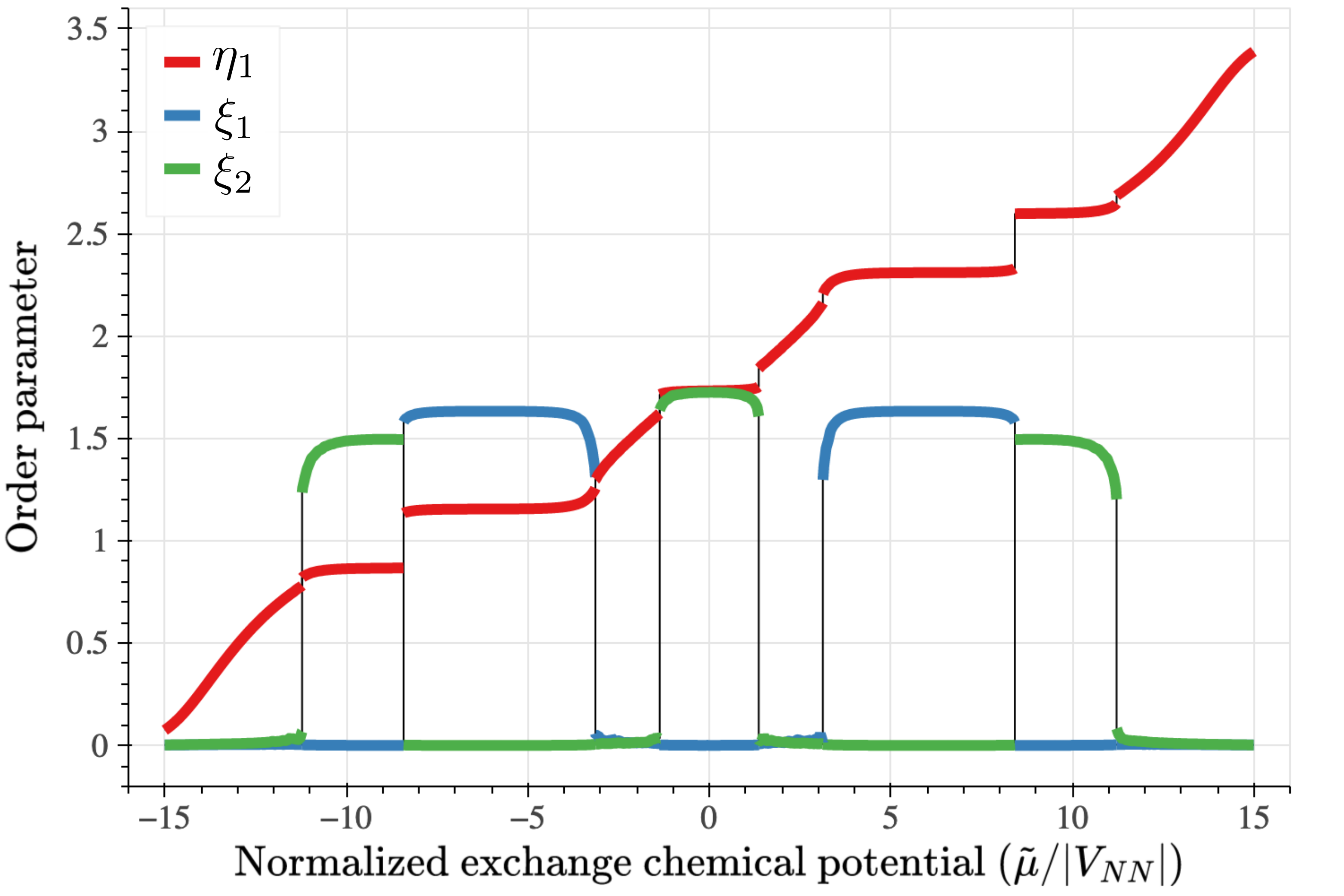}
    \label{fig:op_mu_by_irrep}
    \vspace{-0.5cm}
    \caption{}
\end{subfigure}
\caption{\label{fig:order_parameter} 
The magnitude of vectors formed from the symmetry adapted order parameters in each irreducible subspace for the fully commensurate supercell of the ground state orderings. 
Results correspond to line B in Figure~\ref{fig:pd_hex}.
}
\end{figure}

While the full set of 6 order parameters are necessary to distinguish all the symmetrically equivalent variants of each ordered phase in Monte Carlo simulations, it is often only necessary to detect which ordering is present, irrespective of the particular variant. 
The number of order parameters that needs to be tracked can then be reduced. 
For example, in order to determine if a $\sqrt{3}a\times\sqrt{3}a$ ordering is present within a Monte Carlo cell, it is sufficient to only track the length of the order-parameter vector in $\eta_2$-$\eta_3$ space, $\xi_1=\sqrt{\eta_2^2+\eta_3^2}$.
This length measure in the $\eta_2$-$\eta_3$ space, together with the overall concentration as measured by $\eta_1$, is sufficient to detect which particular $\sqrt{3}a\times\sqrt{3}a$ ordering is present within a Monte Carlo simulation, independent of the particular translational variant. 
A similar length metric can be defined in the space spanned by $\eta_4$, $\eta_5$ and $\eta_6$ according $\xi_2=\sqrt{\eta_4^2+\eta_5^2+\eta_6^2}$.
Here again, the combination of $\eta_1$, which is a measure of the overall composition, and $\xi_2$ is sufficient to detect whether either the $x=1/4$, $x=1/2$ or the $x=3/4$ orderings are present within a Monte Carlo simulation. 
Figure \ref{fig:order_parameter} plots $\eta_1$, $\xi_1$ and $\xi_2$ as a function of the composition and  exchange chemical potential along line B in the phase diagrams of Figure \ref{fig:pd_hex}(c) and (d).
The order parameters were calculated using Monte Carlo simulations in a supercell commensurate with the $2\sqrt{3}\times2\sqrt{3}a$ supercell in which the order parmeters are defined, and otherwise identical calculation parameters as in Section~\ref{sec:triangular_lattice_first_order_transition}.

\subsection{Example 4: Diffusion on a triangular lattice}

Atomic transport properties within a crystal can be estimated with kinetic Monte Carlo simulations. 
To illustrate this, we consider a binary atom-vacancy system on the two-dimensional triangular lattice and use the same model cluster expansion as was used to calculate the phase diagrams of Figure \ref{fig:pd_hex}(c) and (d). 
This model predicts a variety of stable ordered phases at low temperature and a solid solution at high temperature and is thereby able to illustrate the combined effects of concentration and different degrees of long-range order on atomic diffusion. 

A common mechanism of diffusion in a crystal is through atom-vacancy exchanges in distinct atomic hop events.
Each atomic hop event takes the system from one microstate to another. 
In the kinetic Monte Carlo simulation supercell, when an atom at site $i$ hops to an adjacent vacant site $j$, for example, the system evolves from the current configurational microstate $\mathbb{C}=(\sigma_1,\dots,\sigma_i=1,\dots,\sigma_j=-1,\dots,\sigma_N)$ to a new microstate $\mathbb{C}'=(\sigma_1,\dots,\sigma_i=-1,\dots,\sigma_j=1,\dots,\sigma_N)$. 
The frequency with which such an event occurs, $\Gamma_{\mathbb{C}\mathbb{C}'}$, and its dependence on local configurational order can be parameterized with local cluster expansion surrogate models as described in Section~\ref{sec:atomistic_desc_diffusion}.
However, for our model system, we choose a constant value of $\Delta E_{\mathbb{C}\mathbb{C}'}^{KRA} = 3|V_{NN}|$ and a constant value for $\nu^{*}_{\mathbb{C}\mathbb{C}'} = 10^{12}$ s$^{-1}$.
The fraction of vacancies, $x_{Va}$, on a triangular lattice that can host atoms from the dilute limit at $x\approx 0$ to the fully saturated limit at $x=1$ will vary according to $x_v=1-x$. 
Hence the number of hop paths available to diffusing atoms depends strongly on the overall concentration of the diffusing atoms.
Along with the configuration dependence on the end state energies and therefore also migration energies, diffusion will depend strongly on both concentration and ordering.

The model system has much in common with layered lithium intercalation compounds such as Li$_x$CoO$_2$ and Li$_x$TiS$_2$, where Li ions can diffuse over octahedrally coordinated interstitial sites that form a triangular lattice between two-dimensional sheets of CoO$_2$ and TiS$_2$.\cite{van2001first,van2008nondilute} 
It also shares similarities with adatom diffusion (e.g. adsorbed oxygen atoms) on a surface of a close-packed metal (e.g. Pt). 
In both examples the intercalation compound host crystal structure or the surface substrate imposes a constraint on the total number of triangular lattice sites. 
As derived in \ref{sec:Appendix_flux_examples}, the flux expression for this binary A-B system (where A are vacancies and B are the diffusing atoms) takes the form
\begin{equation}
    \tilde{J}=-\tilde{L}\nabla\tilde{\mu}
\end{equation}
where $\tilde{\mu} = \mu_{B}-\mu_A$ is the parametric chemical potential, $\tilde{J}=J_B$ and where $\tilde{L}=L_{BB}$ is the Onsager transport coefficient. 
The above flux expression can be converted to Fick's first law using the chain rule of differentiation
\begin{equation}
    \tilde{J}=-v_{u}\tilde{L}\frac{\partial \tilde{\mu}}{\partial x}\nabla c=-D\nabla c
\end{equation}
where $x = N_B/\Nunit = n_B$ is the parametric composition, $v_u$ is the volume per unit cell and $c=x/v_u$, the number of diffusing atoms per unit volume. 
The chemical diffusion coefficient $D$ is the product of a kinetic factor ($v_u\tilde{L}$) and a thermodynamic factor ($\partial \tilde{\mu}/\partial x$). 

For a simple atom-vacancy system, where there are no off-diagonal transport coefficients, the chemical diffusion coefficient $D$ is often factored into a product of the collective diffusion coefficient $D_B$ and a thermodynamic factor $\tilde{\Theta}$ (i.e. $D=D_B\tilde{\Theta}$) according to \cite{gomer1990diffusion,van2020rechargeable}
\begin{equation}
    D_B=\frac{K_{BB}}{x}
\end{equation}
and 
\begin{equation}
    \tilde{\Theta}=\frac{x}{k_{B}T}\frac{\partial\tilde{\mu}}{\partial x}. 
    \label{eqn:thermo_fac}
\end{equation}
The resulting diffusion coefficient, $D_B$, can then be directly compared to the tracer diffusion coefficient.

\begin{figure*}
\centering
\begin{subfigure}{6in}
    \centering
    \vspace{0.2cm}
    \includegraphics[width=5.5in]{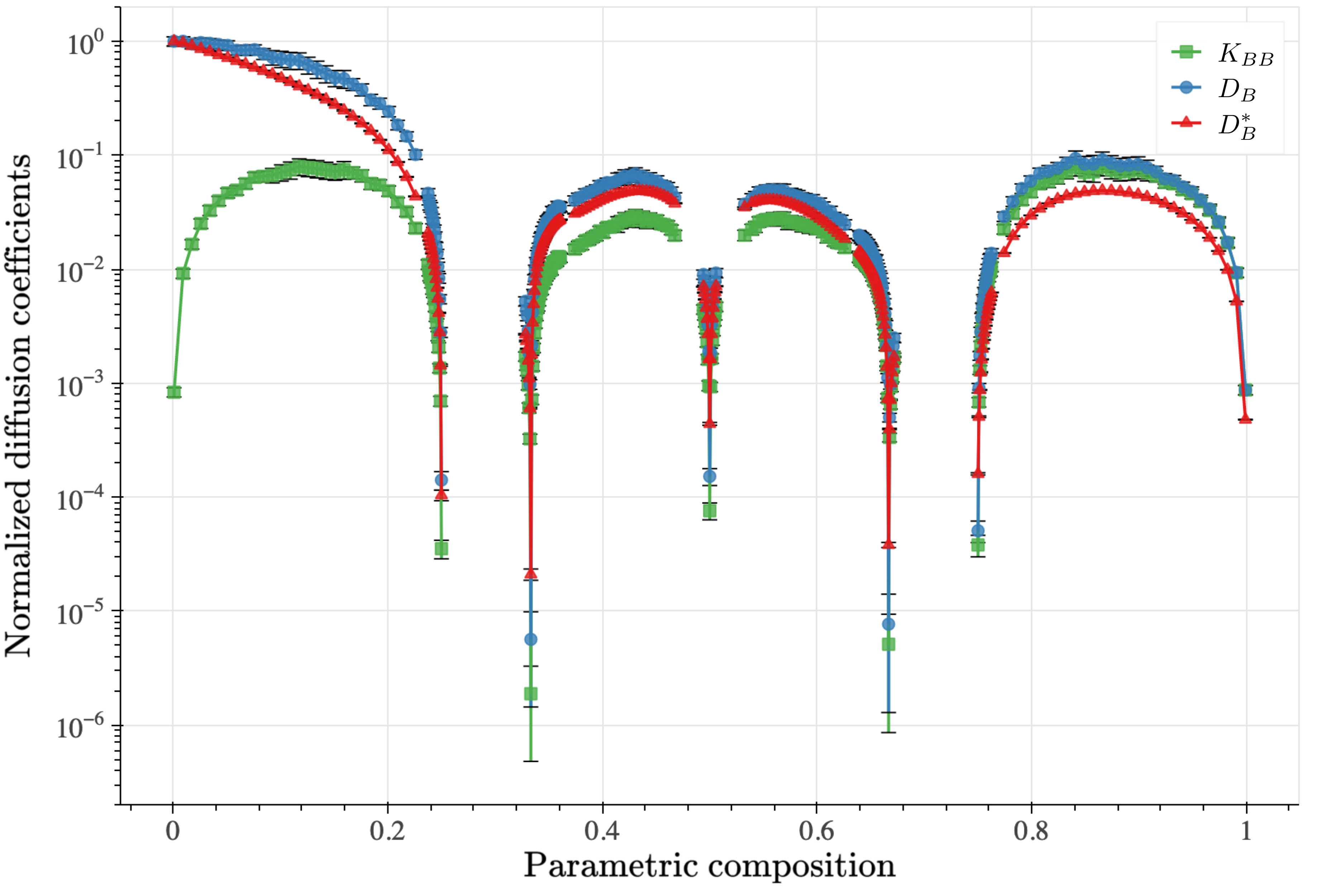}
    \label{fig:diff_coeff}
    \caption{}
\end{subfigure}
\begin{subfigure}{6in}
    \centering
    \vspace{0.2cm}
    \includegraphics[width=5.5in]{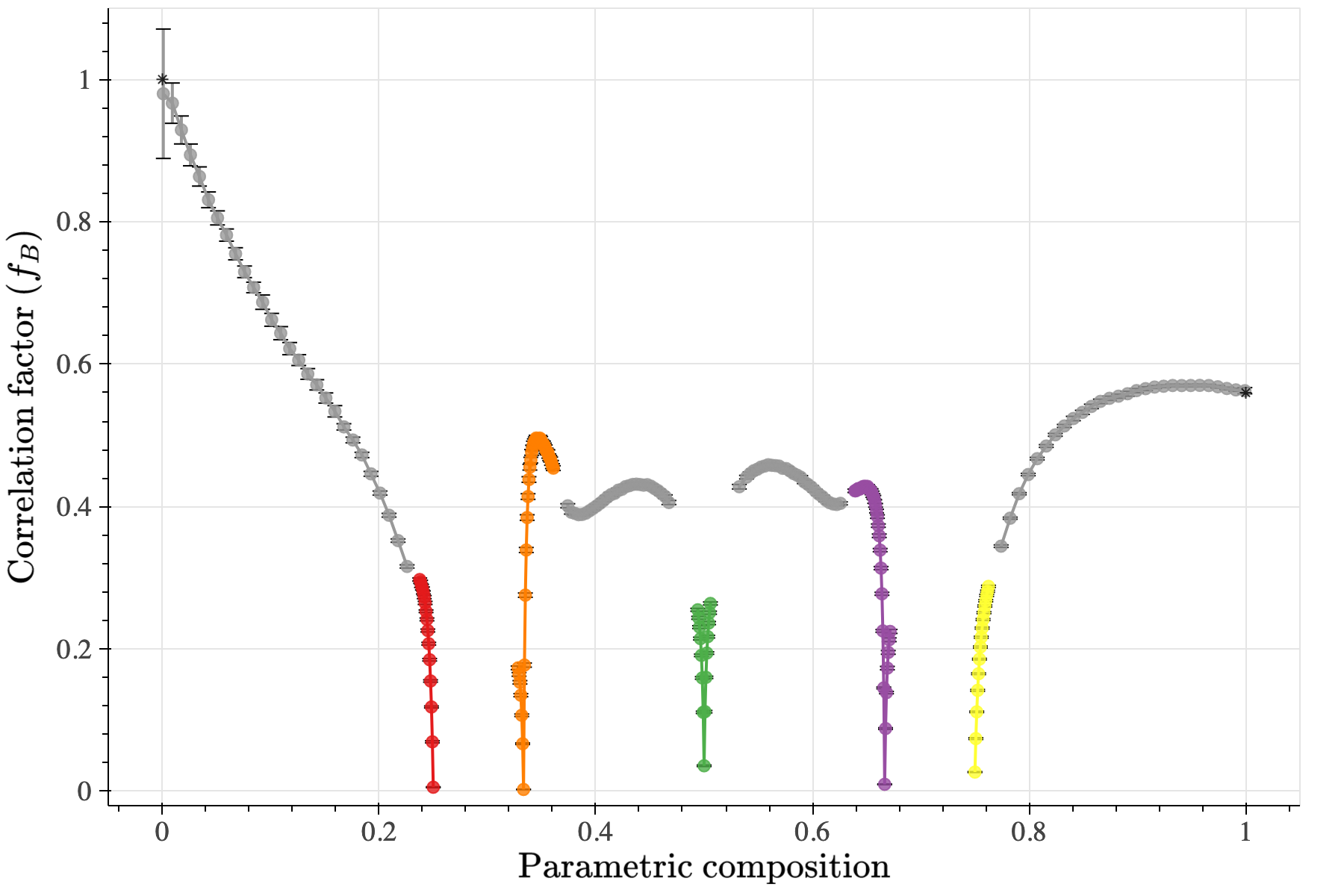}
    \label{fig:correlation_factor}
    \caption{}
\end{subfigure}
\caption{\label{fig:kmc} 
The (a) kinetic coefficients, and (b) correlation factor, calculated using kinetic Monte Carlo are shown as a function of composition at constant temperature.
Standard error of the mean is estimated using Eq. \ref{todo}.
In (b) exact results for the correlation factor on a triangle lattice, $f_{B}(x=0) = 1.0$ and $f_{B}(x=1.0) = \frac{\pi + 6\sqrt{3}}{11\pi-6\sqrt{3}}$ \cite{Montet1973}, are shown as ($*$).
Results correspond to line B in Figure~\ref{fig:pd_hex}, and colors indicate the equilibrium phase.
}
\end{figure*}

Figure \ref{fig:kmc}(a) shows calculated values for the collective and tracer diffusion coefficients, $D_B$ and $D^{*}_B$, respectively, along with the kinetic coefficient, $K_{BB}$, at a reduced temperature of $k_{B}T/V_{NN}=0.5$ (line B in Figure \ref{fig:pd_hex}(c) and (d)). 
The various diffusion coefficients were calculated by sampling atomic trajectories with kinetic Monte Carlo simulations and inserting them in Eqs.~\ref{eqn:collective_diffusion} and \ref{eqn:tracer_diffusion}.
The kinetic Monte Carlo simulations were performed in a $N=1200$ site supercell of the $2\sqrt{3}a \times 2\sqrt{3}a$ supercell shown in Figure \ref{fig:orderparameter_sublattices}.
An initial set of kinetic Monte Carlo calculations in which sampling was performed at a range of simulated time steps showed convergence in the kinetic coefficients when observations were taken every $10^{-2}$ simulated seconds.
Calculations were run with a requested absolute precision of $\pm10^{-2}|V_{NN}|$ for the average semi-grand canonical energy $\langle\Omega\rangle$, and a requested relative precision of $\pm10^{-1}$ for $D^{*}_B$ and $K_{BB}$.
A cutoff was enforced to stop the calculations if a maximum of $1$ hr of computational time was reached.
The estimated error bars, which are generally on the same order of magnitude as the markers, are included in Figure \ref{fig:kmc}.

In the dilute limit, $x \rightarrow 0$, $D_B$ and $D_B^*$ converge to a common value as the interactions between different atoms become negligible. 
The diffusion coefficients in Figure~\ref{fig:kmc} are normalized by the diffusion coefficient at $x=0$.
The diffusion coefficients exhibit strong dips at $x=1/4$, $1/3$, $1/2$, $2/3$ and $3/4$, the compositions at which the diffusing atoms adopt a state of long-range order. 
This is because atoms are locked into their sublattice positions and will have a strong thermodynamic bias to hop back to their preferred sublattice positions. 

A useful metric to understand diffusion is the correlation factor, which measures the degree with which successive hops of a diffusing atom are correlated. 
An uncorrelated random walker has a correlation factor $f=1$. 
This is only reached in the dilute limit where interstitial diffusers do not interact with each other. 
Anything less than 1 indicates correlated diffusion. 
The correlation factor as calculated for this model system is shown in Figure~\ref{fig:kmc}(b).
Starting at 1 in the dilute limit, it decreases with increasing concentration $x$, to values between 0.3 and 0.6 except for strong dips to near 0.0 at the stoichiometric compositions of the ordered phases.
This is consistent with reverse hops being highly likely after an atom hops out of perfect ordering.

\begin{figure}
\centering
\begin{subfigure}{3in}
    \centering
    \vspace{0.2cm}
    \includegraphics[width=3in]{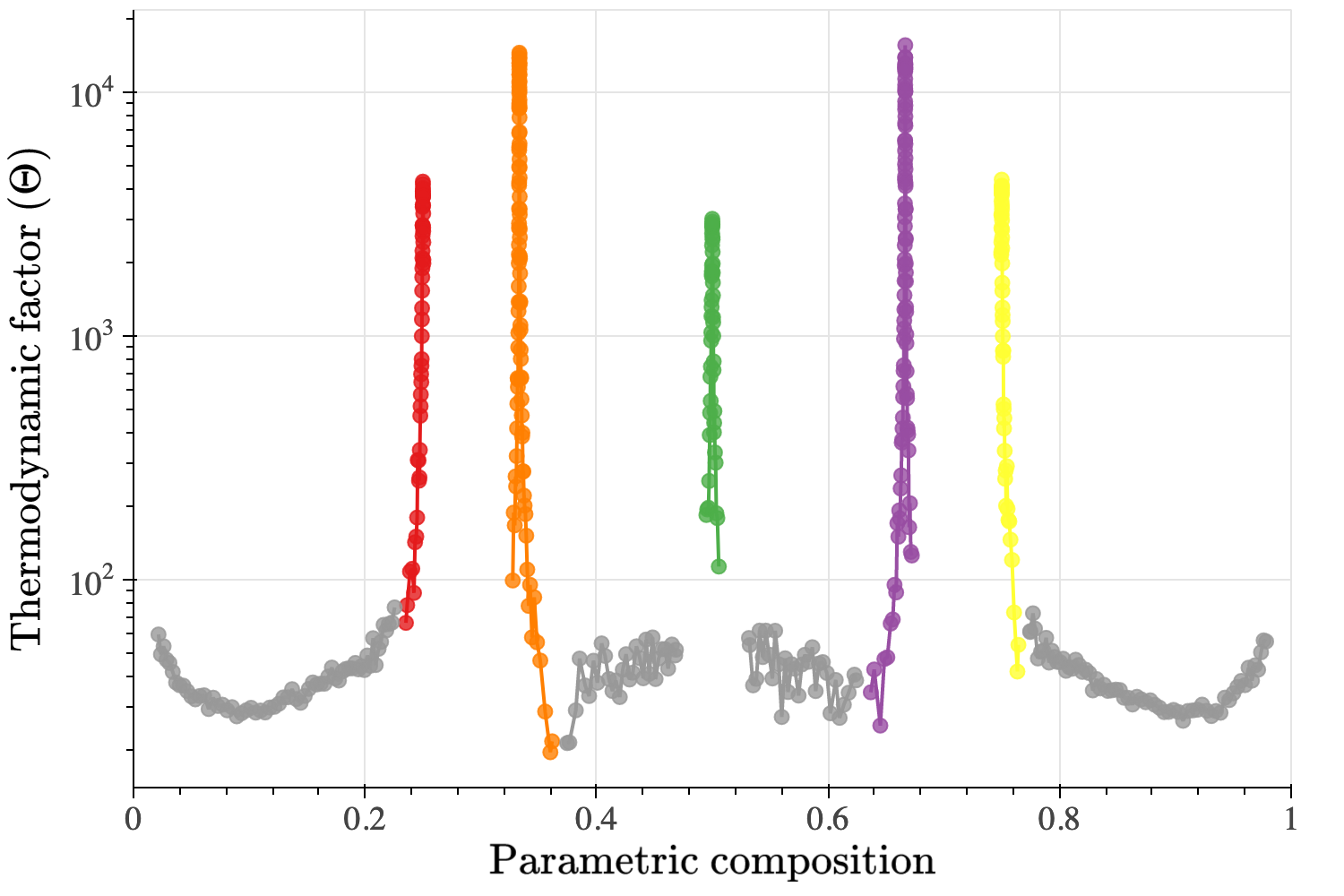}
    \label{fig:thermofactor_vs_x}
    \vspace{-0.5cm}
    \caption{}
\end{subfigure}
\begin{subfigure}{3in}
    \centering
    \vspace{0.2cm}
    \includegraphics[width=3in]{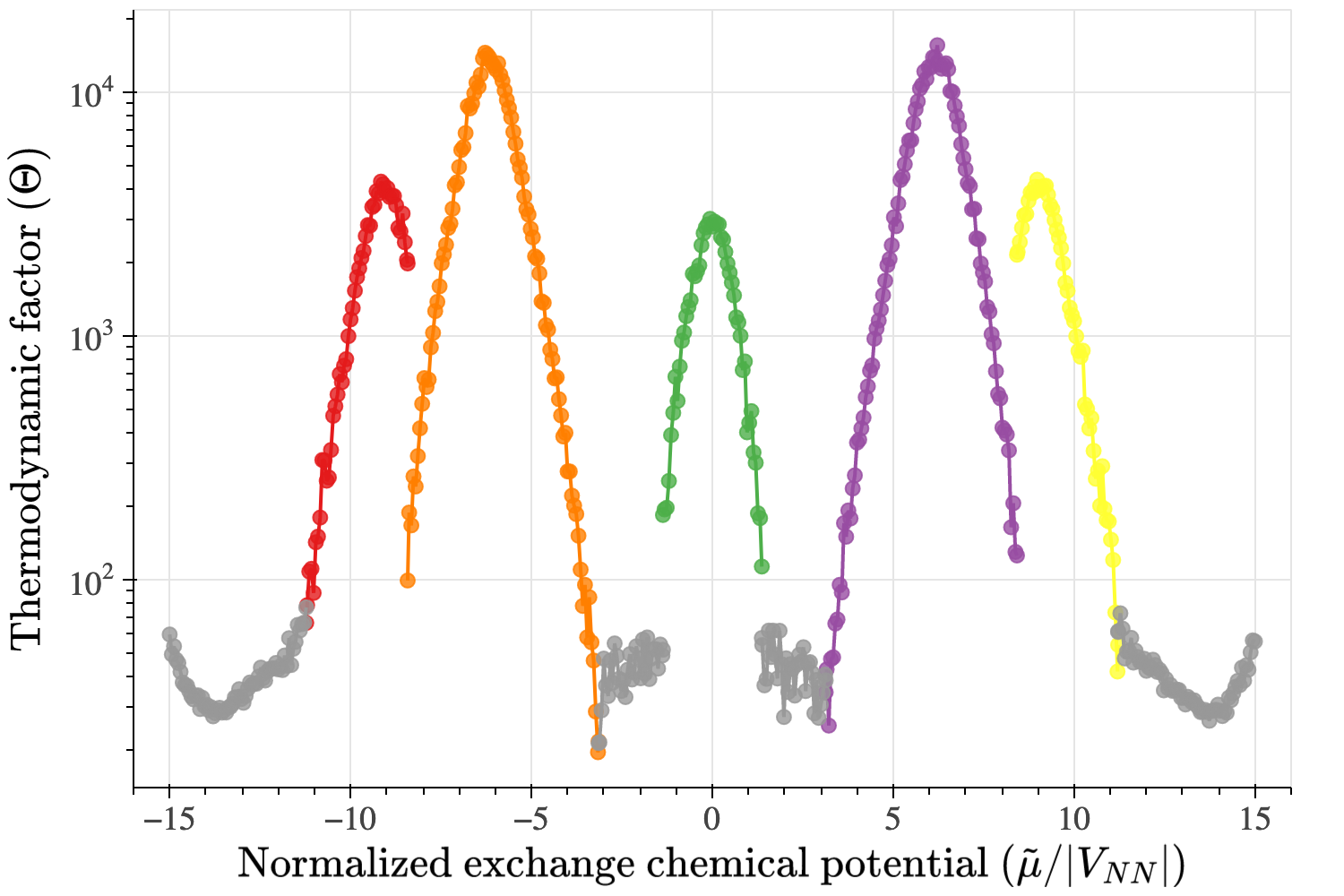}
    \label{fig:thermofactor_vs_mu}
    \vspace{-0.5cm}
    \caption{}
\end{subfigure}
\begin{subfigure}{3in}
    \centering
    \vspace{0.2cm}
    \includegraphics[width=3in]{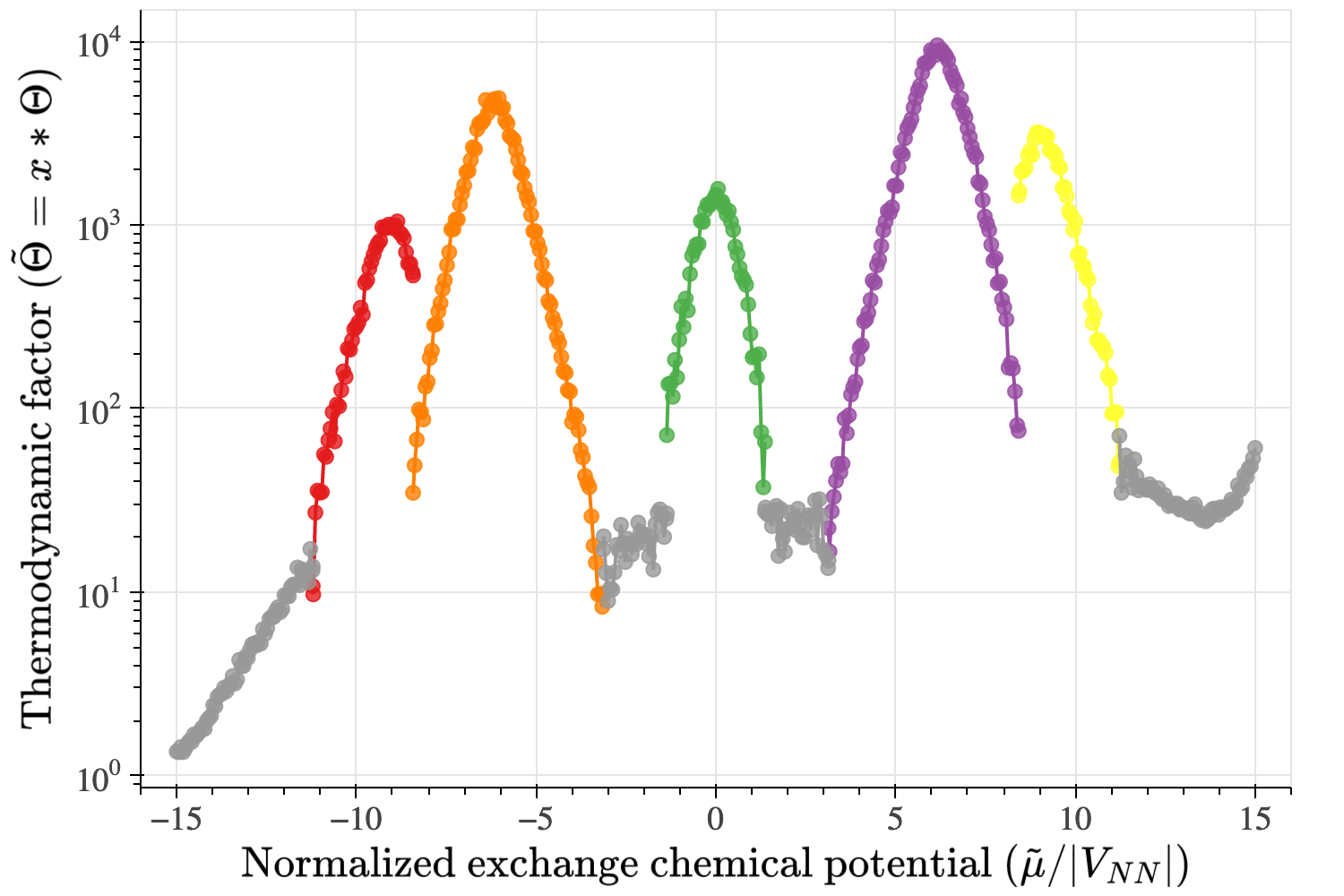}
    \label{fig:tilde_thermofactor_vs_mu}
    \vspace{-0.5cm}
    \caption{}
\end{subfigure}
\caption{\label{fig:thermofactor} 
The thermodynamic factor, $\Theta$, as defined according to Equation \ref{eq:thermofactor} as a function of (a) composition $x$ and (b) the exchange chemical potential $\tilde{\mu}$, and (c) the thermodynamic factor $\tilde{\Theta}$ as defined according to Equation  \ref{eqn:thermo_fac} as a function of the exchange chemical potential $\tilde{\mu}$. 
$\Theta$ is calculated according to Eq.~\ref{eq:thermofactor_from_chi} using the inverse of the chemical susceptibility, $\chi$, measured according to Eq. \ref{eq:susc_x}.  
Results correspond to line B in Figure~\ref{fig:pd_hex}, and colors indicate the equilibrium phase.
}
\end{figure}

Figure \ref{fig:thermofactor}(a) and (b) show the calculated thermodynamic factor, $\Theta$, as a function of the concentration $x$ and as a function of the exchange chemical potential $\tilde{\mu}$, respectively, as defined according to Equation \ref{eq:thermofactor}. 
Figure \ref{fig:thermofactor}(c) shows the thermodynamic factor $\tilde{\Theta}$ as defined according to Equation \ref{eqn:thermo_fac} versus the exchange chemical potential $\tilde{\mu}$.
The thermodynamic factor as defined according to Equation \ref{eqn:thermo_fac} measures the deviation from thermodynamic ideality, which is only realized in the dilute limit as $x \rightarrow 0$ where $\tilde{\Theta}$ is equal to 1. 
The system deviates strongly from thermodynamic ideality when the atoms order and by either definition the thermodynamic factor tends to diverge at the stoichiometric compositions of the equilibrium ordered phases.

\section{Free energy integration}
\label{sec:free_energy_integration}

While Monte Carlo methods do not provide direct access to free energies, it is possible to calculate them indirectly using the results of Monte Carlo simulations. 
For example, the relationship between the parametric composition $x_i$ of a multicomponent crystal and its conjugate exchange chemical potential $\tilde{\mu}_i$ as generated with semi-grand canonical Monte Carlo simulations can be used to calculate the Gibbs free energy, $g=G/N_u$, by integrating the differential form
\begin{equation}
    dg=-sdT+vdP+\sum_{i=1}^{k}\tilde{\mu}_idx_i
\end{equation}
along a constant temperature and constant pressure path that starts at ($\vec{x}_0$,$\vec{\tilde{\mu}}_0$) and ends at ($\vec{x}$,$\vec{\tilde{\mu}}$) to yield
\begin{equation}
    g(T,\vec{x})=g(T,\vec{x}_0)+\int_{\vec{x}_0}^{\vec{x}}\sum_{i=1}^{k}\tilde{\mu}_idx_i
    \label{eq:gibbs_free_energy_integration}
\end{equation}
A difficulty with free energy integration techniques is that an evaluation of $g(T,\vec{x})$ requires knowledge of $g(T,\vec{x}_0)$ in a particular reference state. 
Usually a reference state is chosen in which the free energy can be calculated easily. 
For example, it is common to choose a composition $\vec{x}_0$ in which all sites of the crystal are occupied by exclusively one chemical species such that there is no configurational entropy. 
In a binary A-B alloy, this could for example be a crystal of pure A atoms or a crystal of pure B atoms. 
In these reference states, there is only one configurational microstate and $g(T,\vec{x}_0)$ is then simply equal to $e(\vec{x}_0)$, the energy of the solid per number of unit cells at $\vec{x}_0$.
To evaluate the integral of Eq. \ref{eq:gibbs_free_energy_integration}, Monte Carlo simulations must be performed along a continuous path from the reference state to the composition $\vec{x}$ to collect the relationship between $\vec{x}$ and $\vec{\tilde{\mu}}$. 
This is most conveniently generated with semi-grand canonical Monte Carlo simulations at constant temperature, where a discrete, but finely spaced, grid of $\vec{\tilde{\mu}}$ values are imposed as boundary conditions and a corresponding set of values for $\langle\vec{x}\rangle$ are then calculated. 

A similar scheme can be derived for the semi-grand canonical free energy $\phi=\Phi/N_u$ at constant temperature and pressure by integrating the differential form
\begin{equation}
    d\phi=-sdT+vdP-\sum_{i}^{k}x_{i}d\tilde{\mu}_i
\end{equation}
yielding the expression
\begin{equation}
    \phi(T,\vec{\tilde{\mu}})=\phi(T,\vec{\tilde{\mu}}_0)-\int_{\vec{\mu}_0}^{\vec{\mu}}\vec{x}d\vec{\tilde{\mu}}
    \label{eq:semi_grand_free_energy_integration}
\end{equation}
where again the free energy of a reference state at $(T,\vec{\tilde{\mu}}_0)$ is required. 

Expressions can also be derived to relate free energies at different temperatures.
The derivatives of $g$ and $\phi$ with respect to temperature are related to the entropy, which is not directly calculated with Monte Carlo simulations. 
It is, therefore, more convenient to start with the total differentials (at constant $P$, usually set to zero) of $\beta g$ and $\beta \phi$,\cite{van2002self}  which take the form
\begin{equation}
    d(\beta g)=hd\beta+\beta\sum_{i}^{k}\tilde{\mu}_i dx_i
\end{equation}
\begin{equation}
    d(\beta\phi)=\omega d\beta-\beta\sum_{i=1}^{k}x_{i}d\tilde{\mu}_{i}
\end{equation}
where $\beta=1/k_{B}T$, $h$ is the enthalpy per unit cell of the crystal ($h=e+Pv$) and $\omega$ is the semi-grand canonical energy per unit cell ($\omega=e+Pv-\vec{\tilde{\mu}}\vec{x}$).
Both $h$ and $\omega$ can be calculated as $\langle \Omega \rangle/\Nunit$, using canonical and semi-grand canonical Monte Carlo simulations, respectively. 
Integrating these expressions along a path at constant $\vec{x}$ (as in canonical Monte Carlo simulations) or at constant $\vec{\tilde{\mu}}$ (as in semi-grand canonical Monte Carlo simulations) yields expressions that relate free energies at two different temperatures
\begin{equation}
    \beta g(T,\vec{x})=\beta_{0}g(T_0,\vec{x})+\int_{\beta_0}^{\beta}hd\beta
    \label{eq:gibbs_temperature_integration}
\end{equation}
\begin{equation}
    \beta \phi(T,\vec{\tilde{\mu}})=\beta_{0}\phi(T_0,\vec{\tilde{\mu}})+\int_{\beta_0}^{\beta}\omega d\beta
    \label{eq:grand_temperature_integration}
\end{equation}
where $\beta_0=1/k_{B}T_0$. 
A reference state where the free energy can easily be calculated is again needed to evaluate both expressions. 
Common reference states for paths that traverse different temperatures are low temperature ordered ground states or a high temperature disordered solid solution. 
The free energy of stable ordered phases can be approximated at low temperatures using a low temperature expansion of the partition function.\cite{kohan1998computation,van2002self}
The free energy of the solid solution at a particular composition or set of chemical potentials can be calculated with Eq. \ref{eq:gibbs_free_energy_integration} or Eq. \ref{eq:semi_grand_free_energy_integration} using a reference state at one of the corners of composition space at a temperature that is sufficiently high to be above all order-disorder transition temperatures. 

\begin{figure}
    \centering
    \includegraphics[width=8cm]{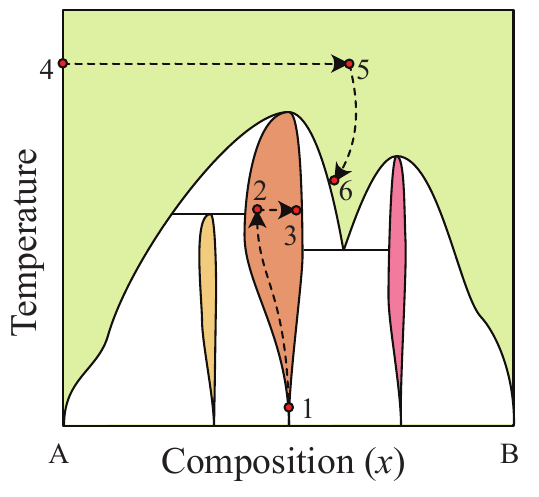}
    \caption{Schematic diagram of integration pathways for the calculation of free energies,  $g$ or $\phi$, from semi-grand canonical Monte Carlo simulation of the mean composition, $\vec{x}$, and semi-grand canonical energy, $\omega$, as a function of chemical potential, $\tilde{\mu}$, and tempature, $T$. The first pathway $(1 \rightarrow 2 \rightarrow 3)$ uses a low-temperature expansion calculation $(1)$ for a free energy reference and then integrates along a constant $\tilde{\mu}$ and increasing $T$ pathway $(1 \rightarrow 2)$ followed by an increasing $\tilde{\mu}$ and constant $T$ pathway $(2 \rightarrow 3)$. The second pathway $(4 \rightarrow 5 \rightarrow 6)$ uses the energy at zero configurational entropy $(4)$ for an initial free energy reference and then integrates along an increasing $\tilde{\mu}$ and constant $T$ pathway $(4 \rightarrow 5)$ followed by a constant $\tilde{\mu}$ and decreasing $T$ pathway $(5 \rightarrow 6)$.}
    \label{fig:integration_pathways}
\end{figure}

Common pathways to calculate free energies as a function of composition (or exchange chemical potentials) and temperature are schematically illustrated in Figure \ref{fig:integration_pathways}. 
The path starting at point 1 and ending at 2 has a constant chemical potential $\tilde{\mu}$ and can be traversed with semi-grand canonical Monte Carlo simulations. 
If the temperature of point 1 is sufficiently low, then the free energy at point 1 can be calculated with a low temperature expansion.\cite{kohan1998computation,van2002self} 
The free energy at point 2 can then be calculated with Eq. \ref{eq:grand_temperature_integration} by integrating over the calculated values of $\omega$ with respect to temperature (or equivalently the inverse temperature, $\beta$).
Free energy integration can also be performed to link the free energy calculated at point 2 to that at point 3 using either Eq. \ref{eq:gibbs_free_energy_integration} or Eq. \ref{eq:grand_temperature_integration}. 
The data along the path from 2 to 3 would be collected with semi-grand canonical Monte Carlo simulations. 
Figure \ref{fig:integration_pathways} also shows a path to calculate free energies in high temperature solid solutions. 
A convenient reference state for free energy integration is point 4, for pure A, where the absence of any configurational entropy makes it possible to equate the free energy at this point to the energy. 

\begin{figure}
    \centering
    \includegraphics[width=8cm]{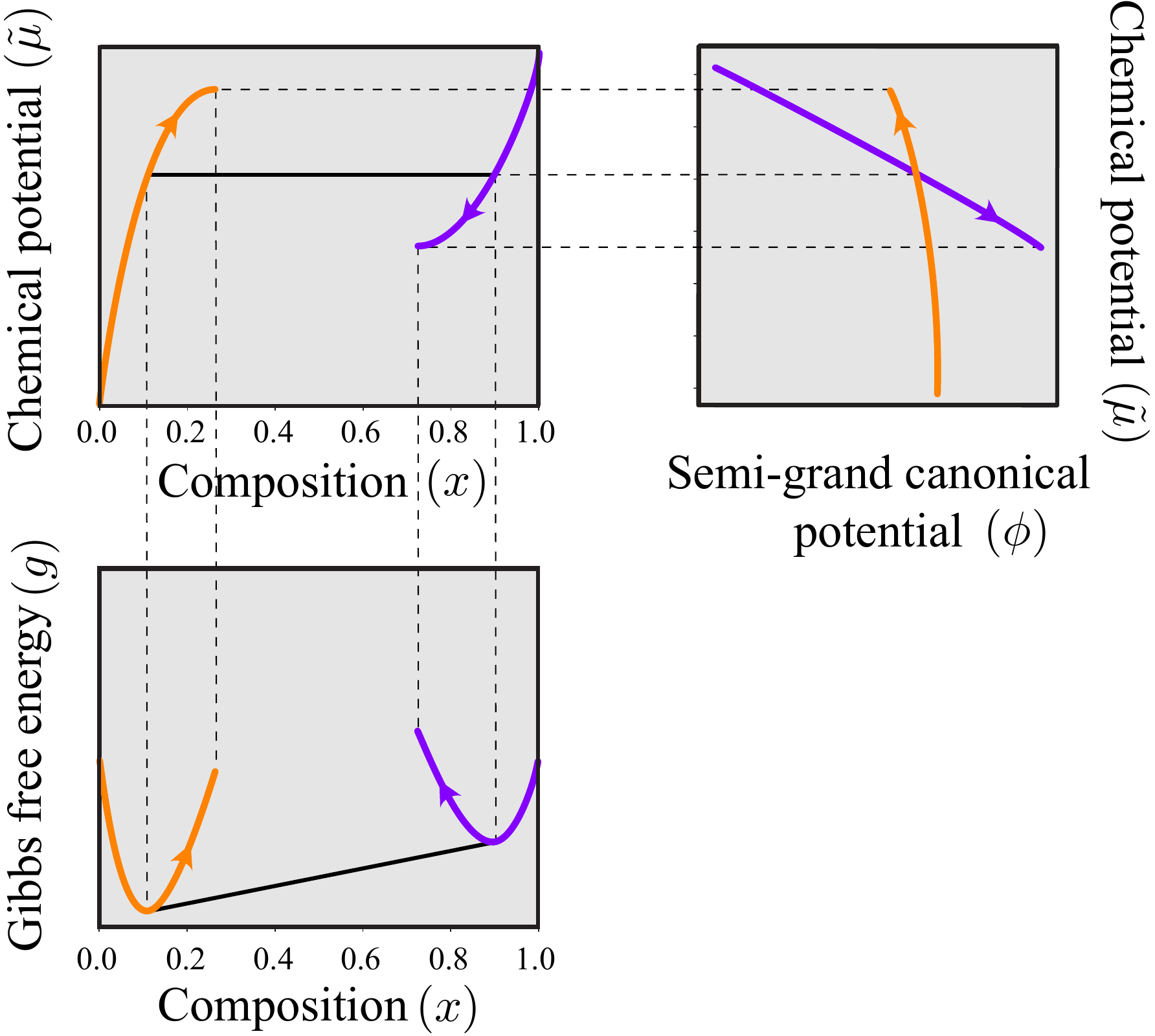}
    \caption{Schematic of the hysteresis behavior of the calculated mean composition as a function of chemical potential in semi-grand canonical ensemble Monte Carlo simulations. The composition at which the two phases are at equilibrium can be equivalently determined by finding the chemical potential at which the semi-grand canonical potentials are equal, or by using the common tangent construction to identify the compositions at which the Gibbs free energies are equal.}
    \label{fig:hysteresis}
\end{figure}

\begin{figure*}
    \centering
    \includegraphics[width=16cm]{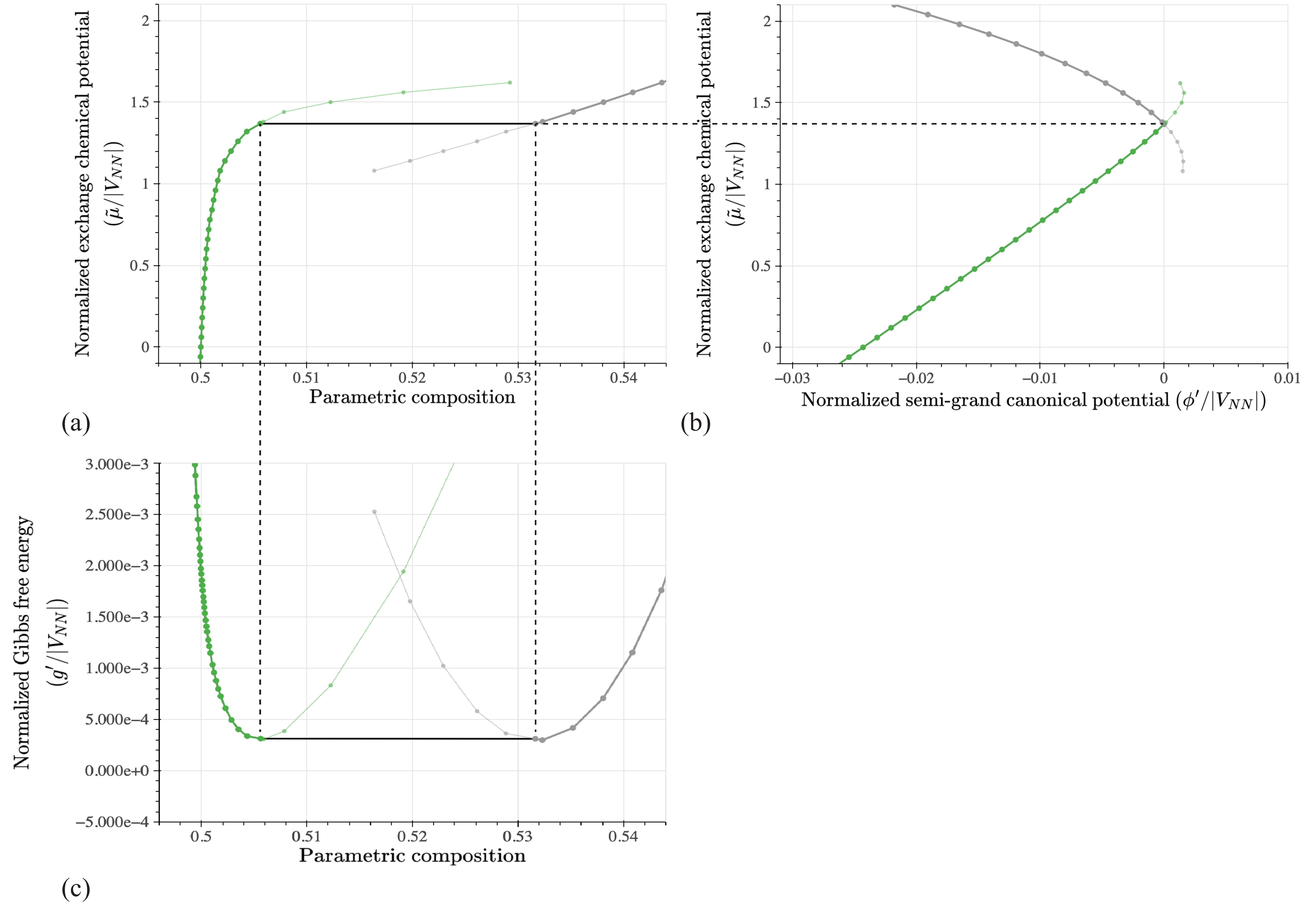}
    \caption{The relationships calculated from Monte Carlo simulations along line B in Figure~\ref{fig:pd_hex} near the phase boundary between the ordered phase with composition $x= 1/2$ (green) and the disordered state (gray) of (a) the exchange chemical potential, $\tilde{\mu}$, and composition $x$, (b) the semi-grand canonical potential, $\phi$ and $\tilde{\mu}$, and (c) the Gibbs free energy, $g$, and $x$.
    Solid black lines in (b) and (c) indicate phase equilibria, and dashed lines are guides for the eye at the corresponding values of $x$ and $\tilde{\mu}$.
    In this figure, $g$ and $\phi$, as calculated using Eqs.~\ref{eq:gibbs_temperature_integration} and \ref{eq:grand_temperature_integration}, are re-referenced, indicated by $g^{\prime}$ and $\phi^{\prime}$, respectively, to more clearly show energy differences which are small relative to the overall changes in $g$ and $\phi$ over the relevant ranges of $x$ and $\tilde{\mu}$.}
    \label{fig:pd_construction_hysteresis}
\end{figure*}

\begin{figure*}
    \centering
    \includegraphics[width=16cm]{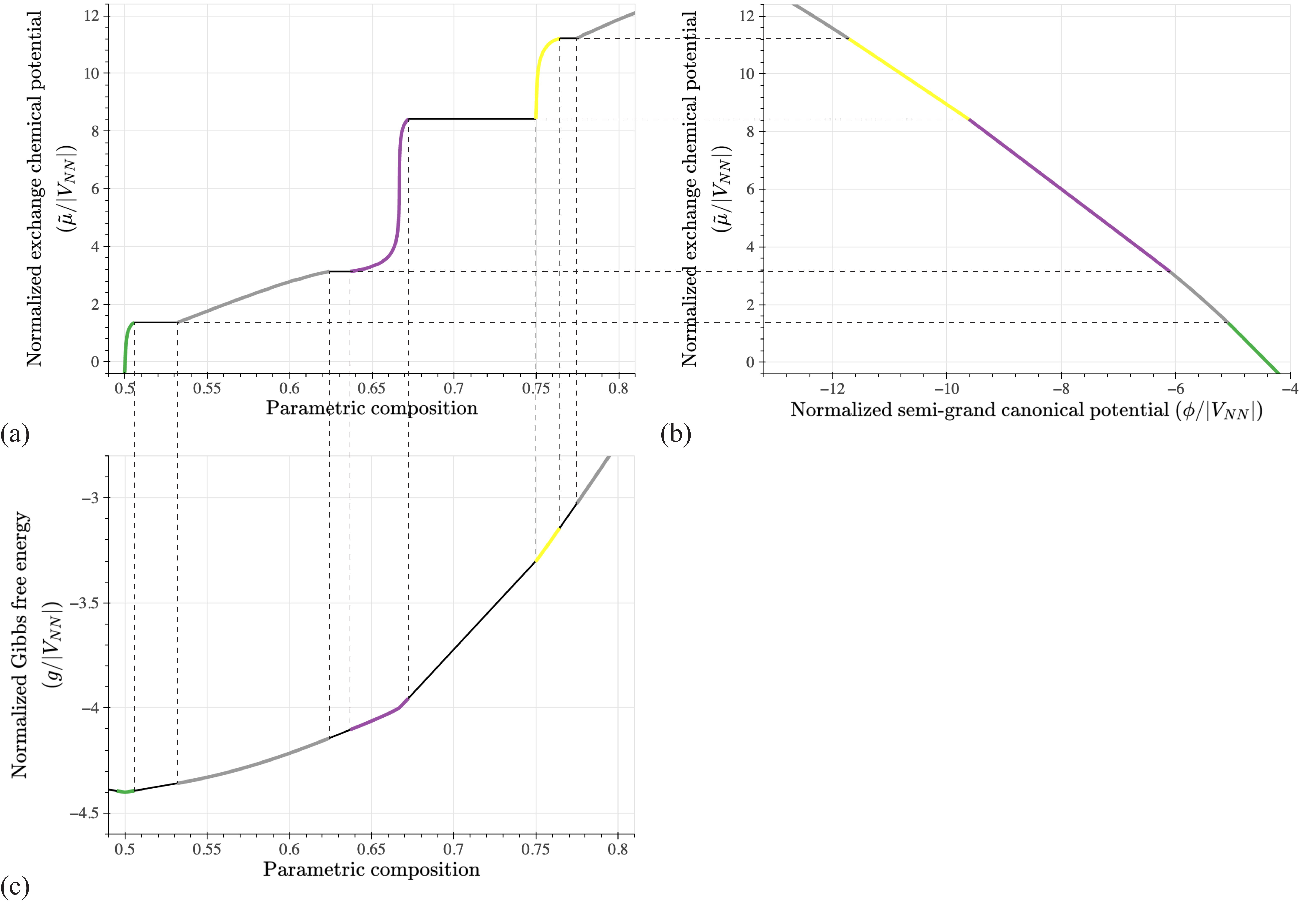}
    \caption{Phase boundary construction along line B in Figure~\ref{fig:pd_hex} over the range $x > 0.5$ ($\tilde{\mu} > 0.0$), showing the relationships calculated from Monte Carlo simulations of (a) the exchange chemical potential, $\tilde{\mu}$, and composition $x$, (b) the semi-grand canonical potential, $\phi$ and $\tilde{\mu}$, and (c) the Gibbs free energy, $g$, and $x$.
    Solid black lines in (b) and (c) indicate phase equilibria, and dashed lines are guides for the eye at the corresponding values of $x$ and $\tilde{\mu}$.
    Colors indicate the equilibrium phase as in Figure~\ref{fig:pd_hex}.}
    \label{fig:pd_construction}
\end{figure*}

Free energies calculated with free energy integration techniques as described above are only reliable as long as the traversed path does not cross any phase boundaries. 
This is because of the phenomenon of hysteresis. 
Typically, Metropolis or n-fold-way Monte Carlo algorithms use the last sampled microstate at one point along a path as the initial microstate for the next point of the path. 
This introduces a memory effect where microstates continue to be sampled corresponding to a metastable free energy. 
This is schematically illustrated in Figure \ref{fig:hysteresis} for a semi-grand canonical Monte Carlo simulation. 
The orange curve represents the relationship between $\tilde{\mu}$ and $x$ as calculated with semi-grand canonical Monte Carlo simulations for increasing values of $\tilde{\mu}$. 
The solid black horizontal line in the $x$-$\tilde{\mu}$ plot corresponds to the equilibrium chemical potential of a first-order phase transition at which the concentration increases discontinuously. 
Above the black line, the equilibrium $\tilde{\mu}$ versus $x$ curve should follow the purple curve. 
When the final sampled microstate of a Monte Carlo simulation at fixed $T$ and $\tilde{\mu}$ is inherited as the initial microstate at $\tilde{\mu}+\Delta\tilde{\mu}$, the calculated $\tilde{\mu}$ versus $x$ curve will overshoot the equilibrium transition chemical potential $\tilde{\mu}^{*}$. 
The same occurs when performing semi-grand canonical Monte Carlo simulations in the opposite direction (i.e. decreasing chemical potential). 
Also shown in Figure \ref{fig:hysteresis} are the corresponding Gibbs free energies $g$ and semi-grand canonical free energies $\phi$. 
The phenomenon of hysteresis within Monte Carlo simulations makes it possible to estimate metastable free energies that extend some distance into multi-phase coexistence regions, depending on the energetics of the system and the convergence details of the calculations. 

In practice, the energy differences between phases may be small relative to the overall ranges of $g$ and $\phi$.
This can be seen for the model system from Section~\ref{sec:triangular_lattice_first_order_transition} by comparing the energy scales in Figures~\ref{fig:pd_construction_hysteresis} and~\ref{fig:pd_construction}.
Figure~\ref{fig:pd_construction_hysteresis} shows how the phase boundaries along line B in Figure~\ref{fig:pd_hex} between the ordered phase with composition $x= 1/2$  (green) and the disordered state (gray) are obtained.
The compositions of the phases at the phase boundaries are shown in Figure~\ref{fig:pd_construction_hysteresis}(a) at the value of the chemical potential, $\tilde{\mu}$, for which the two phases have equal semi-grand canonical potential, $\phi$, in Figure~\ref{fig:pd_construction_hysteresis}(b).
At these compositions, the Gibbs free energy, $g$, vs $x$ curves have a common tangent, as shown in Figure~\ref{fig:pd_construction_hysteresis}(c).
Figure
\ref{fig:pd_construction} shows the result of repeating the process to identify phase boundaries across the range of positive $\tilde{\mu}$ where calculations were performed.

The free energy integration techniques described in this section enable the calculation of free energies in regions where a phase is stable or metastable. 
They are unable to generate free energies where a phase is unstable with respect to small perturbations in composition. 
To access these free energies using equilibrium methods, umbrella sampling techniques can be used. 
A general approach to calculate free energies in domains where the solid is unstable was introduced by Sadigh and Erhart \cite{Sadigh2012} for miscibility gaps and was extended by Natarajan et al \cite{natarajan2017symmetry} for order-disorder reactions. 
Recent work has shown how an integrable neural network \cite{teichert2019machine} can be used to represent the free energy in high dimensional composition and order-parameter spaces and efficiently parameterized using adaptive learning techniques \cite{teichert2020scale}. 
The variance constrained umbrella sampling techniques of Sadigh and Erhart \cite{Sadigh2012} and Natarajan et al \cite{natarajan2017symmetry} are implemented in the CASM Monte Carlo code suite.

\section{Monte Carlo algorithms}

A variety of Monte Carlo algorithms have been implemented within CASM. 
The Metropolis algorithm is the simplest and at intermediate to high temperatures the most practical method to calculate basic thermodynamic properties. 
At low temperatures the Metropolis algorithm becomes less efficient and the n-fold way algorithm becomes more favorable.  
Kinetic Monte Carlo algorithms enable the sampling of diffusion trajectories necessary to calculate Onsager transport coefficients. 
Finally, when treating continuous degrees of freedom, such as a crystal of interacting non-collinear magnetic moments, Hamiltonian Monte Carlo methods become necessary to ensure efficient sampling of microstates.

Each algorithm is described in more detail in the following sections.
The Monte Carlo methods described are examples of Markov processes, which is defined as a stochastic process in which the probability of the next state only depends on the current state. 
The series of generated states is called a Markov chain.
There is a large body of literature describing the statistics of stochastic processes generally, of Markov chains and Markov chain Monte Carlo simulations, and their application to problems in statistical physics \cite{Hammersley1964, Gelman1992, Geyer1992,  binder2002monte, Geyer2011, landau_binder_2014}.

\subsection{The Metropolis algorithm}

The Metropolis algorithm \cite{Metropolis1953} is the most commonly used Monte Carlo method for the calculation of thermodynamic properties. 
The algorithm produces a series of configurations, $[ \mathbb{C}_1, \mathbb{C}_2, \dots, \mathbb{C}_n ]$. 
In the limit of large $n$, a configuration $\mathbb{C}$ will appear in the series with a frequency that is proportional to the probability predicted by the Boltzmann distribution $P_{\mathbb{C}} = e^{-\beta \Omega_{\mathbb{C}}} / Z$. 
The series of microstates is constructed by proposing an event, $\Delta_{\mathbb{C}\mathbb{C}'}$, which changes the current configuration, $\mathbb{C} = \mathbb{C}_i$, to a different configuration, $\mathbb{C}'$, calculating the associated change in potential energy, $\Delta \Omega_{\mathbb{C}\mathbb{C}'}$, and accepting the event with probability
\begin{align*} 
p(\Delta_{\mathbb{C}\mathbb{C}'}) = \mathrm{min}(1, e^{-\beta \Delta\Omega_{\mathbb{C}\mathbb{C}'}}).
\end{align*}
If the event is accepted, then the next configuration in the series includes the change, $\mathbb{C}_{i+1} = \mathbb{C}'$. 
If the event is rejected, then the next configuration in the series does not include the change, $\mathbb{C}_{i+1} = \mathbb{C}_i$.

Given that the Metropolis algorithm generates configurations in proportion to the probability predicted by the Boltzmann distribution, the ensemble average value of a property can be estimated from Monte Carlo simulations as 
\begin{align}
\langle X \rangle \approx \bar{X} = \frac{\sum_l^N X_l}{N},
\end{align}
where $X_l$ is the $l$-th of $N$ observations of the property (a function of the configuration, $\mathbb{C}_{l}$, and thermodynamic conditions), and $\bar{X}$ is the estimate of $\langle X \rangle$. 

The Metropolis algorithm is effective in many situations, but if the proportion of rejected states becomes large then the method may be less efficient than other methods.
This situation often occurs at low temperatures or when the proposed events result in a large change in the configuration.
Alternative Monte Carlo methods become useful in these situations.

\subsection{The n-fold way algorithm}

At low temperatures the probability of the lowest energy microstates are very high, resulting in exceedingly low acceptance rates of excited microstates when using the Metropolis algorithm. 
Very long computational times are then required to calculate well-converged thermodynamic averages. 
The n-fold way method \cite{bortz1975new} picks a new microstate at each step by considering a list of candidate events, $\Delta_{\mathbb{C}\mathbb{C}'}$, and choosing the next event according to
\begin{align*}
p(\Delta_{\mathbb{C}\mathbb{C}'}) &= q_{\mathbb{C}\mathbb{C}'} / Q_{\mathbb{C}} \\
q_{\mathbb{C}\mathbb{C}'} &= \mathrm{min}(1, e^{ -\beta \Delta\Omega_{\mathbb{C}\mathbb{C}'}}) \\
Q_{\mathbb{C}} &= \sum_{\mathbb{C}'} q_{\mathbb{C}\mathbb{C}'}.
\end{align*}
The weight of the $i$-th configuration in the ensemble average, $w_i$, is calculated as
\begin{align}
w_i = -\frac{1}{Q_{\mathbb{C}_i}}\ln(R),
\end{align}
where $R \in [0, 1)$ is a random number.
This weighting factor provides a statistically equivalent reflection of the number of rejected events that would have occurred before an acceptance if instead the Metropolis algorithm were performed with the same candidate events.
Thus, for n-fold way Monte Carlo simulations the ensemble average estimate is the weighted average
\begin{align}
\langle X \rangle \approx \bar{X} = \frac{\sum_l^N w_l X_l}{\sum_l^N w_l}.
\end{align}
While a new microstate is chosen at every step in the n-fold way, the probabilities of many candidate perturbations to the current state need to be calculated, making the computational cost of each step significantly larger than that required during a Metropolis algorithm step. 
The n-fold way method only becomes computationally more efficient than the Metropolis method when the rejection rate of the Metropolis algorithm becomes exceedingly high, as occurs at low temperatures. 

\subsection{Kinetic Monte Carlo algorithm}

The kinetic Monte Carlo (KMC) algorithm \cite{bortz1975new} is equivalent to the n-fold way algorithm, but the candidate events are chosen based on the physically possible atomic mechanisms for diffusion. 
This allows each trajectory generated by a kinetic Monte Carlo simulation to be interpreted as a likely trajectory that would have been sampled by a diffusing atom in the crystal over time. 
The event probabilities for KMC are calculated from the event rates, Eq.~\ref{eq:kmc_rate}, according to
\begin{align*}
p(\Delta_{\mathbb{C}\mathbb{C}'}) &= q_{\mathbb{C}\mathbb{C}'} / Q_{\mathbb{C}} \\
q_{\mathbb{C}\mathbb{C}'} &= \Gamma_{\mathbb{C}\mathbb{C}'} \\
Q_{\mathbb{C}} &= \sum_{\mathbb{C}'} q_{\mathbb{C}\mathbb{C}'}.
\end{align*}
For KMC calculations, the time spent in the $i$-th configuration is calculated as
\begin{align}
t_i = -\frac{1}{Q_{\mathbb{C}_i}}\ln(R),
\end{align}
where $R \in [0, 1)$ is a random number, exactly as the weight is calculated in the n-fold way algorithm.

The value of kinetic coefficients can be estimated from kinetic Monte Carlo simulations as:
\begin{align}
    X \approx \bar{X} = \frac{\sum_l^N X_l}{N},
\end{align}
where $X_l$ is the observed value over the $l$-th of $N$ time intervals. 
For example, an observation of $K_{ij}$ can be calculated using Eq.~\ref{eq:Kubo_Green_Onsager2} as
\begin{align}
\left(K_{ij}\right)_{l} = \frac{\paren{\sum_{\zeta} \Delta\vec{R}^{\zeta}_{i}} \paren{\sum_{\zeta} \Delta\vec{R}^{\zeta}_{j}} }{2dt_{l}N_u}, 
\end{align}
where $t_l$ is the length of the $l$-th time interval. 
For measurements of steady-state quantities such as $K_{ij}$, the time intervals need to be long enough to ensure sampling of events that contribute to long-range diffusion.
If the time intervals are too short, the estimated kinetic values may be incorrectly inflated by local rearrangements.
Generally, a convergence study should be used to find an appropriate sampling time interval before performing calculations for a new system or at very different thermodynamic conditions.

To take time-interval based samples, CASM checks if a chosen event will occur after the next observation should be taken, and if so, takes a sample before applying the change in configuration due to the event.
In some cases, when the event time is long relative to the sampling interval, this may mean that multiple samples are taken in the same configuration.
As an alternative approach with higher data storage requirements, CASM supports saving the state of the entire Monte Carlo supercell at each sampling time, allowing for a post-run convergence analysis using a range of sampling intervals.


The n-fold way / KMC algorithm is effective in many situations, but becomes inefficient if the probability of transitioning between two configurations, or a small set of configurations, is much more likely than transitioning to other configurations.
In KMC calculations, this can be understood as an energy basin in which the system makes many transitions until escaping.
There exist methods for grouping such states and calculating and sampling the distribution of exit times and configurations in order to accelerate calculations \cite{Novotny1995,Puchala2010,Fichthorn2013}.

\subsection{Hamiltonian Monte Carlo}

For continuous degrees of freedom (DoF), it is often challenging to propose events that are not rejected too often by the Metropolis algorithm. 
This is relevant for finite temperature studies of non-collinear magnetic spin cluster expansions, where the local magnetic moments at each site can adopt a continuum of different orientations \cite{wang2019accelerating,kitchaev2020mapping}. 
The Hamiltonian Monte Carlo method \cite{duane1987hybrid, betancourt2017conceptual} addresses this issue by using the Hamiltonian to combine aspects of molecular dynamics with Metropolis Monte Carlo simulations. 
At each step a random momentum is generated and applied to the continuous degrees of freedom and the configuration is updated by integration (for example, a leapfrog integration) for several integration steps. 
The final state is then used as the proposed next configuration in the Metropolis algorithm. 

The integration steps results in a greater likelihood of acceptance with a larger change in DoF values than a randomly generated change. 
The use of the Metropolis algorithm allows for a random change of momentum and the integration to be done approximately. 
Together this enables the Hamiltonian Monte Carlo method to be more efficient in calculating equilibrium properties than a direct molecular dynamics calculation.

To implement the Hamiltonian Monte Carlo method, forces must be calculated. 
CASM uses the automatic differentiation library FADBAD++  \cite{fadbad} to support the calculation of forces from effective Hamiltonians.

\subsection{Averaging methods, convergence criteria, and simulation output}
\label{sec:avg_converge_output}


When calculating averages $\bar{X}$ using Monte Carlo methods, it is important to know how well converged $\bar{X}$ is, since $\bar{X}$ is only equal to $\langle X \rangle$ as the number of samples tends to infinity ($N \to \infty$). 
The accuracy of $\bar{X}$ is affected by how long it takes the Markov chain to reach its equilibrium probability distribution given its initial configuration.
It is typical to implement some procedure to discard a certain number of initial observations which may be present well outside of their equilibrium proportion.
The accuracy of $\bar{X}$ is also affected by the correlations between observations, the number of samples, and the underlying probability distribution which is a function of the effective Hamiltonian and the supercell size.

Excluding the supercell size effect, the central limit theorem states that the error in the Monte Carlo estimate of the ensemble average, $\bar{X} - \langle X \rangle$, converges to zero as the number of samples increases, with a normal distribution of mean zero \cite{Geyer2011}
\begin{align}
\bar{X}-\langle X \rangle \approx \mathcal{N}(0, \sigma^2/N).
\end{align}

For stationary distributions, such as those generated by the Monte Carlo algorithms considered here after they have been suitably equilibrated, the variance $\sigma^2$ is \cite{Geyer2011}
\begin{align}
    \label{eq:mc_variance}
    \sigma^2&= \gamma_0 + 2 \sum^\infty_{k=1} \gamma_k \\
    \gamma_k &= \mathrm{Cov}\left( X_j, X_{j+k} \right)
\end{align}
Here $\gamma_k$ is the lag $k$ autocovariance, which quantifies correlations between observations $X_j$ and $X_{j+k}$.
An estimate, $\hat{\gamma}_k$, of the autocovariance can be calculated directly from the Monte Carlo simulation observations using
\begin{align}
    \hat{\gamma}_k &= \sum^{N-k}_i\left( X_i - \bar{X} \right) \left( X_{i+k} - \bar{X} \right).
\end{align}
A common assumption in Monte Carlo simulations that is the autocovariance decays like
\begin{align}
    \label{eq:autoregress_covar}
    \gamma_k = \gamma_0 \rho^{-|k|},
\end{align}
where $\rho$ is a system and algorithm dependent autoregression parameter.
In this case, the sum in Eq.~\ref{eq:mc_variance} can be evaluated to give
\begin{align}
    \sigma^2 = \gamma_0 \left(\frac{1+\rho}{1-\rho}\right).
\end{align}

CASM implements the automatic equilibration and convergence checks introduced by Van de Walle and Asta \cite{van2002self} to determine the number of initial samples that should be excluded from the sample average as the system equilibrates, and to calculate from the remaining observations an estimate $\hat{\rho}$ of the autoregression parameter $\rho$.
The error $\bar{X} - \langle X \rangle$ can be then be estimated to a user-provided confidence level and calculations extended until the requested precision level is reached.

For example, assuming equilibrium has been reached and the covariance structure is well represented by Eq.~\ref{eq:autoregress_covar}, there is a 95\% confidence level in error $\bar{X} \pm p$ where 
\begin{align}
    p = 1.96 \sqrt{ \frac{\hat{\gamma}_0}{N} \left(\frac{1+\hat{\rho}}{1-\hat{\rho}}\right) }.
\end{align}

For n-fold way Monte Carlo simulations, the same approach can be used to estimate the error $\bar{X} - \langle X \rangle$ after performing a re-sampling procedure to generate $N'$ equally weighted observations $X_l'$ from the n-fold way observations $X_l$ with weights $w_l$.
The procedure is to use the n-fold way observations and associated weights to construct a time series $\tilde{X}(t) = X_l \ni \sum^{l-1}_{i=1} w_i < t < \sum^{l}_{i=1} w_i$
and then sample the time series at $N'$ regular intervals of size $\delta t = \sum^N_l w_l / N'$, such that $X_l' = \tilde{X}(l * \delta t)$.

CASM Monte Carlo methods allow setting a target precision level, as either an absolute value or fraction relative to the calculated mean, for any component of a non-scalar sampled quantity.
When multiple quantities are requested to be converged to a target precision level, they must all be equilibrated before samples are used for averaging, and all quantities must be converged to their requested target precision for an automatically converging Monte Carlo simulation to be considered complete.
When applicable to the particular Monte Carlo method being used, CASM allows setting a minimum and maximum number of Monte Carlo steps, Monte Carlo passes (where 1 pass equals 1 step per supercell site with variable degrees of freedom), samples, simulated time, or elapsed computational time.
In order to ensure proper equilibration, it is generally a good practice to set minimums for the number of steps, passes, samples, or simulated time, or to delay sampling until a certain number of steps, passes, or simulated time has passed.
Setting maximums is useful for stopping calculations that are very slow to converge so that a larger set of Monte Carlo simulations may complete.
As long as equilibration is deemed to have been reached for all requested quantities, the estimated means and errors in the mean are reported.

CASM Monte Carlo methods also give users significant control over when sampling takes places, and how much output is generated.
The default is to only output estimated means and standard errors, but users may also request to output all individual observations, or snapshots of the Monte Carlo supercell at each sampling time and in the initial and final states.
In either C++ or Python, it is straightforward to implement sampling functions for new properties and new analysis functions to evaluate quantities such as $c_{P, \tilde{\mu}_{i}}$ and $\chi$ which depend on all observations.

\section{The casm-flow software package}

The CASM Monte Carlo methods may be run directly, but to support high-throughput simulations covering multiple Hamiltonian models or systems, we have also created a Python package, casm-flow, that helps manage running and analyzing Monte Carlo simulations.
It is designed to run matching simulations on multiple sets of model system parameters, taking into account variations across model systems, such as differences in the number of ordered phases, the chemical potential ranges where they are stable, and the transition temperatures.
This makes it possible to compare and analyze results that were generated with multiple models fit to different data or sampled from a probability distribution.

To manage large numbers of Monte Carlo simulations, casm-flow makes integrated use of the software package signac \cite{signac1, signac2}.
Signac allows users to setup customized workflows and manage job submission on a cluster when dealing with large parameter spaces.

The casm-flow Python library includes a number of methods which create one or more paths of calculations in thermodynamic space, and then provides a standard set of features that allow a user to setup input files, run simulations directly or submit them on a cluster, and then help analyze and plot results.
Example methods include:
\begin{itemize}
\item
\texttt{PathFlowMethod}, which generates a single Monte Carlo path, with linearly or logarithmically spaced thermodynamic conditions. 
\item
\texttt{SeriesFlowMethod}, which generates a series of Monte Carlo paths covering a two-dimensional thermodynamic space.
\item
\texttt{GridFlowMethod}, which generates a $n$-dimensional grid of Monte Carlo paths.
This may be useful when running heating or cooling paths on a grid in an $n$-dimensional parametric chemical potential or composition space.
The grid may be explicitly set by the user, or automatically determined from the predicted convex hull of a model.
\item
\texttt{TreeFlowMethod}, which generates a tree of Monte Carlo paths in $n$-dimensions.
A tree starts with a Monte Carlo run along a single path in thermodynamic space and then branches out to a second dimension, then a third, etc.
This is useful for integrating free energies across temperature and chemical potential space.
It is also useful for varying other parameters such as the variance-constrained Monte Carlo bias.
\item
\texttt{SparseFlowMethod}, which generates a single Monte Carlo path at the approximate center of stability for each predicted ground state of a model.
This is useful for efficiently determining approximate transition temperatures before running more detailed calculations.
\item
\texttt{SparseTreesFlowMethod}, which generates a set of Monte Carlo path trees, with one starting from the approximate center of stability for each predicted ground state of a model.
The extent of the tree branches can be explicitly set by the user, or automatically determined from the predicted convex hull of a model.
\item
\texttt{ParamGridMethod}, which generates a set of Monte Carlo paths with identical thermodynamic conditions for a grid of simulation parameters, such as supercell size or target precision level.
\end{itemize}
Each of the above methods can be run in parallel for multiple models, and Monte Carlo simulation parameters can be customized on a per-model basis.

The casm-flow package also includes a number of features for analyzing model systems, generating Monte Carlo simulation parameters, and visualizing Monte Carlo results.
It can be used to calculate the predicted convex hull for a cluster expansion model and use the hull to determine the ground state configuration at a particular value of the chemical potential.
The package includes options to automatically determine Monte Carlo simulation supercells that match size or shape criteria, such as being commensurate with a particular ordered phase or supercell in which order parameters have been defined, having a minimum volume, or being restricted to one- or two-dimensional superlattices of a particular unit cell.
The package also includes functions to perform numerical free energy integration along paths in the canonical, semi-grand canonical, and variance-constrained ensembles.
It includes integration with the Python software package Bokeh \cite{bokeh} to generate interactive plots of the convex hull, Monte Carlo simulation results, and integrated free energies, in the style used in this paper.

As an example workflow, casm-flow can construct convex hulls, identify the ground state ordered phases, and generate appropriate order parameters for a given set of cluster expansion models.
Next, \texttt{SparseFlowMethod} can be used to generate constant chemical potential and increasing temperature Monte Carlo paths at the center of the stability region for each ground state phase of each cluster expansion model. 
The results of those runs can be used to identify approximate transition temperatures based on the values of the order parameters.
Then, \texttt{SparseTreesFlowMethod} can be used to construct Monte Carlo path trees for each ground state, with maximum temperatures set by the previous result, and free energies can be calculated for each phase by integration of the results.
Additionally, \texttt{GridFlowMethod} can be used to generate cooling Monte Carlo paths which can be integrated to calculate the free energy of the high temperature disordered phase.
Finally, casm-flow can plot the minimum free energy phase along a series of paths to construct phase diagrams such as Figure~\ref{fig:pd_hex}(a) and (b).

Besides the Python library which forms the core of casm-flow, a command line program is included which allows convenient use of many of its features, including the job control features available through integration with signac.
An initial public release of casm-flow is planned for fall 2023.

\section{Conclusions}

In this paper, we described the implementation of Monte Carlo techniques for the study multicomponent crystalline materials within the Clusters Approach to Statistical Mechanics (CASM) software suite.
The framework used by CASM to formulate thermodynamic potentials and kinetic transport coefficients accounting for arbitrarily complex crystal structures was presented and demonstrated with examples applying it to crystal systems of increasing complexity.
The cluster expansion method used by CASM to parameterize formation energies and the local cluster expansion method used to parameterize kinetic barriers was introduced.
Application of the methods implemented in CASM was demonstrated with the use of semi-grand canonical Metropolis Monte Carlo simulations to characterize first and second order phase transitions, calculate free energies, and construct a phase diagrams.
Additionally, the use of kinetic Monte Carlo (KMC) simulations to calculate kinetic coefficients was demonstrated.
Finally, a new software package has been introduced, casm-flow, which helps automate the setup, submission, management, and analysis of Monte Carlo simulations performed using CASM.

\section{Data availability}

Input files and results will be made available on Materials Commons \cite{Puchala2016mc}. Detailed usage instructions for CASM, including installation instructions are available online \cite{casmdocs}.

\section{Acknowledgments}

This work was supported by the U.S. Department of Energy, Office of Basic Energy Sciences, Division of Materials Sciences and Engineering under Award \#DE-SC0008637 as part of the Center for Predictive Integrated Structural Materials Science (PRISMS Center) at University of Michigan.

\appendix

\section{Parametric compositions and the semi-grand canonical potential}

\subsection{Parametric compositions in CASM}

The overall chemical composition of a multicomponent crystal can be tracked with a variety of different variables. 
Consider a crystal that can host $s$ species over different sublattices, where one of the species can be a vacancy. 
The total number of atoms of each type $i=1,...,s$ is $N_i$.
It is often desirable to normalize $N_i$ by the number of primitive unit cells, $N_u$ or by the total number of atoms in the crystal.
The number of each species per primitive cell can be tracked with the variables $n_i$ = $N_i/N_u$. 
These variables, however, are not independent for a crystal with a fixed number of primitive unit cells $N_u$ due to the constraint that $\sum_i n_i=n$, where $n$ is the number of crystal sites per primitive cell.
Furthermore, additional constraints may exist if certain species are restricted to a subset of sublattices.
These constraints reduce the number of independent composition variables and make it convenient to work with parametric composition variables.

\begin{figure}
    \centering
    \includegraphics[width=7cm]{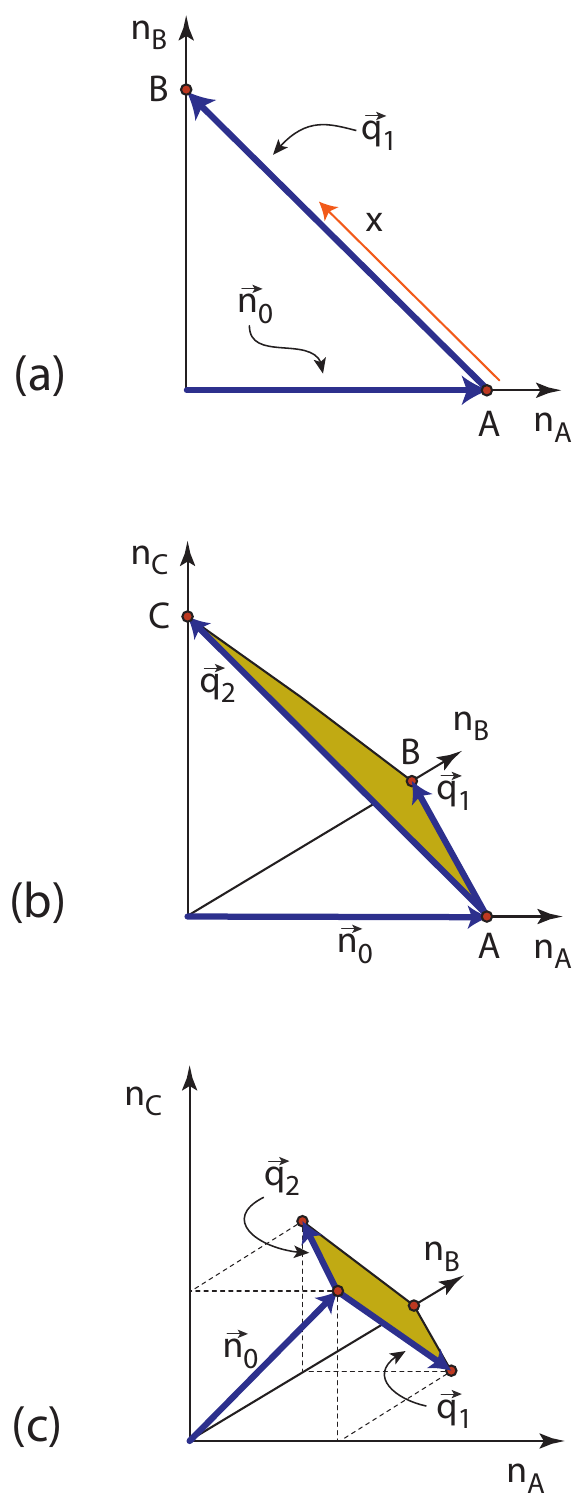}
    \caption{Examples of the allowed composition space for (a) an A-B binary alloy with a single sublattice, (b) an A-B-C ternary alloy with a single sublattice, and (c) an A-B-C ternary alloy with two sublattices in which A and B can occupy the first sublattice and B and C can occupy the second sublattice. Each system is shown with a possible choice of composition axes $\vec{q}^{\mathsf{T}}_1$ and $\vec{q}^{\mathsf{T}}_2$ and origin $\vec{n}^{\mathsf{T}}_0$.}
    \label{fig:composition_axes}
\end{figure}

The geometric meaning of parametric composition is illustrated in Figure \ref{fig:composition_axes}(a) for the simple example of a binary A-B alloy with one sublattice per unit cell. 
In this example, $n_A$ and $n_B$ track the number of A and B atoms per unit cell. 
They are not independent, however, since A and B share the same sites of the crystal, leading to the constraint that $n_A+n_B=1$. 
There are two extremes in the subspace of allowed compositions: a crystal containing only A atoms at $n_A=1$ and $n_B=0$ and a crystal containing only B atoms with $n_A=0$ and $n_B=1$. 
These two extremes are illustrated in Figure \ref{fig:composition_axes}(a) and are labeled as A and B. 
Any intermediate concentration of the alloy will reside on the line connecting points A and B. The introduction of independent parametric compositions requires the choice of an origin, for example, pure A.
In $n_A$ and $n_B$ space, this origin is represented with the vector $\vec{n}_0$ as illustrated in Figure \ref{fig:composition_axes}(a). 
The space of allowed concentrations can be spanned with a second vector, $\vec{q}_1$, as illustrated in Figure \ref{fig:composition_axes}(a). 
Any allowed composition of A and B in this crystal can then be expressed in vector form as 
\begin{equation}
    \vec{n}=\vec{n}_0+x\vec{q}_1
\end{equation}
where $\vec{n}^{\mathsf{T}}=[n_A, n_B]$ and $x$ is a parametric composition within the subspace of allowed compositions. 
If the length of $\vec{q}_1$ is equal to the distance between points 1 and 2 (the two extreme compositions) in $n_A$ and $n_B$ space, then the parametric composition will take on any value between 0 and 1. 

Another example is shown in Figure \ref{fig:composition_axes}(b) for a ternary A-B-C alloy having a primitive cell containing one sublattice (i.e. $n=1$). 
There are now three composition axes with variables $n_A$, $n_B$ and $n_C$.
The constraint that $n_A+n_B+n_C=1$ restricts the space of allowed compositions to a two-dimensional subspace.
In Figure \ref{fig:composition_axes}(b) the origin is chosen to be pure A. 
Two vectors are needed to span the two-dimensional subspace of allowed compositions. 
These are denoted as $\vec{q}_1$ and $\vec{q}_2$ in Figure \ref{fig:composition_axes}(b). 
These vectors connect the chosen origin, $\vec{n}^{\mathsf{T}}_0=[1,0,0]$ to $\vec{n}^{\mathsf{T}}=[0,1,0]$ and $\vec{n}^T=[0,0,1]$, respectively. 
Any concentration $\vec{n}^{\mathsf{T}}=[n_A,n_B,n_C]$ can then be expressed as 
\begin{equation}
    \vec{n}=\vec{n}_0+x_1\vec{q}_1+x_2\vec{q}_2
\end{equation}
where $x_1$ and $x_2$ are parametric compositions.

Many crystals are more complex than the two examples above. 
For example, a crystal with two sites per primitive cell may host A and B atoms on the first sublattice and B and C on the second sublattice. 
The number of each species per primitive cell can again be tracked in a three dimensional space spanned by the variables $n_A$, $n_B$ and $n_C$. 
However, not all values of $n_A$, $n_B$ and $n_C$ are allowed due to the constraint of a fixed number of sites per primitive cell and the additional constraint that $A$ and $C$ atoms can only occupy one of the two sublattice, while $B$ can occupy both sublattices. 

The subspace of allowed values of $n_A$, $n_B$ and $n_C$ can be identified by enumerating the extreme compositions. 
One extremum in the subspace of allowed compositions corresponds to AC, with $\vec{n}^{\mathsf{T}}=[1,0,1]$. 
Two other extrema in the subspace of allowed compositions include the chemical formula AB with $\vec{n}^{\mathsf{T}}=[1,1,0]$ and BC with $\vec{n}^{\mathsf{T}}=[0,1,1]$.
A fourth extremum corresponds to the chemical formula BB in which both sublattices are occupied by B atoms, with $\vec{n}^{\mathsf{T}}=[0,2,0]$.
Figure \ref{fig:composition_axes}(c) illustrates the extrema of allowed compositions for this particular crystal and highlights the two-dimensional subspace of allowed values of $n_A$, $n_B$ and $n_C$.

Parametric compositions can again be introduced to navigate within the two-dimensional subspace of allowed compositions. 
First an origin needs to be chosen. 
For example, $\vec{n}_0^{\mathsf{T}}=[1,0,1]$ corresponding to the chemistry AC could be a possible origin. 
Two spanning vectors for the two-dimensional subspace are $\vec{q}_{1}$ and $\vec{q}_2$ as illustrated in Figure \ref{fig:composition_axes}(c). 
The concentration of the compound in terms of parametric compositions is again of the form
\begin{equation}
    \vec{n}=\vec{n}_0+x_1\vec{q}_1+x_2\vec{q}_2
\end{equation}
where $x_1$ and $x_2$ are parametric compositions. 

In general, for a solid containing $s$ species, the number of each species per primitive cell as collected in the $s$-dimensional vector, $\vec{n}$, can be expressed in terms of parametric compositions $\vec{x}$ according to
\begin{equation}
    \vec{n}=\vec{n}_0+\pmb{Q}\vec{x}
    \label{eq:general_composition_relation}
\end{equation}
where as before, $\vec{n}_0$ points to a chosen origin and the matrix $\pmb{Q}=[\vec{q}_1,...,\vec{q}_k]$ collects as columns $k$ independent vectors that span the subspace of allowed values of $\vec{n}$, consistent with the constraint of a fixed number of sites per primitive cell and any additional sublattice constraints. 
The matrix $\pmb{Q}$, therefore, has dimensions $s\times k$, leading to $k < s$ parametric compositions, $\vec{x}^{\mathsf{T}}=[x_1,...,x_k]$.

The $k$ vectors $\vec{q}_i$ that span the subspace of allowed values of $\vec{n}$ do not necessarily form an orthogonal set. 
It therefore becomes useful to introduce a second set of vectors, $[\vec{r}_{1},\dots,\vec{r}_k]$, that span the same subspace and that satisfy $\vec{r}^{\mathsf{T}}_i\vec{q}_{j}=\delta_{i,j}$ for $i,j=1,\dots,k$, with $\delta_{i,j}$ being the Kronecker delta. 
The matrix
\begin{equation}
    \pmb{R}=[\vec{r}_{1},\dots,\vec{r}_k],
\end{equation}
whose transpose is the left pseudoinverse of $\pmb{Q}$ (i.e. ${\pmb{R}}^\mathsf{T}\pmb{Q}=I$ where $I$ is a $k\times k$ identity matrix) makes it possible to determine the parametric compositions $\vec{x}$ given the concentration variables per unit cell, $\vec{n}$, according to 
\begin{align} \label{eq:concentration_equation_R}
    \vec{x} = {\pmb{R}}^\mathsf{T}(\vec{n} - \vec{n}_0)
\end{align}
Since $\pmb{Q}$ describes the allowed composition space, it has full column rank and $\pmb{R}^\mathsf{T}$ can be calculated using:
\begin{align} \label{composition_space_basis_R} 
{\pmb{R}}^\mathsf{T} &= ({\pmb{Q}}^\mathsf{T} {\pmb{Q}})^{-1} {\pmb{Q}}^\mathsf{T}.
\end{align}
The set of $\vec{r}_i$ vectors become equal to the set of $\vec{q}_i$ vectors when the latter form an orthonormal set. 

When imposing the constraints of the crystal on diffusional flux expressions in Appendix B, it will be convenient to utilize the orthogonal projection operator defined as 
\begin{equation}
    \mathbf{P}=\sum_{i=1}^{k}\vec{r}_{i}\vec{q}_{i}^\mathsf{T} = \pmb{R}{\pmb{Q}}^\mathsf{T},
    \label{eq:projection_operator1}
\end{equation}
which is a $s\times s$ matrix with rank $k$. 
Any vector $\vec{v}=\vec{n}-\vec{n}_0$ in the full $s$ dimensional composition space will be projected onto the subspace of allowed compositions spanned by the sets $\vec{q}_i$ or $\vec{r}_i$ when multiplied by $\mathbf{P}$. 
Any vector $\vec{v}$ that is already in the space spanned by the column vectors of $\pmb{Q}$, (i.e. $\vec{v}=\pmb{Q}\vec{x}$) is unaffected by $\mathbf{P}$ (i.e. $\mathbf{P}\vec{v}=\vec{v}$), which can be verified by substituting Eq. \ref{composition_space_basis_R} into Eq. \ref{eq:projection_operator1} and using the fact that $({\pmb{Q}}^\mathsf{T} {\pmb{Q}})^{-1}$ is a symmetric matrix and is equal to its transpose.

\begin{figure}
    \centering
    \includegraphics[width=7cm]{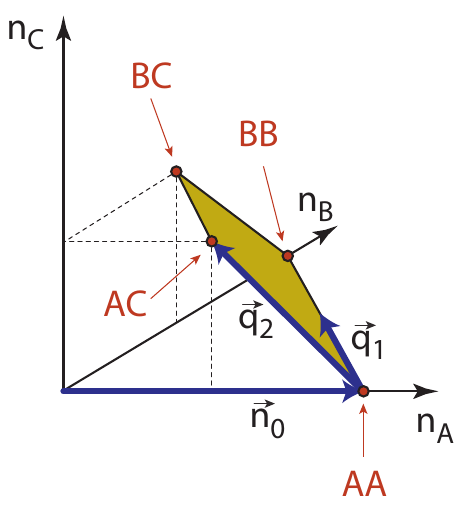}
    \caption{The space of allowed compositions for a crystal with two sublattices in which A and B can occupy the first sublattice and A, B and C can occupy the second sublattice. A choice of composition axes $\vec{q}^{\mathsf{T}}_1=[-1,1,0]$ and $\vec{q}^{\mathsf{T}}_2=[-1,0,1]$ with origin $\vec{n}^{\mathsf{T}}_0=[2,0,0]$ is shown along with the stoichiometic compositions at the extrema of the allowed composition space: AA, AC, BB and BC.}
    \label{fig:composition_axes2}
\end{figure}

In general, it is most convenient to choose the length of each $\vec{q}_i$ such that it corresponds to one unit exchange of atoms per unit cell, i.e. the resulting vector is of the form $[0,...,1,...,-1,...0]$.
In the examples treated so far, this has always been the case for the chosen $\vec{q}$ vectors. For crystals with more than one sublattice per primitive unit cell, this will generally lead to parametric compositions that are not restricted to the interval [0,1].
To illustrate this, consider the example of a crystal with two sublattices in which A and B can occupy the first sublattice and A, B and C can occupy the second sublattice. 
The composition space for this example is illustrated in Figure \ref{fig:composition_axes2}. 

As the origin of the parametric composition space, one possible choice is pure A, in which both sublattices are completely occupied by A atoms, i.e. $\vec{n}^{\mathsf{T}}_0=[2,0,0]$. 
The subspace of allowed compositions can be spanned by $\vec{q}^{\mathsf{T}}_1=[-1,1,0]$ and $\vec{q}^{\mathsf{T}}_2=[-1,0,1]$ as illustrated in Figure \ref{fig:composition_axes2}. 
Notice that $\vec{q}_1$ does not extend all the way from the AA composition to the BB composition, but has a length that is half that distance. 
This is to ensure that moving along that axis by a unit distance of $\vec{q}_1$ corresponds to an exchange of one A atom with one B atom. 
By using this particular vector, the parametric composition corresponding to this axis, $x_1$, ranges from 0 to 2, as two A atoms must be replaced by 2 B atoms to go from the AA composition to the BB composition. 
This convention ensures that the exchange chemical potentials that are conjugate to the resulting parametric compositions are quantities of an atom as opposed to that of multiple atoms.

\subsection{Examples of semi-grand canonical potentials}

It is instructive to inspect the form of semi-grand canonical potentials and the exchange chemical potentials $\vec{\tilde{\mu}}$ for specific examples. 
As defined in Section \ref{sec:characteristic_potentials}, the semi-grand canonical potential takes the form
\begin{equation}
    \phi=g-\vec{\tilde{\mu}}^{\mathsf{T}}\vec{x}=\vec{\mu}^{\mathsf{T}}\vec{n}_0
\end{equation}
where $\phi=\Phi/N_u$ and $g=G/N_u$ and where $\vec{\tilde{\mu}}=\pmb{Q}^{\mathsf{T}}\vec{\mu}$. 
Another important thermodynamic quantity is the semi-grand canonical enthalpy defined as
\begin{equation}
    \omega=e-\vec{\tilde{\mu}}^{\mathsf{T}}\vec{x}
\end{equation}
where $e$ is the average energy per primitive cell. 
The next sections illustrate these potentials for concrete examples.

\subsubsection{Binary and ternary alloys on a single sublattice}
\label{sec:app_A_ex_simple_binary_alloys}

For the simple binary alloy with the choice of origin as shown in Figure \ref{fig:composition_axes}(a), 

\begin{align}
    \vec{n}_0 & = \left[ \begin{array}{c} 1 \\ 0  \end{array} \right] & \vec{q}_1 & = \left[\begin{array}{c} -1 \\ 1  \end{array} \right]
\end{align}
Since the subspace of allowed compositions is one dimensional, $\pmb{Q}$ is a $2\times 1$ matrix
\begin{align}
    \pmb{Q} & =\left[ \begin{array}{c} -1 \\ 1  \end{array} \right], 
\end{align}
According to Equation \ref{eq:exchange_chemical_potentials}, the exchange chemical potential is the equal to
\begin{align}
    \tilde{\mu}_1 & = \mu_B - \mu_A,
\end{align}
and is the intensive variable that is conjugate to the parametric composition $x_1$.
The semi-grand canonical free energy becomes equal to the chemical potential of A since, using Equation \ref{eq:normalized_semi-grand_canonical_potential}
\begin{align}
    \phi=\vec{\mu}^{\mathsf{T}}\vec{n}_0=[\mu_A, \mu_B]\left[ \begin{array}{c} 1 \\ 0  \end{array} \right]= \mu_A
\end{align}

In semi-grand canonical Monte Carlo simulations, $\tilde{\mu}_1$ together with temperature T are the input variables, while quantities such as the average parametric composition $\langle x_1\rangle$ and the average semi-grand canonical enthalpy, $\langle\omega\rangle=\langle e -\tilde{\mu}_1 x_1\rangle$ are outputs. 
Since $\tilde{\mu}_1$ is a difference in the chemical potentials of B and A, it is necessary to know both $\tilde{\mu}_1$ and the semi-grand canonical potential $\phi=\mu_A$ in order to determine the individual chemical potentials $\mu_A$ and $\mu_B$. 
This requires a free energy integration step described in Section \ref{sec:free_energy_integration}. 

Similar results are obtained for the ternary A-B-C alloy with origin and parametric composition axes as chosen in Figure \ref{fig:composition_axes}(b). 
Now there are two exchange chemical potentials, with 
\begin{align}
    \tilde{\mu}_1 & = \mu_{B}-\mu_{A}, \\
    \tilde{\mu}_2 & = \mu_{C}-\mu_{A}
\end{align}
while the semi-grand canonical potential becomes
\begin{equation}
    \phi=g-\tilde{\mu}_{1}x_1-\tilde{\mu}_{2}x_2=\mu_{A}
\end{equation}
and the semi-grand canonical enthalpy (at zero pressure) per primitive cell becomes
\begin{equation}
    \omega=e-\tilde{\mu}_{1}x_1-\tilde{\mu}_{2}x_2
\end{equation}

\subsubsection{Two binary sublattices sharing a common species}

Consider the example of Figure \ref{fig:composition_axes}(c) again for a crystal with two sites in the primitive cell, with the first sublattice hosting A and B and the second sublattice hosting B and C. 
For the choice of origin in Figure \ref{fig:composition_axes}(c) and the choice of parametric composition axes
\begin{align}
    \vec{n}_0 & = \left[ \begin{array}{c} 1 \\ 0 \\ 1  \end{array} \right] & \vec{q}_1 & = \left[\begin{array}{c} 0 \\ 1  \\ -1 \end{array} \right] & \vec{q}_2 & = \left[\begin{array}{c} -1 \\ 1  \\ 0 \end{array} \right]
\end{align}
The two-dimensional composition subspace leads to a $3\times 2$ $\pmb{Q}$ matrix
\begin{align}
    \pmb{Q} & =\left[ \begin{array}{cc} 0 & -1 \\ 1 & 1 \\ -1 & 0   \end{array} \right], 
\end{align}
This matrix yields for the exchange chemical potentials
\begin{align}
    \tilde{\mu}_1 & = \mu_{B}-\mu_{C}, \\
    \tilde{\mu}_2 & = \mu_{B}-\mu_{A}
\end{align}
while the semi-grand canonical potential becomes
\begin{equation}
    \phi=g-\tilde{\mu}_{1}x_1-\tilde{\mu}_{2}x_2=\mu_A+\mu_C
\end{equation}

\subsubsection{Binary and ternary sublattices sharing multiple common species}
As a final example, consider the more complex crystal with two sublattices per primitive cell with the first sublattice hosting A and B and the second sublattice hosting A, B and C. 
The composition space for this example is shown in Figure \ref{fig:composition_axes2}.
In Figure \ref{fig:composition_axes2}, the origin and parametriccomposition axes are chosen as 
\begin{align}
    \vec{n}_0 & = \left[ \begin{array}{c} 2 \\ 0 \\ 0  \end{array} \right] & \vec{q}_1 & = \left[\begin{array}{c} -1 \\ 1  \\ 0 \end{array} \right] & \vec{q}_2 & = \left[\begin{array}{c} -1 \\ 0  \\ 1 \end{array} \right]
\end{align}
This leads to
\begin{align}
    \pmb{Q} & =\left[ \begin{array}{cc} -1 & -1 \\ 1 & 0 \\ 0 & 1   \end{array} \right], 
\end{align}
The exchange chemical potentials then become 
\begin{align}
    \tilde{\mu}_1 & = \mu_{B}-\mu_{A}, \\
    \tilde{\mu}_2 & = \mu_{C}-\mu_{A}
\end{align}
The semi-grand canonical potential becomes
\begin{equation}
    \phi=g-\tilde{\mu}_{1}x_1-\tilde{\mu}_{2}x_2=2\mu_A
\end{equation}
The factor of two in front of the chemical potential of A emerges since the primitive unit cell contains two sites, which are both occupied by A atoms at the origin of the parametric composition space.

\section{Fluxes and driving forces for diffusion in crystals}
\label{sec:diffusion_appendix}

\subsection{Crystallographic constraints on flux expressions}

Diffusion within a crystal that occurs far away from extended defects proceeds under crystallographic constraints that affect the form of the flux expressions, thermodynamic driving forces and the Onsager transport coefficients. 
Indeed, diffusion within single crystal regions of a solid can only lead to a spatial redistribution of different chemical species, but cannot result in the extension of the crystal. 
It, therefore, occurs at constant $N_u$, the number of unit cells. 
This imposes constraints on diffusional fluxes and on the form of diffusional potentials (driving forces). 
To establish these constraints, it is necessary to revisit the composition variables of the previous section. 
For a crystal having a fixed number of unit cells, $N_u$, the concentrations of $s$ species per unit cell, $\vec{n}^{\mathsf{T}}$, reside in a lower dimensional subspace spanned by a set of vectors $[\vec{q}_1,\dots,\vec{q}_k]$.
Parametric compositions $\vec{x}$ serve as coordinates in this subspace. 
Diffusion can only result in composition changes of $\vec{n}$ that stay in the subspace spanned by $[\vec{q}_1,\dots,\vec{q}_k]$. 
This means that any changes in $\vec{n}$ that leave this subspace are forbidden. 

The subspace that is orthogonal to $[\vec{q}_1,\dots,\vec{q}_k]$ has a dimension $s-k$ and can be spanned by a set of orthogonal vectors $[\vec{t}_{k+1},\dots,\vec{t}_s]$. 
Such a set of vectors can always be found by finding the null space of $\pmb{Q}=[\vec{q}_1,\dots,\vec{q}_k]$.
The spanning vectors of the subspace orthogonal to $\pmb{Q}$ can be collected in the $s\times (s-k)$ matrix
\begin{equation}
    \pmb{T}=[\vec{t}_{k+1},\dots,\vec{t}_s].
\end{equation}
The constraint of a constant number of unit cells $N_u$ on the composition variable $\vec{n}$ can be expressed as
\begin{equation}
    \pmb{T}^{\mathsf{T}}(\vec{n}-\vec{n}_0)=\vec{0}
    \label{eq:orthogonal_constraints1}
\end{equation}
which ensures that variations in the concentration $\vec{n}$ do not stray into the subspace spanned by $[\vec{t}_{k+1},\dots,\vec{t}_s]$.
Equation \ref{eq:orthogonal_constraints1} form a set of $s-k$ linear constraints on the concentration variables $\vec{n}^{\mathsf{T}}=[n_{A},n_{B},\dots]$.

Constraints similar to Eq. \ref{eq:orthogonal_constraints1} also apply to the diffusional fluxes of the different species within a perfect crystal having a fixed number of unit cells $N_u$. 
This is a consequence of the conservation of particles equation, that relates each concentration variable $n_{m}$ for species $m$ to a corresponding flux $J_m$ according to
\begin{equation}
    \frac{\partial (n_m/v_{u})}{\partial t} = -\nabla{J_m}
\end{equation}
where $v_u$ represents the volume of the unit cell of the crystal. 
Due to the linearity of time and space derivatives, the constraints of Eq. \ref{eq:orthogonal_constraints1} that apply to $\vec{n}$ transfer to the fluxes $\vec{J}$ and can be expressed as
\begin{equation}
    \pmb{T}^{\mathsf{T}}\vec{J}=\vec{0}
    \label{eq:orthogonal_constraints2}
\end{equation}
where $\vec{J}$ collects the fluxes of the $s$ species of the crystal.
There are therefore $s-k$ linear constraints on the fluxes of each element within the crystal when holding the number of unit cells constant. 

The constraints on the fluxes also affect the Onsager transport coefficients that appear in the flux expressions
\begin{align}
\vec{J} = -\Lonsager \nabla \vec{\mu}, 
\label{eq:flux_expressions}
\end{align}
Since the constraints of Eq. \ref{eq:orthogonal_constraints2} hold independent of the values of the driving forces, $\nabla\vec{\mu}$ appearing in Eq. \ref{eq:flux_expressions}, the following must hold for the Onsager matrix
\begin{equation}
    {\pmb{T}}^{\mathsf{T}}\Lonsager= \mathbf{0}
    \label{eq:L_constraints1}
\end{equation}
where the $\mathbf{0}$ represents a $(s-k)\times s$ matrix of zeros. 
These equations, therefore, constitute a set of $(s-k)\times s$ linear relationships on the Onsager transport coefficients that emerge due to the constraint of a fixed number of unit cells. 
Due to the Onsager reciprocity relationships ($\Lonsager=\Lonsager^{T}$), the above equation can also be rewritten as 
\begin{equation}
    \Lonsager{\pmb{T}}=\mathbf{0}
    \label{eq:L_constraints2}
\end{equation}
where the $\mathbf{0}$ now represents a $s\times (s-k)$ matrix of zeros. 

The linear relationships between the Onsager transport coefficients can be accounted for by rewriting the flux expressions $\ref{eq:flux_expressions}$. 
To this end, it is useful to rely on projection operators for the subspace of allowed compositions when fixing the number of unit cells, which was introduced in Appendix A, Eq. \ref{eq:projection_operator1}, as
\begin{equation}
    \mathbf{P}_{c}=\pmb{R}{\pmb{Q}}^{\mathsf{T}}
\end{equation}
and for the subspace that is orthogonal to it, which can be expressed in terms of the matrix $\pmb{T}$ as
\begin{equation}
     \mathbf{P}_{o}=\sum_{i=k+1}^{s}\vec{t}_i{\vec{t}_i}^{\mathsf{T}}=\pmb{T}{\pmb{T}}^{\mathsf{T}}
\end{equation}
The projection operators, $\mathbf{P}_{c}$ projecting onto the allowed composition space and $\mathbf{P}_{o}$ projecting onto its kernel, are complementary, and sum to the identity matrix
\begin{equation}
    \mathbf{I}=\mathbf{P}_{c}+\mathbf{P}_{o} = \pmb{R}{\pmb{Q}}^{\mathsf{T}}+\pmb{T}{\pmb{T}}^{\mathsf{T}}.
    \label{eq:identity}
\end{equation}

The identity operator, Eq. \ref{eq:identity}, can be inserted between the Onsager matrix and the gradients in chemical potentials in the flux expressions, Eq. \ref{eq:flux_expressions}, to yield 
\begin{equation}
    \vec{J}=-\Lonsager\pmb{R}\nabla({\pmb{Q}}^{\mathsf{T}}\vec{\mu})
\end{equation}
where Eq. \ref{eq:L_constraints2} was used. 
By further multiplying both sides by ${\pmb{R}}^{\mathsf{T}}$, the flux expressions can be rewritten as 
\begin{equation}
    \vec{\tilde{J}}=-\tilde{\Lonsager}\nabla\vec{\tilde{\mu}}
    \label{eq:projected_flux_equations}
\end{equation}
where the $k$ independent exchange fluxes are defined as
\begin{equation}
    \vec{\tilde{J}}={\pmb{R}}^{\mathsf{T}}\vec{J} 
    \label{eq:flux_projection}
\end{equation}
and where 
\begin{equation}
    \tilde{\Lonsager}={\pmb{R}}^{\mathsf{T}}\Lonsager\pmb{R}
    \label{eq:projected_L_matrix}
\end{equation}
is a $k\times k$ Onsager transport coefficient matrix that is consistent with the constraints of a constant number of unit cells. 
The $\vec{\tilde{\mu}}={\pmb{Q}}^{\mathsf{T}}\vec{\mu}$ are the exchange chemical potentials defined in Appendix A. 
These expressions describe the fluxes, transport coefficients and thermodynamic driving forces for diffusion in a crystal with a fixed number of unit cells $N_u$.

The transformed fluxes, $\vec{\tilde{J}}$ reside in the dual space of the exchange chemical potentials, $\vec{\tilde{\mu}}$.

\subsection{Examples of flux expressions}
\label{sec:Appendix_flux_examples}

\subsubsection{Diffusion in a binary alloy with a single sublattice by direct exchange}

Consider the simple binary alloy in which diffusion occurs through direct exchanges between A and B atoms on a single sublattice. As detailed in~\ref{sec:app_A_ex_simple_binary_alloys}, for such a system
\begin{align}
    \pmb{Q} & =\left[ \begin{array}{c} -1 \\ 1  \end{array} \right],
\end{align}
is consistent with the composition formula A$_{1-{x_1}}$B$_{x_1}$.
The transpose of the left pseudo inverse of $\pmb{Q}$ is
\begin{align}
    \pmb{R} & =\frac{1}{2}\left[ \begin{array}{c} -1 \\ 1  \end{array} \right].
\end{align}

The space that is orthogonal to the space of allowed compositions is the column vector space
\begin{align}
    \pmb{T} & =\frac{1}{\sqrt{2}}\left[ \begin{array}{c} 1 \\ 1  \end{array} \right]. 
\end{align}
Hence according to Eq. \ref{eq:orthogonal_constraints2}, the crystallographic constraints require that
\begin{equation}
    \frac{1}{\sqrt{2}}[1 , 1]\left[ \begin{array}{c} J_{A} \\ J_{B}  \end{array} \right]=\frac{1}{\sqrt{2}}(J_{A}+J_{B})=0,
    \label{eq:binary_alloy_flux_constraint}
\end{equation}
which imposes the constraint that a flux in A atoms must be compensated by an opposite flux of B atoms (i.e. $J_{A}=-J_{B}$) in order to preserve crystal sites. 

For a binary system with direct exchange hops, the flux expressions take the form
\begin{align}
    \left[ \begin{array}{c} J_{A} \\ J_{B}  \end{array}\right] & = -\left[ \begin{array}{cc} L_{AA} & L_{AB} \\ L_{AB} & L_{BB}  \end{array} \right] \left[ \begin{array}{c} \nabla\mu_{A} \\ \nabla\mu_{B}  \end{array}\right], 
\end{align}
Projecting the flux expressions to the subspace of allowed concentrations consistent with a fixed number of unit cells yields
\begin{equation}
    \tilde{J}_{1}=\tilde{L}\nabla\tilde{\mu}_{1}
    \label{eq:}
\end{equation}
where according to Eq. \ref{eq:projected_L_matrix},  
\begin{align}
    \tilde{L}=
    \left[ \begin{array}{cc} -1/2 & 1/2  \end{array} \right]
    \left[ \begin{array}{cc} L_{AA} & L_{AB} \\ L_{AB} & L_{BB}  \end{array} \right]
    \left[ \begin{array}{c} -1/2 \\ 1/2  \end{array} \right]
\end{align}
which yields
\begin{equation}
    \tilde{L}=\frac{1}{4}(L_{AA}+L_{BB}-2L_{AB})
    \label{eq:tildeL_binary_alloy}
\end{equation}
This expression can be further simplified by exploiting the constraints on the $\Lonsager$ matrix as codified by Eq. \ref{eq:L_constraints2}, which for this example can be expressed explicitly as
\begin{align}
    \left[ \begin{array}{cc} L_{AA} & L_{AB} \\ L_{AB} & L_{BB}  \end{array} \right] \left[ \begin{array}{c} 1/\sqrt{2} \\ 1/\sqrt{2}  \end{array}\right] = \left[ \begin{array}{c} 0 \\ 0  \end{array}\right] 
\end{align}
These two linear equations make it possible to eliminate two of the three Onsager coefficients such that Eq. \ref{eq:tildeL_binary_alloy} can be rewritten as
\begin{equation}
    \tilde{L}=L_{BB}
    \label{eq:L=LBB}
\end{equation}

The exchange chemical potential is equal to
\begin{align}
    \tilde{\mu}_1 & = \left[ \begin{array}{cc} -1 & 1  \end{array} \right]
    \left[ \begin{array}{c} \mu_{A} \\ \mu_{B}  \end{array} \right]=\mu_B - \mu_A,
\end{align}
while the exchange flux is
\begin{equation}
    \tilde{J}_{1}= \left[ \begin{array}{cc} -1/2 & 1/2  \end{array} \right]
    \left[ \begin{array}{c} J_{A} \\ J_{B}  \end{array} \right]=\frac{1}{2}(J_{B}-J_{A}) = J_{B}
\end{equation}
where the third equality follows by using Eq. \ref{eq:binary_alloy_flux_constraint}.
The flux expression projected into the constant number of crystal unit cells subspace can therefore be expressed as 
\begin{equation}
    J_{B}=-L_{BB}\nabla (\mu_{B}-\mu_{A})
\end{equation}

\subsubsection{Diffusion in a binary alloy with a single sublattice mediated by a vacancy mechanism}

\begin{figure}
    \centering
    \includegraphics[width=7cm]{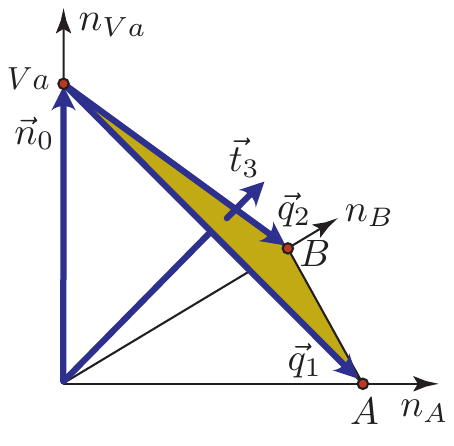}
    \caption{The space of allowed compositions for a crystal with a single sublattice that can be occupied by A atoms, B atoms and vacancies (Va). A choice of composition axes $\vec{q}^{\mathsf{T}}_1=[1,0,-1]$ and $\vec{q}^{\mathsf{T}}_2=[1,0,-1]$ with origin $\vec{n}^{\mathsf{T}}_0=[0,0,1]$ is shown along with the stoichiometic compositions at the extrema of the allowed composition space: A, B and Va. This is consistent with the composition formula A$_{x_1}$B$_{x_2}$Va$_{1-{x_1}-{x_2}}$.}
    \label{fig:ternary_composition_axes_ABVa_1}
\end{figure}

A common scenario is diffusion in a binary A-B alloy that is mediated by a vacancy. 
The crystal contains three species, A, B and vacancies denoted Va and the concentration per unit cell $\vec{n}^{\mathsf{T}}=[n_{A},n_{B},n_{Va}]$. 
For the purpose of setting up flux expressions, it is convenient to choose as the origin for the parametric composition variables the crystal in which all sites are fully vacant, i.e.
\begin{align}
    \vec{n}_0 & = \left[ \begin{array}{c} 0 \\ 0 \\ 1  \end{array} \right] & \vec{q}_1 & = \left[\begin{array}{c}  1 \\ 0  \\ -1 \end{array} \right] & \vec{q}_2 & = \left[\begin{array}{c}  0 \\ 1  \\ -1 \end{array} \right].
\end{align}
The space of allowed compositions consistent with the constraints of the crystal, as illustrated in Figure \ref{fig:ternary_composition_axes_ABVa_1}, is
\begin{align}
    \pmb{Q} & = \left[\begin{array}{cc} 1 & 0 \\ 0 & 1 \\ -1  & -1 \end{array} \right],  & \pmb{R} & = \left[\begin{array}{cc} 2/3 & -1/3 \\ -1/3 & 2/3  \\ -1/3 & -1/3 \end{array} \right]
\end{align}

The space that is orthogonal to space of allowed compositions is $\vec{t}^{\mathsf{T}}_3=1/\sqrt{3}[1,1,1]$, such that 
\begin{align}
    \pmb{T} & = \left[ \begin{array}{c} 1/\sqrt{3} \\ 1/\sqrt{3} \\ 1/\sqrt{3}  \end{array} \right].
\end{align}

According to Eq. \ref{eq:orthogonal_constraints2}, the crystallographic constraints require that
\begin{equation}
    \frac{1}{\sqrt{3}}\left[ \begin{array}{ccc} 1 & 1 & 1 \end{array} \right]\left[ \begin{array}{c} J_{A} \\ J_{B} \\ J_{Va}  \end{array} \right]=\frac{1}{\sqrt{3}}(J_{A}+J_{B}+J_{Va})=0,
    \label{eq:ternary_alloy_flux_constraint}
\end{equation}

The flux expressions take the form
\begin{align}
    \left[ \begin{array}{c} J_{A} \\ J_{B} \\ J_{Va} \end{array}\right] & = -\left[ \begin{array}{ccc} L_{AA} & L_{AB} & L_{AVa} \\ L_{AB} & L_{BB} & L_{BVa} \\ L_{AVa} & L_{BVa} & L_{VaVa}\end{array} \right] \left[ \begin{array}{c} \nabla\mu_{A} \\ \nabla\mu_{B} \\ \nabla\mu_{Va} \end{array}\right], 
\end{align}

Linear constraints on the Onsager coefficients due to Eq. \ref{eq:L_constraints2} makes it possible to eliminate the Onsager coefficients involving vacancies (i.e. $L_{AVa}=-(L_{AA}+L_{AB})$, $L_{BVa}=-(L_{AB}+L_{BB})$ and $L_{VV}=-(L_{AVa}+L_{BVa})$). 
Projecting the Onsager coefficient matrix to the fixed crystal subspace according to Eq. \ref{eq:projected_L_matrix} leads to 
\begin{align}
    \tilde{\Lonsager} & = \left[ \begin{array}{cc} L_{AA} & L_{AB} \\ L_{AB} & L_{BB}  \end{array} \right], 
\end{align}
where the above linear relationships between the Onsager coefficients were used. 

Projecting the fluxes according to Eq. \ref{eq:flux_projection} yields

\begin{align}
    \left[ \begin{array}{c} \tilde{J}_{1} \\ \tilde{J}_{2} \end{array}\right] & = \frac{1}{3}\left[ \begin{array}{ccc} 2 & -1 & -1 \\ -1 & 2 & -1 \end{array} \right] \left[ \begin{array}{c} J_{A} \\ J_{B} \\ J_{Va} \end{array}\right]= \left[ \begin{array}{c} J_{A} \\ J_{B} \end{array}\right]
\end{align}
where the constraint on the fluxes, Eq. \ref{eq:ternary_alloy_flux_constraint}, was used to eliminate the vacancy flux. 
The diffusion potentials, which are the exchange chemical potentials, become
\begin{align}
    \left[ \begin{array}{c} \tilde{\mu}_{1} \\ \tilde{\mu}_{2} \end{array}\right] & = \left[ \begin{array}{ccc} 1 & 0 & -1 \\ 0 & 1 & -1 \end{array} \right] \left[ \begin{array}{c} \mu_{A} \\ \mu_{B} \\ \mu_{Va} \end{array}\right]= \left[ \begin{array}{c} \mu_{A}-\mu_{Va} \\ \mu_{B}-\mu_{Va} \end{array}\right]
\end{align}

The resulting flux expressions in the fixed crystal frame becomes
\begin{align}
    \left[ \begin{array}{c} J_{A} \\ J_{B} \end{array}\right] & = -\left[ \begin{array}{cc} L_{AA} & L_{AB}  \\ L_{AB} & L_{BB} \end{array} \right] \left[ \begin{array}{c} \nabla(\mu_{A}-\mu_{Va}) \\ \nabla(\mu_{B}-\mu_{Va}) \end{array}\right], 
\end{align}




\subsubsection{Interstitial diffusion with a fixed host lattice}

\begin{figure}
    \centering
    \includegraphics[width=7cm]{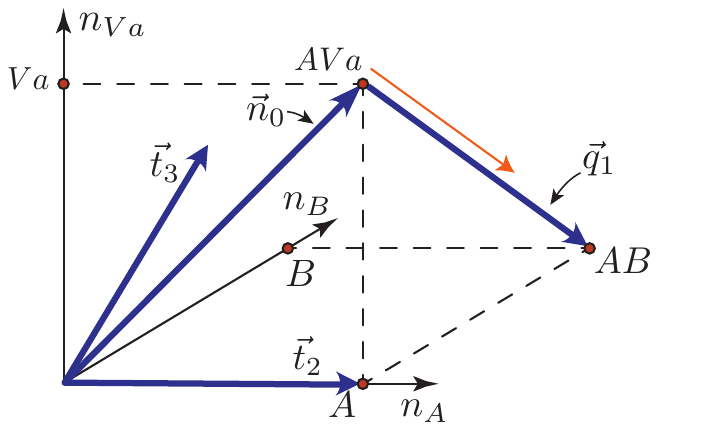}
    \caption{The space of allowed compositions for a crystal with one fixed host sublattice occupied by A atoms, and an interstitial sublattice occupied by B atoms and vacancies (Va). A choice of composition axis $\vec{q}^{\mathsf{T}}_1=[0,1,-1]$ with origin $\vec{n}^{\mathsf{T}}_0=[1,0,1]$ is shown along with the stoichiometic compositions at the extrema of the allowed composition space: AVa and AB, along with compositions A, B and Va as guides for the eye. This is consistent with the composition formula AB$_{x_1}$Va$_{1-{x_1}}$.}
    \label{fig:binary_interstitial_composition_axes_ABVa_1}
\end{figure}

Another common system is one in which there is a fixed sublattice of A atoms and an interstitial sublattice that allows vacancies and B atoms.
This crystal also contains three species, A, B and Va, and the concentration per unit cell can be expressed as the vector $\vec{n}^{\mathsf{T}}=[n_{A},n_{B},n_{Va}]$.
Clearly, this system behaves like a binary alloy on the interstitial sublattice, but with a fixed concentration of A atoms.
This makes for a useful system to use as an exercise because the results can be easily checked against intuition.

The choice of composition origin and axis vectors consistent with the composition formula AB$_{{x_1}}$Va$_{1-{x_1}}$ is
\begin{align}
    \vec{n}_0 & = \left[ \begin{array}{c} 1 \\ 0 \\ 1  \end{array} \right] & 
    \vec{q}_1 & = \left[\begin{array}{c}  0 \\ 1  \\ -1 \end{array} \right],
\end{align}
as illustrated in Figure \ref{fig:binary_interstitial_composition_axes_ABVa_1}.

The equations relating parametric composition, $\vec{x}$, and concentration per unit cell, $\vec{n}$, (Eqs.~\ref{eq:concentration_equation2}, \ref{eq:concentration_equation_R}, and \ref{composition_space_basis_R}) are 
\begin{align*}
    \vec{n} & =\vec{n}_0+\pmb{Q}\vec{x}, \\
    \vec{x} & = {\pmb{R}}^\mathsf{T}(\vec{n} - \vec{n}_0), \\
    {\pmb{R}}^\mathsf{T} & = ({\pmb{Q}}^\mathsf{T} {\pmb{Q}})^{-1} {\pmb{Q}}^\mathsf{T},
\end{align*}
which for this system and choice of composition axes take the values
\begin{align}
    \pmb{Q} & =\left[ \begin{array}{c} 0 \\ 1 \\ -1 \end{array} \right] &
    \pmb{R} & =\frac{1}{2}\left[ \begin{array}{c} 0 \\ 1 \\ -1 \\  \end{array} \right].
\end{align}

The null space of \pmb{Q} is the space of disallowed compositions, 
\begin{align}
    \pmb{T} & = \left[ \begin{array}{cc} 1 & 0 \\ 0 & \frac{1}{\sqrt{2}} \\ 0 & \frac{1}{\sqrt{2}} \end{array} \right], 
\end{align}
which can be used to define the constraints on the the concentration per unit cell (Eq.~\ref{eq:orthogonal_constraints1}), flux (Eq.~\ref{eq:orthogonal_constraints2}), and Onsager coefficients (Eq.~\ref{eq:L_constraints2}) according to
\begin{align*}
    \pmb{T}^{\mathsf{T}}(\vec{n}-\vec{n}_0) & =\vec{0}, \\
    \pmb{T}^{\mathsf{T}}\vec{J} & =\vec{0}, \\
    \Lonsager{\pmb{T}} & =\mathbf{0}. \\
\end{align*}

For this system and choice of composition axes, the concentration constraints are 
\begin{align}
    \left[ \begin{array}{ccc} 1 & 0 & 0 \\ 0 & \frac{1}{\sqrt{2}} & \frac{1}{\sqrt{2}} \end{array} \right]\left(\left[ \begin{array}{c} n_{A} \\ n_{B} \\ n_{Va} \end{array} \right]- \left[\begin{array}{c} 1 \\ 0 \\ 1 \end{array}\right]\right) & = \left[\begin{array}{c} 0 \\ 0 \end{array}\right],
\end{align}
which gives the expected result
\begin{equation}
  \begin{aligned}
    n_{A} & = 1, \\
    n_{B} + n_{Va} & = 1,
  \end{aligned}
\end{equation}

The flux constraints are
\begin{align}
    \left[ \begin{array}{ccc} 1 & 0 & 0 \\ 0 & \frac{1}{\sqrt{2}} & \frac{1}{\sqrt{2}} \end{array} \right] \left[ \begin{array}{c} J_{A} \\ J_{B} \\ J_{Va} \end{array} \right] & =\left[ \begin{array}{c} 0 \\ 0 \end{array} \right],
\end{align}
which gives the expected result
\begin{equation}
  \begin{aligned}
    J_{A} & = 0, \\
    J_{B} & = -J_{Va},
  \end{aligned}
\end{equation}

The Onsager coefficients constraints are
\begin{align}
    \left[ \begin{array}{ccc} L_{AA} & L_{AB} & L_{AVa} \\ L_{AB} & L_{BB} & L_{BVa} \\ L_{AVa} & L_{BVa} & L_{VaVa} \end{array} \right] \left[ \begin{array}{cc} 1 & 0 \\ 0 & \frac{1}{\sqrt{2}} \\ 0 & \frac{1}{\sqrt{2}} \end{array} \right] & = \left[ \begin{array}{cc} 0 & 0 \\ 0 & 0 \\ 0 & 0 \end{array} \right],
\end{align}
which gives the obvious result 
\begin{align}
    L_{AA} = L_{AB} = L_{AVa} = L_{AB} & = 0,
\end{align}
and also the expected result
\begin{align}
    L_{BB} = L_{VaVa} = -L_{BVa}.
\end{align}

The choice of composition space results in exchange chemical potentials (Eq.~\ref{eq:exchange_chemical_potentials}) defined as
\begin{align*}
\vec{\tilde{\mu}}=\pmb{Q}^{T}\vec{\mu}
\end{align*}
which take the values
\begin{align}
    \left[ \begin{array}{c} \tilde{\mu}_1 \end{array} \right] & = \left[ \begin{array}{c} 0 \\ 1 \\ -1 \end{array} \right] \left[ \begin{array}{c} \tilde{\mu}_{A} \\ \tilde{\mu}_{B} \\ \tilde{\mu}_{Va} \end{array} \right],
\end{align}
or equivalently
\begin{align}
    \tilde{\mu}_1 & = \mu_{B}-\mu_{Va}.
\end{align}

The flux expressions and Onsager coefficients can be projected onto the allowed composition space (Eqs.~\ref{eq:projected_flux_equations}, \ref{eq:flux_projection} and \ref{eq:projected_L_matrix}),
\begin{align*}
    \vec{\tilde{J}} & =-\tilde{\Lonsager}\nabla\vec{\tilde{\mu}}, \\ 
    \vec{\tilde{J}} & ={\pmb{R}}^{\mathsf{T}}\vec{J}, \\
    \tilde{\Lonsager} & ={\pmb{R}}^{\mathsf{T}}\Lonsager\pmb{R}.
\end{align*}
which for this system and choice of composition axes gives
\begin{align}
    \left[ \begin{array}{c} \tilde{J}_1 \end{array} \right] & = \frac{1}{2}\left[ \begin{array}{ccc} 0 & 1 & -1 \end{array} \right] \left[ \begin{array}{c} J_{A} \\ J_{B} \\ J_{Va} \end{array} \right],
\end{align}
or equivalently
\begin{align}
    \tilde{J}_1 & = \frac{1}{2}\left( J_{B} - J_{Va} \right),
\end{align}
and
\begin{align}
    \left[ \begin{array}{c} \tilde{L}_{11} \end{array} \right] & = \frac{1}{4}\left[ \begin{array}{ccc} 0 & 1 & -1 \end{array} \right] \left[ \begin{array}{ccc} L_{AA} & L_{AB} & L_{AVa} \\ L_{AB} & L_{BB} & L_{BVa} \\ L_{AVa} & L_{BVa} & L_{VaVa} \end{array} \right] \left[ \begin{array}{c} 0 \\ 1 \\ -1 \end{array} \right],
\end{align}
or equivalently
\begin{align}
    \tilde{L}_{11} & = \frac{1}{4}\left( L_{BB} + L_{VaVa} - 2L_{BVa} \right).
\end{align}

The exchange chemical potentials, projected fluxes, and projected Onsager coefficients can be combined with the constraints found previously to yield the simplified one-dimensional results
\begin{align}
    \tilde{J} & = -\tilde{L}\nabla \tilde{\mu} \\ 
    \tilde{J} & = \tilde{J}_1 = J_{B}, \nonumber \\
    \tilde{L} & = \tilde{L}_{11} = L_{BB}. \nonumber \\
    \tilde{\mu} & = \tilde{\mu}_1 = \mu_{B}-\mu_{Va}. \nonumber
\end{align}

\bibliography{references}

\end{document}